\documentclass[11pt]{article}
\usepackage{graphicx}
\usepackage{amsmath}
\usepackage{amsfonts}
\usepackage{amssymb}
\usepackage{epsfig}
\usepackage{times}
\def\tw{\textwidth}
\begin{document}


\begin{titlepage}

May 2006 \hfill

\vskip 4cm

\centerline{\bf TWO PARTICLE CORRELATIONS INSIDE ONE JET}
\medskip
\centerline{\bf  AT ``MODIFIED LEADING LOGARITHMIC APPROXIMATION''}
\medskip
\centerline{\bf  OF QUANTUM CHROMODYNAMICS}
\medskip
\centerline{\bf  I: EXACT SOLUTION OF THE EVOLUTION EQUATIONS AT SMALL
        $\boldsymbol{X}$}

\vskip 1cm

\centerline{Redamy Perez-Ramos
\footnote{E-mail: perez@lpthe.jussieu.fr}
}

\baselineskip=15pt

\smallskip
\centerline{\em Laboratoire de Physique Th\'eorique et Hautes Energies
\footnote{LPTHE, tour 24-25, 5\raise 3pt \hbox{\tiny \`eme} \'etage,
Universit\'e P. et M. Curie, BP 126, 4 place Jussieu,
F-75252 Paris Cedex 05 (France)}}
\centerline{\em Unit\'e Mixte de Recherche UMR 7589}
\centerline{\em Universit\'e Pierre et Marie Curie-Paris6; CNRS;
Universit\'e Denis Diderot-Paris7}

\vskip 2cm

{\bf Abstract}: We discuss correlations between two particles in jets 
at high energy colliders and exactly solve the MLLA evolution equations
in the small $x$ limit. We thus extend the Fong-Webber analysis 
to the region away from the hump of the single inclusive energy spectrum.
We give our results for LEP, Tevatron and LHC energies, and compare with 
existing experimental data.

\vskip 1 cm


\vfill

\null\hfil\epsfig{file=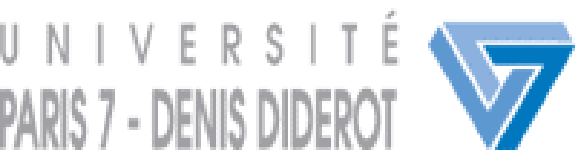, height=1.2cm,width=4.5cm}

\vskip .5cm


\end{titlepage}


\tableofcontents

\newpage

\section{INTRODUCTION}
\label{section:intro}

Perturbative QCD (pQCD) successfully predicts inclusive energy
spectra of particles in jets.  To this end it was enough to make one
step beyond the leading ``Double Logarithmic Approximation'' (DLA) which
is known to overestimate soft gluon multiplication, and to describe
parton cascades with account of first sub-leading single logarithmic
(SL) effects.
Essential SL corrections to DLA arise from:

$\ast$\  the running coupling $\alpha_s(k_\perp^2)$;

$\ast$\  decays of a parton into two with comparable
energies, $z\sim 1$ (the so called ``hard corrections'', taken care of
by employing exact DGLAP \cite{DGLAP} splitting functions);

$\ast$\  kinematical regions of successive parton decay angles of the same
order of magnitude, $\Theta_{i+1} \sim \Theta_i$. The solution
to the latter problem turned
out to be extremely simple namely, the replacement of the {\em
strong} angular ordering (AO), $\Theta_{i+1} \ll \Theta_i$,
imposed by gluon coherence in  DLA , by the {\em
exact} AO condition $\Theta_{i+1} \le \Theta_i$ (see \cite{EvEq}
and references therein).
The corresponding approximation is known as MLLA (Modified Leading
Logarithm Approximation) and embodies the next-to-leading correction, of 
order $\gamma_0^2$, to the parton evolution ``Hamiltonian'', 
$\gamma_0\propto\sqrt{\alpha_s}$ being the DLA multiplicity anomalous
dimension \cite{EvEq}.

So doing, single inclusive charged hadron spectra (dominated by pions)
were found to be mathematically similar to that of the MLLA parton
spectrum, with an overall proportionality coefficient ${\cal K}^{ch}$
normalizing partonic distributions to the ones of charged hadrons; ${\cal K}^{ch}$ 
depends neither on the jet hardness nor on the particle energy. This finding was
interpreted as an experimental confirmation of the Local
Parton--Hadron Duality hypothesis (LPHD) (for a review
see  \cite{DKTM}\cite{KO} and references therein). However, in the ratio of two particle distribution and the product of two single particle distributions that determine the correlation, this non-perturbative parameter cancels. Therefore, one expects this 
observable to provide a more stringent test of parton dynamics.
At the same time, it constitutes much harder a problem for the naive perturbative
QCD (pQCD) approach.
 
The correlation between two soft gluons  was tackled in DLA 
in \cite{DLA}.  The first realistic prediction with account
of next-to-leading (SL) effects was derived by Fong and Webber in
1990 \cite{FW}. They obtained the expression for the two
particle correlator in the kinematical region where both particles
were close in energy to the maximum ("hump") of the single inclusive
distribution.  In \cite{OPAL} this pQCD result was compared with the
OPAL $e^+e^-$ annihilation data at the $Z^0$ peak: the analytical
calculations were found to have largely overestimated the measured
correlations.

In this paper we use the formalism of jet generating functionals \cite{KUV}
to derive the MLLA evolution equations for particle
correlators (two particle inclusive distributions). We then use the
soft approximation for the energies of the two particle by neglecting terms
proportional to powers of $x_1,x_2\ll1$ ($x$ is the fraction of the jet
energy carried away by the corresponding particle).  Thus simplified, the
evolution equations can be solved  iteratively and their  solutions 
are given explicitly in terms of logarithmic derivatives of single particle
distributions.

This allows us to achieve two goals. First, we generalize the
Fong--Webber result by extending its domain of application to the full
kinematical range of soft particle energies.  Secondly, by doing this, we
follow the same logic as was applied in describing inclusive spectra
namely, treating {\em exactly} {\em approximate} evolution
equations. Strictly speaking, such a solution, when formally expanded,
inevitably bears sub-sub-leading terms that exceed the accuracy with
which the equations themselves were derived.  This logic, however, was
proved successful in the case of single inclusive spectra \cite{OPALTASSO}, 
which demonstrated that MLLA equations, though approximate, fully take into
account essential physical ingredients of parton cascading:
energy conservation, coherence, running coupling constant.
Applying the same logic to double inclusive distributions
should help to elucidate the problem of particle correlations in QCD
jets.

The paper is organized as follows. 

$\bullet$\quad in section \ref{section:evol} we recall the formalism of jet
generating functionals and their evolution equations; we specialize first
 to inclusive energy spectrum, and then to 2-particle correlations; 

$\bullet$\quad in section \ref{section:soft}, we solve exactly the evolution
equations in the low energy (small $x$) limit; how various
corrections are estimated and controlled is specially emphasized;

$\bullet$\quad section \ref{section:corglu} is dedicated to correlations in a
gluon jet; the equation to be solved iteratively is exhibited, and an
estimate of the order of magnitudes of various contributions is given;

$\bullet$\quad section \ref{section:quark} is dedicated to correlations in a
quark jet, and follows the same lines as section \ref{section:corglu};

$\bullet$\quad in section \ref{section:numer} we give all  numerical
results, for LEP-I,  Tevatron and LHC. They are commented, compared with
Fong-Webber for OPAL, but all detailed numerical investigations concerning
 the size of various corrections is postponed, for the sake of clarity,
to appendix \ref{section:numcorr};

$\bullet$\quad a conclusion summarizes this work.

Six appendices provide all necessary theoretical demonstrations and
numerical investigations.

$\bullet$\quad in appendix \ref{section:Gcorr} and \ref{section:Qcorr} 
we derive the exact solution of the evolution equations for the gluon and 
quark jet correlators;

$\bullet$\quad appendix \ref{section:inspiredDLA} is a technical
complement to subsection \ref{subsection:estimate};

$\bullet$\quad in appendix \ref{section:ESEE} we demonstrate the exact
solution of the MLLA evolution equation for the inclusive spectrum and
give analytic expressions for its derivatives;

$\bullet$\quad appendix \ref{section:numcorr} is dedicated to a
numerical analysis of all corrections that occur in the
iterative solutions of the evolution equations;

$\bullet$\quad in appendix \ref{section:DLAcomp} we perform a comparison
between DLA and MLLA correlators.

\section{EVOLUTION EQUATIONS FOR JET GENERATING FUNCTIONALS}
\label{section:evol}

\begin{figure}
\vbox{
\begin{center}
\epsfig{file=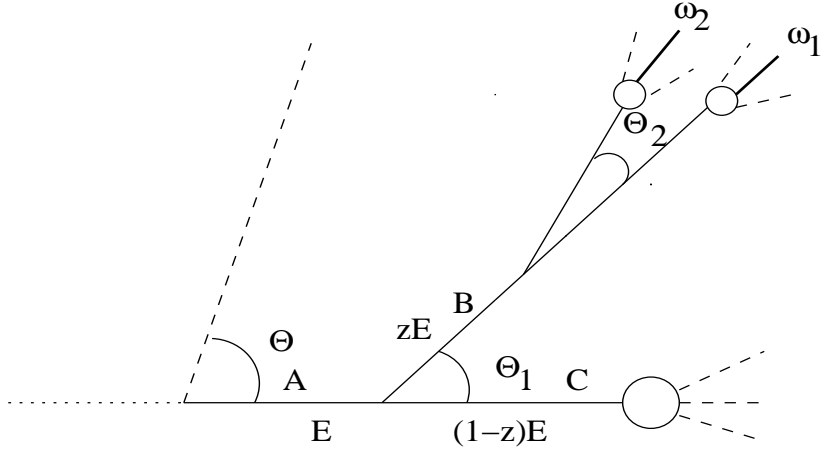, height=6truecm,width=11truecm}
\vskip .5cm
\caption{Two-particle correlations and Angular Ordering}
\label{fig:process}
\end{center}
}
\end{figure}

Consider (see Fig.~\ref{fig:process})
 a jet generated by a parton of type $A$ (quark or gluon)
with 4-momentum $p=(p_0\equiv E,\vec p)$.

 A generating functional $Z(E,\Theta;
\{u\})$ can be constructed \cite{KUV} that describes the azimuth averaged
 parton content of a jet of energy $E$ with a given opening half-angle $\Theta$;
by virtue of the exact angular ordering (MLLA),
it satisfies the following integro-differential evolution equation \cite{EvEq}

\medskip
\vbox{
\begin{eqnarray}
\frac{d}{d\ln\Theta}Z_A\left(p,\Theta;\{u\}\right)
\!\!&\!\!=\!\!&\!\!\frac{1}{2}\sum_{B,C}\int_{0}^{1}dz\>
\Phi_A^{B[C]}(z)\ \frac{\alpha_s\left(k^2_{\perp}\right)}{\pi} \cr
&&
\Big(Z_B\big(zp,\Theta;\{u\}\big)\ Z_C\big((1-z)p,\Theta;\{u\}\big)
-Z_A\big(p,\Theta;\{u\}\big)\Big); 
\label{eq:red1}
\end{eqnarray}
}
\medskip

in (\ref{eq:red1}), $z$ and $(1-z)$ are the energy-momentum fractions
carried away by the two
offspring of the $A\to B C$ parton decay described by the standard
one loop splitting functions

\medskip

\begin{eqnarray}
\label{eq:split}
&& \Phi_q^{q[g]}(z) = C_F\, \frac{1+z^2}{1-z} , \quad
\Phi_q^{g[q]}(z)=C_F\, \frac{1+(1-z)^2}{z} , \\\notag\\
\label{eq:cst}
&& \Phi_g^{q[\bar{q}]}(z) = T_R\left( z^2+(1-z)^2\right) , \quad
\Phi_g^{g[g]}(z)= 2C_A\left(\frac{1-z}{z}+ \frac{z}{1-z} +
z(1-z)\right),\\\notag\\
&& C_A=N_c,\quad C_F=(N_c^2-1)/2N_c,\quad T_R=1/2,
\end{eqnarray}

\medskip

where $N_c$ is the number of colors;
$Z_A$ in the integral in the r.h.s. of (\ref{eq:red1})
 accounts for 1-loop virtual corrections, which exponentiate into
Sudakov form factors.

\vskip 0.5cm

$\alpha_s(q^2)$ is the running coupling constant of QCD
\begin{equation}
\alpha_s(q^2)= \frac{4\pi}{4N_c\beta
\ln\displaystyle\frac{q^2}{\Lambda^2_{QCD}}},
\label{eq:alphas}
\end{equation}
where $\Lambda_{QCD} \approx$ a few hundred $MeV$'s is the intrinsic scale
of QCD, and
\begin{equation}
\beta = \frac{1}{4N_c}\Big(\frac{11}{3}N_c - \frac{4}{3} n_f T_R\Big)
\label{eq:beta}
\end{equation}
is the first term in the perturbative expansion of the $\beta$ function,
$n_f$ the number of light quark flavors.

If the radiated parton with 4-momentum $k=(k_0,\vec k)$
 is emitted with an angle $\Theta$ with respect to
the direction of the jet, one has ($k_\perp$ is the modulus of the
transverse trivector $\vec k_\perp$ orthogonal to the direction of the jet)
$k_\perp\simeq  |\vec k| \Theta \approx k_0 \Theta \approx 
        z E\Theta$ when $z \ll 1$  or $(1-z)E\Theta$ when $z\to 1$,
and a collinear cutoff $k_\perp\geq Q_0$ is imposed. 

In (\ref{eq:red1}) the symbol $\{u\}$ denotes a set of {\em probing
functions} $u_a(k)$ with $k$ the 4-momentum of a secondary parton of
type $a$.
The jet functional is normalized to the total jet production cross
section such that 
\begin{equation}
Z_A(p,\Theta;u\equiv1)=1;
\label{eq:norm}
\end{equation}
for vanishingly small opening
angle it reduces to the probing function of the single initial parton
\begin{equation}
Z_A(p,\Theta\to 0;\{u\})= u_A(k\equiv p).
\end{equation}

To obtain {\em exclusive} $n$-particle distributions one takes $n$
variational derivatives of $Z_A$ over $u(k_i)$ with appropriate
particle momenta, $i=1 \ldots n$, and sets $u\equiv 0$ after wards;
{\em inclusive} distributions are generated by taking variational
derivatives around $u\equiv 1$.

\subsection{Inclusive particle energy spectrum}
\label{subsection:incluspec}

The probability of soft gluon radiation off a color charge
(moving in the $z$ direction) has the polar angle dependence 

\medskip

\begin{equation*}
\frac{\sin\theta\,d\theta}{2(1-\cos\theta)}
= \frac{d\sin(\theta/2)}{\sin(\theta/2)}
 \simeq\frac{d\theta}{\theta};
\end{equation*}

\medskip

therefore, we choose the angular evolution parameter to be

\medskip

\begin{equation}
Y = \ln \frac{2E\sin(\Theta/2)}{Q_0}
\Rightarrow dY = \frac{d\sin(\Theta/2)}{\sin(\Theta/2)};
\label{eq:Ydef}
\end{equation}

\medskip

this choice accounts for finite angles ${\cal O}(1)$ up to
the full opening half-angle $\Theta=\pi$, at which

\medskip

\begin{equation*}
 Y_{\Theta=\pi}=\ln\frac{2E}{Q_0}
,
\end{equation*}

\medskip

where $2E$ is the center-of-mass annihilation energy of the process
$e^+e^- \to q\bar{q}$. 
For small angles (\ref{eq:Ydef}) reduces to

\medskip

\begin{equation}
Y\simeq \ln\frac{E\Theta}{Q_0}, \quad \Theta\ll 1,
\quad \frac{d}{dY} = \frac{d}{d\ln\Theta},
\label{eq:Ydef2}
\end{equation}

\medskip

where $E\Theta$ is the maximal transverse momentum of a parton inside the jet 
with opening half-angle $\Theta$. 

To obtain the inclusive energy distribution of parton $a$ emitted at angles smaller 
than $\Theta$ with momentum $k_a$, energy $E_a = x E$  in a jet $A$, {\em i.e.}
the fragmentation function
${D}_A^{\,a}(x,Y)$, we take the variational derivative of 
(\ref{eq:red1}) over $u_a(k)$ and set $u\equiv1$ (which also corresponds to
 $Z=1$) according to

\medskip

\begin{equation}
x{D_A^a}(x,Y)
= E_a\frac{\delta}{\delta u(k_a)}
Z_A\left(k,\Theta;\left\{u\right\}\right) \Big\vert_{u=1},
\end{equation}

\medskip

where we have chosen the variables $x$ and $Y$ rather than $k_a$ and $\Theta$.

Two configurations must be accounted for:
$B$ carrying away the fraction $z$ and $C$
the fraction $(1-z)$ of the jet energy, and the symmetric one in which the
role of $B$ and $C$ is exchanged. Upon functional differentiation they
give the same result, which cancels the factor $1/2$. The
system of coupled linear integro-differential equations that comes out is

\medskip

\begin{equation}
\frac{d}{dY} \,
x{D}^a_A(x,Y) = \int_0^1 dz
\sum_B \Phi_A^B(z)\,\frac{\alpha_s}{\pi}
 \left[
\frac{x}{z}  {D}^a_B\left(\frac{x}{z},Y+\ln z\right)
- \frac1{2} x{D}^a_A(x,Y) \right].
\label{eq:SIeveq}
\end{equation}

\medskip

We will be interested in the region of small $x$ where fragmentation functions behave as 
\begin{equation}
x{D} (x) \stackrel{x\ll1}{\sim} \rho(\ln x),
\label{eq:rho}
\end{equation}
with $\rho$ a smooth function of $\ln x$.
Introducing logarithmic {\em parton densities} 
\begin{equation}
Q= x {D}^{\,a}_Q (x,Y),\quad G=x {D}^{\,a}_G (x,Y),
\label{eq:QGdef}
\end{equation}
respectively for quark and gluon jets, we obtain from (\ref{eq:SIeveq}) 

\medskip

\begin{eqnarray}
Q_{y}\equiv \frac{dQ}{dy}\!\!&\!\!=\!\!&\!\!\int_0^1 dz\>  \frac{\alpha_s}{\pi} \>\Phi_q^g(z)\>
\bigg[ \Big(Q(1-z)-Q\Big) + G(z) \bigg],
\label{eq:qpr}\\
G_{y}\equiv \frac{dG}{dy}\!\!&\!\!=\!\!&\!\!\int_0^1 dz\> \frac{\alpha_s}{\pi} \>
\bigg[\Phi_g^g(z) \Big(G(z)-zG\Big)
+n_f\; \Phi_g^q(z)\, \Big(2Q(z)-G\Big) \bigg],
\label{eq:gpr}
\end{eqnarray}

\medskip

where, for the sake of clarity,
  we have suppressed $x$ and $Y$ and only kept
the dependence on the integration variable $z$, e.g.,
\begin{equation}
G(z) \equiv  \frac{x}{z}  {D}^a_G\bigg(\frac{x}{z}, Y + \ln z\bigg),
\label{eq:shortnot}
\end{equation}
such that
\begin{equation}
G=G(1),\quad Q=Q(1).
\label{eq:GQ1}
\end{equation}
Some comments are in order concerning these equations. 
\begin{itemize}
\item
We chose to express the derivative with respect to the jet opening angle
$\Theta$ on the l.h.s.'s of equations (\ref{eq:qpr})(\ref{eq:gpr})
 in terms of 
\begin{equation}
y\equiv  Y-\ell=\ln\frac{xE\Theta}{Q_0}=\ln\frac{E_a\Theta}{Q_0}, 
\quad \ell\equiv\ln\frac1x=\ln\frac{E}{E_a},
\label{eq:yelldef}
\end{equation}
instead of $Y$ defined in (\ref{eq:Ydef}). The variable $y$
is convenient for imposing the collinear cutoff condition
$k_\perp\simeq xE\sin\theta \ge Q_0$ since, for small angles, 
it translates simply into $y\ge 0$;

\item
to obtain (\ref{eq:qpr}) one proceeds as follows. When $B$ 
is a quark in (\ref{eq:SIeveq}) , since $A$ is also a quark, 
one gets two contributions:
the real contribution ${D_{B=q}^a}$ and the virtual one 
$-\frac{1}{2}{D_{A=q}^a}$;

\begin{itemize}
\item

in the virtual contribution, since $\Phi_q^q(z) =
\Phi_q^g(1-z)$, the sum over $B$ cancels the factor $1/2$;

\item

in the real contribution,  when it is a
quark, it is associated with $\Phi_q^q(z)$ and, when it is a gluon, with
$\Phi_q^g(z)$; we use like above the symmetry $\Phi_q^q(z) =
\Phi_q^g(1-z)$  to only keep one of the two, namely $\Phi_q^q$, at the
price of changing  the  corresponding $ D(z)$ into
${D}(1-z)$;

\end{itemize}
\item
to obtain (\ref{eq:gpr}), one goes along the following steps;
now $A=g$ and $B=q$ or $g$;

\begin{itemize}

\item
like before, the subtraction term does not depend on $B$ and is summed
over $B=q$ and $B=g$, with the corresponding splitting functions $\Phi_g^q$
and $\Phi_g^g$. In the term $\Phi_g^g$, using the property
$\Phi_g^g(z)=\Phi_g^g(1-z)$ allows us to replace $\frac12\int_0^1dz\Phi_g^g (z)=
\int_0^1z\Phi_g^g(z)$. This yields upon functional differentiation the $-zG$ term in
(\ref{eq:gpr}).
For $B=q$, $2n_f$ flavors 
($n_f$ flavors of quarks and $n_f$ flavors of
anti-quarks) yield identical contributions, which, owing to
the initial factor $1/2$ finally yields $n_f$;

\item
concerning the real terms,
$\Phi_g^g G$ in (\ref{eq:gpr}) comes directly from
$\Phi_g^{g}\frac{x}{z}{D_{g}^a}$ in (\ref{eq:SIeveq}).
For $B=q$, $2n_f$ flavors  of quarks and antiquarks contribute equally since at 
$x\ll1$ sea quarks are produced via gluons
\footnote{accompanied by  a relatively small fraction
 ${\cal O}(\sqrt{\alpha_s})$ of (flavor singlet) sea quark pairs,
 while the valence (non-singlet) quark
distributions are suppressed as ${\cal O}(x)$.}.
This is why we have multiplied
$Q(z)$ by $2n_f$  in (\ref{eq:gpr}).

\end{itemize}
\end{itemize}
Now we recall that both splitting functions $\Phi_q^g(z)$ and $\Phi_g^g$ 
are singular at $z=0$; the symmetric gluon-gluon splitting $\Phi_g^g(z)$ 
is singular at $z=1$ as well.
The latter singularity in (\ref{eq:gpr}) gets regularized by the
factor $\big(G(z)-zG\big)$ which vanishes at $z\to1$.
This regularization can be made explicit as follows

$$
\int_0^1 dz \Phi_g^g(z)\> \Big(G(z)-zG\Big)
\equiv \int_0^1 dz \Phi_g^g(z)\> \bigg[(1-z)G(z) + z\Big(G(z)-G\Big)\bigg];
$$
since $\Phi_g^g(z) = \Phi_g^g(1-z)$, while leaving the first term
$\int_0^1 dz \Phi_g^g(z)(1-z)G(z)$ unchanged, we can rewrite the second
$$
\int_0^1 dz \Phi_g^g(z)z\Big(G(z)-G\Big) =
\int_0^1 dz \Phi_g^g(z)(1-z)\Big(G(1-z)-G\Big),
$$ 
such that, re-summing the two, $(1-z)$ gets factorized and one gets
\begin{equation}
 \int_0^1 dz \Phi_g^g(z)\> \Big(G(z)-zG\Big) =
\int_0^1 dz \Phi_g^g(z)(1-z)\bigg[ G(z) + \Big(G(1-z) -G\Big)\bigg].
\label{eq:regul1}
\end{equation}

Terms proportional to $G(z)$ on r.h.s.'s of equations 
(\ref{eq:qpr})(\ref{eq:gpr})  remain singular at $z\!\to\! 0$ and   
produce enhanced contributions due to the logarithmic integration
over the region $x\ll z\ll 1$. 

\medskip

Before discussing the MLLA evolution equations following from (\ref{eq:qpr})
and (\ref{eq:gpr}), let us derive similar equation for two particle correlations
inside one jet.

\subsection{Two parton correlations}
\label{subsection:2pc}

We study correlation between two particles with fixed energies $x_1=\omega_1/E$, $x_2=\omega_2/E$ ($x_1>x_2$) emitted at arbitrary angles $\Theta_1$ and $\Theta_2$
smaller than the jet opening angle $\Theta$. If these partons are emitted in a 
cascading process, then $\Theta_1\geq\Theta_2$ by the AO property; see 
Fig.~\ref{fig:process}.

\subsubsection{Equations}
\label{subsub:eqs}

Taking the second variational derivative of (\ref{eq:red1})
with respect to $u(k_1)$ and $u(k_2)$, one gets a system of equations
for the two-particle distributions $G^{(2)}$ and $Q^{(2)}$
in gluon and quark jets, respectively:

\vbox{
\begin{eqnarray}
\label{eq:Q2pr}
Q^{(2)}_{y} \!\!\!&\!\!=\!\!&\!\!\!  \int dz\> \frac{\alpha_s}{\pi}\>
 \Phi_q^g(z)\bigg[ G^{(2)}(z)\!+\! \Big(Q^{(2)}(1\!-\!z)\!\!-Q^{(2)}\Big) 
  + G_1(z)Q_2(1\!-\!z)\! +\! G_2(z)Q_1(1\!-\!z) \bigg],\\\notag\\
\label{eq:G2pr}
 G^{(2)}_{y} \!\!\!&\!\!=\!\!&\!\!\!  \int dz\>  \frac{\alpha_s}{\pi}\> \Phi_g^g(z)
\bigg[ \Big(G^{(2)}(z)\!-\!zG^{(2)}\Big) 
  + G_1(z)G_2(1\!-\!z) \bigg] \cr
 && \hskip 3cm +  \int dz\> \frac{\alpha_s}{\pi}\> n_f\Phi^q_g(z)
\bigg[ \Big(2Q^{(2)}(z)\!-\!G^{(2)}\Big)\! +\! 2Q_1(z)Q_2(1\!-\!z) \bigg] .
\end{eqnarray}
}

Like before, the notations have been lightened to a maximum, such that
 $Q^{(2)} = Q^{(2)}(z=1),\ G^{(2)} = G^{(2)}(z=1)$.  More details
about the variables on which $Q^{(2)}$ depend are given in subsection
\ref{subsection:MLLAcor}. 
Now using (\ref{eq:qpr}) we construct the $y$-derivative of the
product of single inclusive spectra. Symbolically, 

\vbox{
\begin{eqnarray}
(Q_1Q_2)_{y}\!\! &\!\!=\!\!&\!\! Q_2\int_0^1 dz \frac{\alpha_s}{\pi}\Phi_q^g(x)
\Big[\big(Q_1(1-z)-Q_1\big)+G_1(z)\Big] \cr
         && \hskip 3cm +  Q_1 \int_0^1 dz \frac{\alpha_s}{\pi}\Phi_q^g(x)
\Big[\big(Q_2(1-z)-Q_2\big)+G_2(z)\Big].
\label{eq:Qprod}
\end{eqnarray}
}

Subtracting this expression from (\ref{eq:Q2pr}) we get

\medskip

\vbox{
\begin{eqnarray}
\label{eq:Q2prsub}
(Q^{(2)}-Q_1Q_2)_{y}  \!\!&\!\!=\!\!& \!\! \int \!dz\>\frac{\alpha_s}{\pi}\> 
 \Phi_q^g(z)\>\bigg[ G^{(2)}(z)+ \Big(Q^{(2)}(1-z)-Q^{(2)}\Big) \cr
&& \hskip -1cm + \Big(G_1(z)-Q_1\Big)\Big(Q_2(1-z) - Q_2\Big) +  \Big(G_2(z)-
Q_2\Big)\Big(Q_1(1-z)- Q_1\Big)\bigg].
\end{eqnarray}
}

\medskip

For the gluon jet,  making use of (\ref{eq:gpr})  we  analogously obtain
from (\ref{eq:G2pr})

\medskip

\vbox{
\begin{eqnarray}
\label{eq:G2prsub}
(G^{(2)}-G_1G_2)_{y} \!\!&\!\!=\!\!&\!\!
\int dz\> \frac{\alpha_s}{\pi}
\Phi_g^g(z)\>\bigg[ \Big(G^{(2)}(z)-zG^{(2)}\Big) 
  + \Big(G_1(z)-G_1\Big)\Big(G_2(1-z)-G_2\Big) \bigg] \cr
\!\!&\!\!+\!\!&\!\!  \int dz\>\frac{\alpha_s}{\pi}\, n_f\Phi_g^q(z)\>
\bigg[ 2\Big(Q^{(2)}(z)-Q_1(z)Q_2(z)\Big) - \Big(G^{(2)}-G_1G_2\Big)\cr
\!\!&\!\! +\!\!&\!\! \Big(2Q_1(z)-G_1\Big)\Big(2Q_2(1-z)-G_2\Big) \bigg].
\end{eqnarray}
}
The combinations on the l.h.s.'s of (\ref{eq:Q2prsub}) and (\ref{eq:G2prsub}) form {\em correlation functions} which vanish when particles 1 and 2 are produced independently.
They represent the combined probability of emitting particle 2 with $\ell_2, y_2,\ldots$ when particle 1 with $\ell_1,y_1,\ldots$ is emitted, too.
This way of representing the r.h.s.'s of the equations is convenient
for estimating the magnitude of the various terms.

\section{SOFT PARTICLE APPROXIMATION}
\label{section:soft}

In the standard DGLAP region $x={\cal O}(1)$
($\ell={\cal O}(0)$), the $x$ dependence of parton distributions is fast while scaling
violation is small

\begin{equation}
\frac{\partial_\ell {D}_{G,Q}(\ell,y)}{D_{G,Q}}
\equiv \psi_{\ell} = {\cal O}(1),\qquad
\frac{\partial_y
{D}_{G,Q}(\ell,y)}{D_{G,Q}} \equiv \psi_{y} =
{\cal O}(\alpha_s).
\end{equation}

\medskip

With $x$ decreasing, the running coupling gets enhanced while the
$x$-dependence slows down so that, in the kinematical region of the
{\em maximum} ("hump") of the inclusive spectrum the two logarithmic
derivatives become of the same order:
\begin{equation}
 \psi_{y} \sim \psi_{\ell} = {\cal O}(\sqrt{\alpha_s}), \quad
y\simeq \ell \simeq \textstyle {\frac12} Y.
\end{equation}
This allows to  significantly simplify  the equations for inclusive
spectra (\ref{eq:qpr})(\ref{eq:gpr}) and two particle correlations
(\ref{eq:Q2prsub})(\ref{eq:G2prsub}) for soft particles, $x_i\ll 1$,
which determine the bulk of parton multiplicity in jets. 
We shall estimate various contributions to evolution equations 
in order to single out the leading and first sub-leading terms in $\sqrt{\alpha_s}$
to construct the MLLA equations.

\subsection{MLLA spectrum}
\label{subsection:MLLAspec}

We start by recalling the logic of the MLLA analysis of the inclusive
spectrum. In fact (\ref{eq:qpr})(\ref{eq:gpr}) are
identical to the DGLAP evolution equations but for one detail: the shift
 $\ln z$ in the variable $Y$ characterizing the evolution of the jet
hardness $Q$.  Being the consequence of
exact angular ordering, this modification is negligible, within leading log
accuracy in $\alpha_s Y$, for energetic partons when 
 $|\ln z| < |\ln x| ={\cal O}(1)$. For soft particles, however,
ignoring this effect amounts to 
corrections of order ${\cal O}((\alpha_s\ln^2x)^n)$ that drastically
modify the character of the parton yield in time-like jets as compared
with space-like deep inelastic scattering (DIS) parton distributions. 

The MLLA logic consists of keeping the leading term  and the first
next-to-leading term in the right hand sides of  evolution equations
(\ref{eq:qpr})(\ref{eq:gpr}). 
Meanwhile, the combinations $\Big(Q(1-z)-Q\Big)$ in (\ref{eq:qpr}) and
$\Big(G(1-z)-G\Big)$ in (\ref{eq:regul1}) produce next-to-MLLA
corrections that can be omitted; 
indeed, in the small-$x$ region the parton densities $G(x)$ and $Q(x)$
are smooth functions (see \ref{eq:rho}) of $\ln x$ and we can estimate, say, $G(1-z)-G$,
using (\ref{eq:rho}), as
\begin{equation*}
 G(1-z)-G \equiv
 G\Big(\frac{x}{1-z}, Y+\ln(1-z) \Big) -  G\big(x,Y \big)
 \simeq   \psi_{\ell}\;  G\; \ln(1-z).
\end{equation*}
Since $\psi_{\ell} \sim \sqrt{\alpha_s}$ (see \ref{eq:psimodel}), combined with $\alpha_s$
 this gives a next-to-MLLA  correction 
${\cal O}(\gamma_0^3)$ to the r.h.s.\ of (\ref{eq:gpr}).
Neglecting these corrections we arrive at 
\begin{eqnarray}
\label{eq:qappr}
Q_{y}  \!\!&\!\!=\!\!&\!\!  \int_x^1 dz\>  \frac{\alpha_s}{\pi} \>\Phi_q^g(z)  G(z), \\
\label{eq:gappr}
G_{y}\!\!&\!\!=\!\!&\!\! \int_x^1 dz\> \frac{\alpha_s}{\pi} \bigg[(1-z)\Phi_g^g(z) G(z)
+ n_f \Phi_g^q(z) \Big(2Q(z)-G\Big)\bigg].
\end{eqnarray}

To evaluate (\ref{eq:qappr}), we rewrite (see (\ref{eq:split}))
$$
\Phi_q^g(z) = C_F\left(\frac{2}{z} + z-2\right);
$$
the singularity in $1/z$
yields the leading (DLA) term; since $G(z)$ is a smoothly varying function
of $\ln z$ (see (\ref{eq:rho})(\ref{eq:QGdef})),
the main $z$ dependence of this non-singular
part of the integrand we only slightly alter by replacing
$(z-2) G(z)$ by $(z-2)G$, which yields
\footnote{ since $x\ll1$, the lower bound of integration is set to ``$0$'' in the 
sub-leading pieces of (\ref{eq:qappr}) and (\ref{eq:gappr})}
\begin{eqnarray}
Q_{y}=\int_x^1 dz\>  \frac{\alpha_s}{\pi} C_F
\bigg(\frac{2}{z} G(z) + (z-2)G\bigg)
\!\!&\!\!=\!\!&\!\! \frac{C_F}{N_c}\int_x^1 \frac{dz}z\> \frac{2N_c\alpha_s}{\pi}
G(z) -\frac{3}{4}\frac{C_F}{N_c}\frac{2N_c\alpha_s}{\pi} G
\label{eq:qap2}
\end{eqnarray}
where $\alpha_s=\alpha_s(\ln z)$ in the integral term while in the second, 
it is just a constant. To get the last term in (\ref{eq:qap2}) we used
\begin{equation}\label{eq:int32}
\int_0^1 dz(z-2) = -\frac{3}{2}.
\end{equation}

To evaluate (\ref{eq:gappr}) we go along similar steps. $\Phi_g^q$ being a
regular function of $z$, we replace $2Q(z)-G$ with $2Q-G$;
$\Phi_g^g(z)$ also reads (see (\ref{eq:split}))

$$
\Phi_g^g(z)=2C_A\bigg(\frac{1}{z(1-z)} -2 + z(1-z)\bigg);
$$

the singularity in $1/(1-z)$ disappears, the one in $1/z$ we leave unchanged,
and in the regular part we replace $G(z)$ with $G$. This yields

\begin{eqnarray}\nonumber
G_{y}\!\!&\!\!=\!\!&\!\! \int_x^1 dz\> \frac{\alpha_s}{\pi} \bigg[2C_A \bigg(
\frac1z G(z) +(1-z)\Big(-2 + z(1-z)\Big)G\bigg)
+ n_f T_R\Big(z^2 + (1-z)^2\Big)\Big(2Q-G\Big)\bigg]\\\nonumber\\
\!\!&\!\!=\!\!&\!\!2C_A \int_x^1 \frac{dz}z\> \frac{\alpha_s}{\pi}
G(z) - \bigg(\frac{11}{6}C_A + \frac{2}{3}n_f T_R\bigg)\frac{\alpha_s}{\pi}G
+\frac{4}{3}n_f T_R\frac{\alpha_s}{\pi}\; Q;\label{eq:gap2}
\end{eqnarray}
the comparison of the singular leading (DLA) terms of (\ref{eq:qap2}) and
(\ref{eq:gap2}) shows that

\begin{equation}
Q \stackrel{DLA}{=} \frac{C_F}{C_A}G,
\label{eq:DLAratio}
\end{equation}

which one uses to replace $Q$ accordingly, in the last (sub-leading)
term of (\ref{eq:gap2}) (the corrections would be next-to-MLLA (see \ref{eq:ratio}) 
and can be neglected).
This yields the MLLA equation for $G$ where we set $C_A=N_c$:
\begin{eqnarray}
\label{eq:gap3}
G_{y}\!\!&\!\!=\!\!&\!\!
\int_x^1 \frac{dz}z\> \;\frac{2N_c\alpha_s}{\pi}G(z) -a\frac{2N_c\alpha_s}{\pi} G
\end{eqnarray}
with
\begin{equation}
a = \frac{11}{12} + \frac{n_f T_R}{3N_c}\bigg(1-\frac{2C_F}{N_c}\bigg)
\>=\> \frac{1}{4N_c}\bigg[\frac{11}{3}N_c + \frac{4}{3}n_f T_R
 \bigg(1-\frac{2C_F}{N_c}\bigg)\bigg]\stackrel{n_f=3}{=}0.935.
\label{eq:adef}
\end{equation}

$a$ parametrizes ``hard'' corrections to soft gluon multiplication and
sub-leading $g\to q\bar{q}$ splittings
\footnote{The present formula for $a$ differs from  (47) in
\cite{PerezMachet} because, there, we defined $T_R=n_f/2$, instead of
$T_R=1/2$ here.}.

We define conveniently the integration variables $z$ and $\Theta'$ satisfying
$x\leq z\leq 1$ and $xE/Q_0\leq\Theta'\leq\Theta$
\footnote{the lower bound on $\Theta'$ follows from the kinematical condition
$k_{\perp}\approx xE\Theta'\geq Q_0$} through
\begin{equation}
\ell'=\ln\frac{z}{x}\quad \text{and}\quad y'=\ln\frac{xE\Theta'}{Q_0}
\end{equation}

The condition $x\leq z\leq 1$ is then equivalent to $0\leq\ell'\leq\ell$ and
$xE/Q_0\leq\Theta'\leq\Theta$ is $0\!\leq\! y'\!\leq\!y$. Therefore,

$$
\int_x^1\frac{dz}z=\int_0^{\ell}d\ell',\qquad\int_{Q_0/xE}^{\Theta}\frac{d\Theta'}
{\Theta'}=\int_0^{y}dy.
$$

\vskip 0.5cm

We end up  with the following system of integral equations of (\ref{eq:qap2})
and (\ref{eq:gap3}) for the
spectrum of one particle inside a quark and a gluon jet

\begin{equation}
Q(\ell,y)= \delta(\ell) + \frac{C_F}{N_c}\bigg[\int_0^\ell d\ell'\int_0^y dy'
\gamma_0^2(\ell'+y')\Big( G(\ell',y')
-\frac34\delta(\ell'-\ell) \Big) G(\ell',y')\bigg],
\label{eq:solq}
\end{equation}

\begin{equation}
G(\ell,y) = \delta(\ell)
+\int_0^{\ell} d\ell'\int_0^{y} dy' \gamma_0^2(\ell'+y')\Big(
 1  -a\delta(\ell'-\ell) \Big) G(\ell',y')
\label{eq:solg}
\end{equation}

\vskip 0.5cm

that we write in terms of the anomalous dimension
\begin{equation}
\gamma_0 = \gamma_0(\alpha_s) = \sqrt{\frac{2N_c\alpha_s}{\pi}}
\label{eq:gammadef}
\end{equation}
which determines the rate of multiplicity growth with energy.
Indeed, using (\ref{eq:alphas}), (\ref{eq:yelldef}) and (\ref{eq:gammadef}) one gets

$$
\gamma_0^2(zE\Theta')=\frac{1}{\beta\ln\left(\displaystyle{\frac{zE\Theta'}
{\Lambda_{QCD}}}\right)}=
\frac1{\beta\left(\ln\displaystyle{\frac{z}{x}}+\displaystyle{\frac{xE\Theta'}{Q_0}}+
\lambda\right)}\equiv\gamma_0^2(\ell'+y')=
\frac{1}{\beta(\ell'+y'+\lambda)}.
$$

with $\lambda=\ln(Q_0/\Lambda_{QCD})$.
In particular, for $z=1$ and $\Theta'=\Theta$ one has

\begin{equation}
\gamma_0^2 = \frac1{\beta(\ell+y+\lambda)}=\frac{1}{\beta(Y+\lambda)},\qquad \ell+y=Y.
\label{eq:gammabeta}
\end{equation}

\medskip

The DLA relation (\ref{eq:DLAratio}) can be refined to

\begin{eqnarray}
Q(\ell, y) =
\frac{C_F}{C_A}
\Big[1 +\left(a-{\textstyle{\frac34}}\right)\Big(\psi_{\ell}+
a\big(\psi_{\ell}^2+\psi_{\ell\,\ell}\big)\Big)
+ {\cal O}(\gamma_0^2)\Big]G(\ell, y),
\label{eq:ratio}
\end{eqnarray}
where
$$\psi_{\ell}=\frac{1}{G(\ell,y)}
\frac{dG(\ell,y)}{d\ell},\quad \psi_{\ell}^2+\psi_{\ell\,\ell}=
\frac1{G(\ell,y)}\frac{d^2G(\ell,y)}{d\ell^2}.
$$ 

Indeed subtracting (\ref{eq:solg}) and (\ref{eq:solq}) gives
\begin{equation}
\label{eq:ratiogq}
Q(\ell,y) - \frac{C_F}{N_c}G(\ell,y) = \frac{C_F}{N_c}\Big(a-\frac34\Big)
\int_0^y dy' \gamma_0^2 G(\ell,y');
\end{equation}

iterating twice (\ref{eq:solg}) yields

$$
\int_0^y dy' \gamma_0^2 G(\ell,y')=G_{\ell} + aG_{\ell\,\ell}+{\cal O}(\gamma_0^2)=
G(\ell,y)\Big(\psi_{\ell}+a\big(\psi_{\ell}^2+\psi_{\ell\,\ell}\big)\Big)+ {\cal O}(\gamma_0^2)
$$
which is then plugged in (\ref{eq:ratiogq}) to get (\ref{eq:ratio}). 
$\psi_{\ell}^2+\psi_{\ell\,\ell}$ can be easily estimated from subsection \ref{subsection:estimate} to be ${\cal O}(\gamma_0^2)$.
In MLLA, (\ref{eq:ratio}) reduces to 

\begin{equation}
Q(\ell, y) =
\frac{C_F}{C_A}
\Big[1 +\left(a-{\textstyle{\frac34}}\right)\psi_{\ell}(\ell,y)+{\cal O}(\gamma_0^2)\Big]
G(\ell, y).
\label{eq:ratioMLLA}
\end{equation}

\subsection{MLLA correlation}
\label{subsection:MLLAcor}

We estimate analogously the magnitude of various terms on the r.h.s.
of (\ref{eq:Q2prsub}) and (\ref{eq:G2prsub}).
Terms proportional to $Q_2(1-z)-Q_2$ and to  $Q_1(1-z)-Q_1$ in the second
line of (\ref{eq:Q2prsub}) will produce next-to-MLLA corrections that we
drop out.  In the first line, $Q^{(2)}(1-z)-Q^{(2)}$ 
($Q^{(2)}(z)$ is also a smooth function of $\ln z$) will also produce
higher order corrections that we neglect. We get

\begin{equation}
(Q^{(2)}-Q_1Q_2)_{y}=
 \int_{x_1}^1 dz\>\frac{\alpha_s}{\pi}\>  \Phi_q^g(z)\> G^{(2)}(z),
\label{eq:QMLLA}
\end{equation}

where we consider $z\!\geq\! x_1\! \geq\! x_2$.
In the first line of (\ref{eq:G2prsub}) we drop for identical reasons the
term proportional to $G_2(1-z)-G_2$, and the term $G^{(2)}(z)-zG^{(2)}$ is
regularized in the same way as we did for $G(z)-zG$ in (\ref{eq:gpr}).
In the second non-singular line, we use the smooth behavior
of $\phi_g^q(z)$ to neglect the $z$ dependence in all $G^{(2)}$, $Q^{(2)}$,
$G$ and $Q$ so that it factorizes and gives

\begin{eqnarray}\nonumber
(G^{(2)}-G_1G_2)_{y} \!\!&\!\!=\!\!&\!\!
\int_{x_1}^1 dz\> \frac{\alpha_s}{\pi}\, (1-z)\Phi_g^g(z)\> G^{(2)}(z)\\\nonumber\\
&&\hskip -3cm +  \int_0^1 dz\>\frac{\alpha_s}{\pi}\, n_f\Phi_g^q(z)\>
\bigg[ 2\big(Q^{(2)}-Q_1Q_2\big) - \big(G^{(2)}-G_1G_2\big) 
 +   (2Q_1-G_1)\,(2Q_2-G_2) \bigg].
\label{eq:GMLLA}
\end{eqnarray}

At the same level of approximation, we use the leading order relations

\begin{equation}
 Q_i = \frac{C_F}{N_c}\, G_i,  \qquad
 Q^{(2)}-Q_1Q_2 = \frac{C_F}{N_c} \,\left(G^{(2)}-G_1G_2\right);
\label{eq:app2}
\end{equation}
the last will be proved consistent in the following.
This makes the equation for the correlation in the gluon jet self
contained, we then get
\begin{eqnarray}\nonumber
(G^{(2)}-G_1G_2)_{y} \!\!&\!\!=\!\!&\!\!
\int_{x_1}^1 dz\> \frac{\alpha_s}{\pi}\, (1-z)\Phi_g^g(z)\> G^{(2)}(z) \\\nonumber\\
&&\hskip -3cm +  \int_0^1 dz\>\frac{\alpha_s}{\pi}\, n_f\Phi_g^q(z)\>
\bigg(2\frac{C_F}{N_c}-1\bigg)\bigg[ \big(G^{(2)}-G_1G_2\big) 
 + \bigg(2\frac{C_F}{N_c}-1\bigg)  G_1G_2) \bigg].
\label{eq:GMLLA2}
\end{eqnarray}

Like for the spectra, we isolate the singular terms $2C_F/z$ and $2C_A/z(1-z)$
of the splitting functions $\phi_q^g$ and $\phi_g^g$ respectively
(see(\ref{eq:split}) and (\ref{eq:cst})). We then write (\ref{eq:QMLLA}) and 
(\ref{eq:GMLLA2}) as follows
\begin{eqnarray}\label{eq:nonsing}
(Q^{(2)}-Q_1Q_2)_{y}=
 \int_{x_1}^1 dz\>\frac{\alpha_s}{\pi}\> 2C_F 
\bigg[\frac1zG^{(2)}(z) + \frac{1}{2}(z-2) G^{(2)}\bigg],
\end{eqnarray}
\begin{eqnarray}
(G^{(2)}-G_1G_2)_{y} \!\!&\!\!=\!\!&\!\!
\int_{x_1}^1 dz\> \frac{\alpha_s}{\pi}\, 
2C_A \bigg[\frac1z G^{(2)}(z) + (1-z)\Big(-2 +z(1-z)\Big)G^{(2)}\bigg]\notag\\\notag\\
&&\hskip -3.5cm +  \int_0^1 dz\>\frac{\alpha_s}{\pi}\, n_f
T_R\Big[z^2+(1-z)^2\Big]
\bigg(2\frac{C_F}{N_c}-1\bigg)\bigg[ \big(G^{(2)}-G_1G_2\big) 
 + \bigg(2\frac{C_F}{N_c}-1\bigg)  G_1G_2 \bigg]\label{eq:nonsingbis},
\end{eqnarray}

which already justifies {\em a posteriori} the last equation in
(\ref{eq:app2}).
One then proceeds with the $z$ integration of the polynomials that occur in
the non-singular terms (that of (\ref{eq:nonsing}) was already written in 
(\ref{eq:int32})). For the term $\propto G^{(2)}$ which we factorize by $2C_A$,
 we find (see (\ref{eq:adef}) for the expression of $a$)

\begin{eqnarray}
\int_0^1 dz\bigg[(1-z)\Big(\!-2 +z(1-z)\Big)
+ \frac{n_fT_R}{2C_A}\Big(z^2+(1-z)^2\Big)
\bigg(2\frac{C_F}{N_c}-1\bigg)\bigg] = -a,
\end{eqnarray}
while in the one $\propto G_1G_2$ we have simply

\begin{eqnarray}\label{eq:g1g2}
\frac{n_fT_R}{C_A}\bigg(1-2\frac{C_F}{N_c}\bigg)\bigg(1-\frac{C_F}{N_c}\bigg)
\int_0^1 dz\left[z^2+(1-z)^2\right] = \frac{2n_fT_R}{3C_A}\bigg(1-2\frac{C_F}{N_c}\bigg)\bigg(1-\frac{C_F}{N_c}\bigg).
\end{eqnarray}
Introducing

\begin{equation}
b = \frac{11}{12} - \frac{n_fT_R}{3N_c}
\left(1-\frac{2C_F}{N_c}\right)^2
= \frac{1}{4N_c}\bigg[\frac{11}{3}N_c -\frac{4}{3}n_f T_R
\bigg(1-2\frac{C_F}{N_c}\bigg)^2\bigg]\stackrel{n_f=3}{=}0.915
\label{eq:bdef}
\end{equation}

allows us to express (\ref{eq:g1g2}) with $C_A=N_c$ as

\begin{equation}
a-b=
\frac{2n_fT_R}{3N_c}\bigg(1-\frac{2C_F}{N_c}\bigg)\bigg(1-\frac{C_F}{N_c}\bigg)
\stackrel{n_f=3}=0.02,
\label{eq:a-b}
\end{equation}

such that (\ref{eq:nonsing}) and (\ref{eq:nonsingbis}) can be easily rewritten 
in the form

\begin{equation}\label{eq:eqqq}
\left(Q^{(2)}-Q_1Q_2\right)_y=\frac{C_F}{N_c}\int_{x_1}^1\frac{dz}z
\frac{2N_c\alpha_s}{\pi}G^{(2)}(z)-\frac34\frac{C_F}{N_c}\frac{2N_c\alpha_s}
{\pi}G^{(2)},
\end{equation}

\begin{equation}\label{eq:eqgg}
\left(G^{(2)}-G_1G_2\right)_y=\int_{x_1}^1\frac{dz}z
\frac{2N_c\alpha_s}{\pi}G^{(2)}(z)-a\frac{2N_c\alpha_s}{\pi}G^{(2)}+
(a-b)\frac{2N_c\alpha_s}{\pi}G_1G_2.
\end{equation}

Again, $\alpha_s=\alpha_s(\ln z)$ in the leading contribution while in the
sub-leading ones it is a constant.
We now introduce the following convenient variables and notations
to rewrite correlation evolution equations

\begin{eqnarray}
\ell_{i} = \ln\frac{1}{x_{i}}= \ln\frac{E}{\omega_{i}},\quad i=1,2
\end{eqnarray}

\begin{equation}
y_{i}= \ln\frac{\omega_{i}\Theta}{Q_0} =
\ln\frac{x_{i}E\Theta}{Q_0}= Y - \ell_{i}\quad \text{and} \quad
\eta=\ln\frac{x_1}{x_2}=\ell_2-\ell_1=y_1-y_2>0.
\label{eq:vars}
\end{equation}

\medskip

The transverse momentum of parton with energy $zE$ is $k_{\perp}\approx zE\Theta_1$. We 
conveniently define the integration variables $z$ and $\Theta_1$ satisfying
$x_1\!\leq\! z\!\leq\!1$ and
$\Theta_2\!\leq\!\Theta_1\!\leq\!\Theta$ with $\Theta_2\!\geq\!(\Theta_2)_{min}\!=\!Q_0/\omega_2$
through

\begin{equation}\label{eq:intvar}
\ell=\ln\frac{z}{x_1},\qquad y=\ln\frac{x_2E\Theta_1}{Q_0},
\end{equation}

then we write

\begin{equation}\label{eq:alpgasbis}
\gamma_0^2(zE\Theta_1)=
\frac{1}{\beta\left(\ln\displaystyle{\frac{z}{x_1}}+
\ln\displaystyle{\frac{x_2E\Theta_1}{Q_0}}+
\ln\displaystyle{\frac{x_1}{x_2}}+\lambda\right)}
\equiv\gamma_0^2(\ell+y)=\frac1{\beta(\ell+y+\eta+\lambda)}.
\end{equation}

\bigskip

In particular, for $z=1$ and $\Theta_1=\Theta$ we have

$$
\gamma_0^2=\frac1{\beta(\ell_1+y_2+\eta+\lambda)}=\frac1{\beta(Y+\lambda)},\qquad
\ell_1+y_2+\eta=Y.
$$

The condition $x_1\!\leq\! z\!\leq\!1$ translates into $0\!\leq\!\ell\!\leq\!\ell_1$, while
$(\Theta_2)_{min}\!\leq\!\Theta_1\!\leq\!\Theta$ becomes $0\!\leq\! y\!\leq\!y_2$.
Therefore, 

$$
\int_{x_1}^1\frac{dz}z=\int_0^{\ell_1}d\ell\quad \text{and}\quad \int_{Q_0/\omega_2}^{\Theta}\frac{d\Theta_1}{\Theta_1}=\int_0^{y_2}dy.
$$

\bigskip

One gets finally the MLLA system of equations of (\ref{eq:eqqq})(\ref{eq:eqgg})
for quark and gluon jets correlations

\bigskip

\vbox{
\begin{eqnarray}
\label{eq:eveeqq}
\hskip -0.9cm Q^{(2)}(\ell_1,y_2,\eta)\!-\! Q_1(\ell_1,y_1)Q_2(\ell_2,y_2)
\!\!\!&\!\!=\!\!&\!\!\! \frac{C_F}{N_c}\!\!
\int_0^{\ell_1}\!\!\! d\ell\!\int_0^{y_2}\!\!\! dy\,
\gamma_0^2(\ell+y) \Big[\!1\!-\!\frac34 \delta(\ell-\ell_1) \!\Big]
G^{(2)}(\ell,y,\eta),\\\notag\\\notag\\
\notag
\hskip -0.9cm G^{(2)}(\ell_1,y_2,\eta) - G_1(\ell_1,y_1)G_2(\ell_2,y_2)
\!\!\!&\!\!\!=\!\!\!&\!\!\!\! 
\int_0^{\ell_1}\!\! d\ell\!\int_0^{y_2}\!\!dy\, \gamma_0^2(\ell+y)
\Big[\!1 - a \delta(\ell-\ell_1) \!\Big] G^{(2)}(\ell,y,\eta)\\\notag\\
\!\!\!&\!\!\!+\!\!\!&\!\!\! (a-b) \int_0^{y_2}dy \> \gamma_0^2(\ell_1+y)
G(\ell_1,y+\eta)G(\ell_1+\eta,y).
 \label{eq:eveeqglu} 
\end{eqnarray}
}

In the last line of (\ref{eq:eveeqglu}) we have made used of (\ref{eq:vars})
to write

\begin{equation}
G_1\equiv G(\ell_1,y_1) = G(\ell_1, y_2+\eta),\quad
G_2\equiv G(\ell_2,y_2) = G(\ell_1+\eta, y_2).
\label{eq:G1G2}
\end{equation}

\bigskip

The first term in (\ref{eq:eveeqq}) and (\ref{eq:eveeqglu}) represents the
DLA contribution;  the terms proportional to $\delta$ functions or to
$a$, $b$, represent MLLA corrections.
$a-b$ appearing in (\ref{eq:eveeqglu}) and defined in (\ref{eq:a-b}) 
is proportional to $n_f$, positive and color suppressed.
%

\section{TWO PARTICLE CORRELATION IN A GLUON JET} 
\label{section:corglu}

\subsection{Iterative solution}
\label{subsection:iterglue}

Since equation (\ref{eq:eveeqglu}) for a gluon  jet is self contained, it
is our starting point.
We define the normalized correlator ${\cal C}_g$ by
\begin{equation}
 G^{(2)} = {\cal C}_g (\ell_1,y_2,\eta)\ G_1\,G_2,
\label{eq:Gnor}
\end{equation}
where $G_1$ and $G_2$ are expressed in (\ref{eq:G1G2}).
Substituting (\ref{eq:Gnor}) into (\ref{eq:eveeqglu}) one gets
(see appendix \ref{section:Gcorr})  the following expression for
the correlator 

\begin{eqnarray}
 {\cal C}_g -1
=\frac{1 -\delta_1 -b\left(\psi_{1,\ell} +\psi_{2,\ell}-
  [\beta\gamma_0^2] \right) - \left[a\chi_{\ell} + \delta_2\right]}
{1+ \Delta
+\delta_1 + \Big[a\left(\chi_{\ell} +{[\beta\gamma_0^2]}\right)+\delta_2\Big]}
\label{eq:CGfull}
\end{eqnarray}
which is to be evaluated numerically. We have introduces
the following notations and variables
\begin{eqnarray}\label{eq:nota4bis}
&& \quad \chi =  \ln {\cal C}_g,\qquad
\chi_{\ell} = \frac{d\chi}{d\ell},\qquad \chi_{y}=\frac{d\chi}{dy};\\\notag\\
&&\label{eq:psi1}\psi_{1} = \ln G_{1},\qquad
\psi_{1,\ell}= \frac{1}{G_1}\frac{dG_1}{d\ell},\qquad
\psi_{1,y}= \frac{1}{G_1}\frac{dG_1}{dy};\\\notag\\
&&\label{eq:psi2}\psi_{2} = \ln G_{2},\qquad
\psi_{2,\ell}= \frac{1}{G_2}\frac{dG_2}{d\ell},\qquad
\psi_{2,y}= \frac{1}{G_2}\frac{dG_2}{dy};\\\notag\\
&&\label{eq:deltabis}\Delta = \gamma_0^{-2}
\Big(\psi_{1,\ell}\psi_{2,y}+\psi_{1,y}\psi_{2,\ell}\Big);\\\notag\\
&&\delta_1 = \gamma_0^{-2}\Big[\chi_{\ell}(\psi_{1,y}+\psi_{2,y}) +
   \chi_{y}(\psi_{1,\ell}+\psi_{2,\ell})\Big];\label{eq:delta1}\\\notag\\
&&\delta_2 = \gamma_0^{-2}\Big(\chi_{\ell}\chi_{y} + \chi_{\ell\,y}\Big).
\label{eq:nota4}
\end{eqnarray}
As long as ${\cal C}_g$ is changing slowly with
$\ell$ and $y$, (\ref{eq:CGfull}) can be solved iteratively. The expressions
of $\psi_{\ell}$ and $\psi_{y}$, as well as the numerical analysis of 
the other quantities are explicitly given in appendices \ref{subsection:Logder} and \ref{section:numcorr} for $\lambda=0$ ($Q_0=\Lambda_{QCD}$), the so call 
``limiting spectrum''. Consequently, (\ref{eq:CGfull}) will be computed in the same limit.

\subsection{Estimate of magnitude of various contributions}
\label{subsection:estimate}

To estimate the relative r\^ole of various terms in (\ref{eq:CGfull}) we
can make use of a simplified model for the MLLA spectrum in which one neglects
the variation of $\alpha_s$, hence of $\gamma_0$ in (\ref{eq:gap3}).
It becomes, after differentiating with respect to $\ell$
\begin{equation}
G_{\ell\,y} = \gamma_0^2\big(G - a\, G_{\ell}\big).
\label{eq:Gsimp}
\end{equation}
The solution of this equation is the function for
$\gamma_0^2=const$ (see appendix \ref{section:inspiredDLA} for details)
\begin{eqnarray}
G(\ell,y) \stackrel{x\ll1}{\simeq} \exp{\Big( 2\gamma_0 \sqrt{\ell\,y} -a\gamma_0^2\,y\Big)}.
\label{eq:Gmod}
\end{eqnarray}

The subtraction term $\propto a$ in (\ref{eq:Gmod}) accounts for hard 
corrections (MLLA) that shifts the position of the maximum of the single inclusive 
distribution toward larger values of $\ell$ (smaller $x$) and partially 
guarantees the energy balance during soft gluons cascading (see \cite{EvEq}\cite{KO}  and
references therein). The position of the maximum follows from (\ref{eq:Gmod})
$$
\ell_{max}=\frac{Y}2(1+a\gamma_0).
$$
From (\ref{eq:Gmod}) one gets
\begin{eqnarray}
\psi_{\ell} = \gamma_0 \sqrt{\frac{y}{\ell}},\qquad  \psi_{y} = \gamma_0
\sqrt{\frac{\ell}{y}} - a\gamma_0^2,\qquad
\psi_{\ell\,y}\sim\psi_{\ell\,\ell}\sim\psi_{y\,y} = {\cal O}(\gamma_0^3),\quad
\ell^{-1}\sim y^{-1}={\cal O}(\gamma_0^2)
\label{eq:psimodel} 
\end{eqnarray}
and the function $\Delta$ in (\ref{eq:deltabis}) becomes
\begin{eqnarray}
\Delta \!\!&\!\!=\!\!&\!\! \left( \sqrt{\frac{y_1\ell_2}{\ell_1 y_2}} +
 \sqrt{\frac{\ell_1 y_2}{y_1 \ell_2}} \right)
-a\gamma_0\left(\sqrt{\frac{y_1}{\ell_1}}+\sqrt{\frac{y_2}{\ell_2}} \right)
\cr
\!\!&\!\!=\!\!&\!\! 2\cosh(\mu_1-\mu_2) - a\gamma_0(e^{\mu_1}+e^{\mu_2}); \qquad
\mu_{i}= \textstyle {\frac12} \displaystyle\ln\frac{y_{i}}{\ell_{i}}.  
\label{eq:Deltamodel} 
\end{eqnarray}
We see that $\Delta={\cal O}(1)$
and depends on the ratio of logarithmic variables $\ell$ and $y$. One step further is needed before we can estimate the order of magnitude of $\chi_{\ell}$, $\chi_{y}$ and $\chi_{\ell\,y}$. Indeed, the leading contribution to these quantities is obtained by taking
the leading (DLA) piece of (\ref{eq:CGfull}), that is

$$
\chi\stackrel{DLA}{\simeq}\ln\left(1+\frac1{1+\Delta}\right);
$$
then, it is easy to get

$$
\chi_{\ell}=-\frac{\Delta_{\ell}}{(1+\Delta)(2+\Delta)},\quad
\chi_{y}=-\frac{\Delta_{y}}{(1+\Delta)(2+\Delta)};
$$
we have roughly
$$
\chi_{\ell}\propto\mu_{\ell},\quad \chi_{y}\propto\mu_{y},\quad\chi_{\ell\,y}
\propto\mu_{\ell}\,\mu_{y};
$$
since $\mu_{i,\ell}=\mu_{i,y}={\cal O}(\gamma_0^2)$ one gets

\begin{equation}
\chi_{\ell} \sim \chi_{y} = {\cal O}(\gamma_0^2), \qquad
\chi_{\ell\,y}\sim\chi_{\ell}\chi_{y}={\cal O}(\gamma_0^4),
\label{eq:magn1}
\end{equation}
which entails for the corrections terms $\delta_1$ and
$\delta_2$ in (\ref{eq:delta1}) (\ref{eq:nota4})
\begin{equation}
\delta_1 ={\cal O}(\gamma_0), \qquad \delta_2={\cal O}(\gamma_0^2).
\label{eq:magn2}
\end{equation}
The term $\delta_1$ constitutes a MLLA correction
while $\delta_2$ as well as other terms that are displayed in square brackets 
in (\ref{eq:CGfull}) are of order $\gamma_0^2$ and are, formally 
speaking, beyond the MLLA accuracy.

\subsection{MLLA reduction of (\ref{eq:CGfull})}

Dropping ${\cal O}(\gamma_0^2)$ terms , the expression for the correlator would simplify to
\begin{equation}
{\cal C}_g-1 \stackrel{MLLA}{\approx} \frac{1-b\left(\psi_{1,\ell}
    +\psi_{2,\ell} \right)-\delta_1} {1+ \Delta + \delta_1}.
\label{eq:CGMLLA}
\end{equation}

\subsection{$\boldsymbol{{\cal C}_g\ge 0}$ in the soft approximation}
\label{subsection:Cgpos}

${\cal C}_g$ must obviously be positive. By looking at ${\cal C}_g\ge 0$
one determines the region of applicability of our soft approximation.
Using (\ref{eq:CGMLLA}), the condition reads
\begin{equation}
   2+\Delta >  b(\psi_{1,\ell}+\psi_{2,\ell}).
\label{eq:Gsoft}
\end{equation}
For the sake of simplicity,  we employ the model
(\ref{eq:Gmod})(\ref{eq:psimodel})(\ref{eq:Deltamodel}), this gives
\begin{equation}
 2\big(1+\cosh(\mu_1-\mu_2)\big)>\gamma_0(a+b) \big(e^{\mu_1}+ e^{\mu_2}\big), 
\end{equation}
which translates into 
\begin{equation}
\label{eq:elliyi}
\sqrt{\frac{\ell_1}{y_1}} + \sqrt{\frac{\ell_2}{y_2}} > \gamma_0\,(a+b).
\end{equation}
For $\ell_1$, $\ell_2 \ll Y$ we can set $y_1\simeq y_2\simeq Y$ and, using  
$\gamma_0^2 \simeq 1/{\beta Y}$
\footnote{for $n_f=3$, $\beta=0.75$},  we get the condition
\begin{equation}
\sqrt{\ell_1} + \sqrt{\ell_2}  >  \frac{a+b}{\sqrt{\beta}} \simeq 2.1,
\end{equation}
which is satisfied as soon as $\ell_1>1$ ($\ell_2>\ell_1$);  
so, for $x_1\lesssim 0.4 , x_2 < x_1$, the correlation $\cal C$ is
positive.

\subsection{The sign of $\boldsymbol{({\cal C}_g-1)}$}
\label{subsection:signG}

In the region of relatively hard particles $({\cal C}_g -1)$ becomes
negative. To find out at which value of $\ell$ it happens, we use the simplified
model and take, for simplicity, $\ell_1=\ell_2=\ell_\pm$.

The condition $1 = \delta_1 + b \big(\psi_{1,\ell} + \psi_{2,\ell}\big)$, using  (\ref{eq:yelldef})(\ref{eq:gammabeta})(\ref{eq:psimodel}) 
and neglecting $\delta_1$ which vanishes 
at $\ell_1\approx\ell_2$ reads
\begin{equation}
1- b\gamma_0\cdot 2 \sqrt{\frac{Y-\ell_\pm}{\ell_\pm}} = 0
\Leftrightarrow
\ell_\pm = \displaystyle\frac{M_g}{1+\frac{M_g}{Y}},\quad
M_g= \frac{4b^2}{\beta} \simeq 4.5. 
\label{eq:pmest} 
\end{equation}
Thus in the $Y\to \infty$ limit the correlation between two equal energy
partons in a gluon jet turns negative at a fixed value, $x> x_\pm
\simeq \exp(4.5)=1/90$.  For finite energies this energy is
essentially larger; in particular, for $Y=5.2$ (which corresponds to
LEP-I energy) (\ref{eq:pmest}) gives $\ell_\pm\simeq 2.4$ ($x_\pm
\simeq 1/11$).

For the Tevatron, let us for instance take the typical value $Y=6.0$, one has 
$\ell_{\pm}\simeq2.6$ and finally, for the LHC we take the typical one, 
$Y=7.5$, one gets the corresponding
$\ell_{\pm}\simeq2.8$. This is confirmed numerically in Figs.~\ref{fig:3gbandsLEP}, \ref{fig:3gbandsTeV} and \ref{fig:3gbandsLHC}.

\section{TWO PARTICLE CORRELATIONS IN A QUARK JET}
\label{section:quark}

\subsection{Iterative solution}
\label{subsection:iterQ}

We define the normalized correlator ${\cal C}_q$ by
\begin{equation}
 Q^{(2)} = {\cal C}_q (\ell_1,y_2,\eta)\ Q_1\,Q_2,
\label{eq:Qnor}
\end{equation}
where $Q_1$ and $Q_2$ are expressed like in (\ref{eq:G1G2}) for $G_1$ and
$G_2$.
By differentiating (\ref{eq:eveeqq}) with respect to $\ell_1$ and $y_2$,
one gets (see appendix \ref{section:Qcorr})

\begin{equation}
{\cal C}_q-1
=\displaystyle\frac{
     \frac {N_c}{C_F} {\cal C}_g
       \Big[ 1-\textstyle {\frac34}\Big(\psi_{1,\ell}
  +\psi_{2,\ell} +[\chi_{\ell}] - [\beta\gamma_0^2]\Big) \Big]
\frac{C_F}{N_c}\frac{G_1}{Q_1} \frac{C_F}{N_c}\frac{G_2}{Q_2}
-\tilde\delta_1 -[\tilde\delta_2]}
{\widetilde\Delta + \Big[1-\textstyle {\frac34}
  \big(\psi_{1,\ell}-[\beta\gamma_0^2]\big)\Big]\frac{C_F}{N_c}\frac{G_1}{Q_1}
+\Big[1-\textstyle {\frac34} \big(\psi_{2,\ell}-[\beta\gamma_0^2]\big)\Big]
\frac{C_F}{N_c}\frac{G_2}{Q_2} 
+ \tilde\delta_{1}+[\tilde\delta_{2}]},  
\label{eq:Qcorr}
\end{equation}

which is used for numerical analysis. $G_i/Q_i$ is computed using (\ref{eq:ratio}).
The terms ${\cal O}(\gamma_0^2)$ are the one that can be neglected
when staying at MLLA (see \ref{subsection:MLLAreduc}).
We have introduced, in addition to (\ref{eq:nota4bis})-(\ref{eq:nota4}), the following notations
\begin{eqnarray}
\widetilde\Delta \!\!&\!\!=\!\!&\!\!
\gamma_0^{-2}\Big(\varphi_{1,\ell}\varphi_{2,y}+\varphi_{1,y}\varphi_{2,\ell}\Big),\\\notag\\
\tilde\delta_1\!\!&\!\!=\!\!&\!\!
\gamma_0^{-2}\Big[\sigma_{\ell}(\varphi_{1,y}+\varphi_{2,y}) +
\sigma_{y}(\varphi_{1,\ell}+\varphi_{2,\ell})\Big],\\\notag\\
\tilde\delta_2 \!\!&\!\!=\!\!&\!\!
\gamma_0^{-2}\Big(\sigma_{\ell}\sigma_{y}+\sigma_{\ell\,y}\Big),
\label{eq:nota5}
\end{eqnarray}
with
\begin{equation}
\varphi_k = \ln Q_k,\quad \sigma= \ln {\cal C}_q.
\label{eq:phisigma}
\end{equation}
Accordingly, (\ref{eq:Qcorr}) will be computed for $\lambda=0$, the analysis of the
previous functions is done in appendix \ref{section:numcorr}.
\subsection{MLLA reduction of (\ref{eq:Qcorr})}
\label{subsection:MLLAreduc}

Using (\ref{eq:ratioMLLA}), which entails
$\frac{C_F}{N_c}\frac{G_i}{Q_i} \simeq 1 - \big(a-\frac34\big)\psi_{i,\ell}
+ {\cal O}(\gamma_0^2)$, reduces (\ref{eq:Qrel}) to

\vbox{
\begin{eqnarray}
{\cal C}_q-1 \!\!&\!\!=\!\!&\!\!\displaystyle\frac {
\frac{N_c}{C_F}{\cal C}_g\Big[1 -a\big(\psi_{1,\ell}+\psi_{2,\ell}\big)
-\frac34[\chi_{\ell} -\beta\gamma_0^2]\Big] -{\cal C}_q(\tilde\delta_1
+[\tilde\delta_2])}
{2 +\widetilde\Delta -a\big(\psi_{1,\ell}+\psi_{2,\ell}\big)
 +[\frac32 \beta\gamma_0^2]}\label{eq:Qcorr3}\cr
\!\!&\!\!=\!\!&\!\!\displaystyle\frac {
\frac{N_c}{C_F}{\cal C}_g\Big[1 -a\big(\psi_{1,\ell}+\psi_{2,\ell}\big)
-\frac34[\chi_{\ell} -\beta\gamma_0^2]\Big] -\tilde\delta_1
-[\tilde\delta_2]}
{2 +\widetilde\Delta -a\big(\psi_{1,\ell}+\psi_{2,\ell}\big)
 +[\frac32 \beta\gamma_0^2] + \tilde\delta_1 + \tilde\delta_2}.
\label{eq:Qcorr1}
\end{eqnarray}
}

As demonstrated in appendix \ref{subsection:corrections},
$\tilde\Delta = \Delta  + {\cal O}(\gamma_0^2)$ and
\begin{equation}
{\cal C}_q (\tilde\delta_1 + \tilde\delta_2) \simeq \frac{N_c}{C_F}{\cal
C}_g (\delta_1 + \delta_2);
\label{eq:numdel}
\end{equation}
and (\ref{eq:Qcorr3}) becomes
\begin{equation}
{\cal C}_q-1 \approx\displaystyle\frac{N_c}{C_F}\,\frac {
{\cal C}_g\Big[1 -a\big(\psi_{1,\ell}+\psi_{2,\ell}\big)
-\frac34[\chi_{\ell} -\beta\gamma_0^2] -\delta_1
-[\delta_2]\Big]}
{2 + \Delta -a\big(\psi_{1,\ell}+\psi_{2,\ell}\big)
 +[\frac32 \beta\gamma_0^2]]}.
\label{eq:Qcorr4}
\end{equation}

\medskip

Would we neglect, according to (\ref{eq:magn1})(\ref{eq:magn2}),
next to MLLA terms, which
amounts to dropping
all ${\cal O}(\gamma_0^2)$ corrections,
(\ref{eq:Qcorr1}) would simply reduce to

\medskip

\begin{equation}
{\cal C}_q-1 \!\!\stackrel{MLLA}{\approx}\!\!\frac{N_c}{C_F}\,\frac {
\displaystyle{\cal C}_g\Big[1 -a\big(\psi_{1,\ell}+\psi_{2,\ell}\big)
\Big] -\delta_1 }
{2 + \Delta -a\big(\psi_{1,\ell}+\psi_{2,\ell}\big) +\delta_1}.
\label{eq:QMLLAap}
\end{equation}

\medskip

Furthermore, comparing (\ref{eq:Qcorr4}) and (\ref{eq:CGMLLA}) and
using the magnitude estimates of subsection \ref{subsection:estimate}
allows to
make an expansion in the small ${\cal O}(\gamma_0)$ corrections $\delta_1$, $\psi_{1,\ell}$ and
$\psi_{2,\ell}$ to get

\medskip

\begin{eqnarray}
\frac{{\cal C}_q-1}{{\cal C}_g-1} &\stackrel{MLLA}{\simeq}&
\frac{N_c}{C_F}\bigg[1+(b-a)(\psi_{1,\ell} + \psi_{2,\ell})
\frac{1+\Delta}{2+\Delta}\bigg]\cr
&\approx&
\frac{N_c}{C_F}\Big[1+(b-a)(\psi_{1,\ell} + \psi_{2,\ell})\Big({\cal C}_g^{DLA}\Big)^{-1}
\Big],
\label{eq:rapMLLA}
\end{eqnarray}

\medskip

where we have consistently used the DLA expression ${\cal C}_g^{DLA} =
\displaystyle\frac{2+\Delta}{1+\Delta}$.
$(a-b)$ is given in (\ref{eq:a-b}).
The deviation of the ratio from the DLA value $N_c/C_F$ is proportional to $n_f$, 
is color suppressed and numerical small.

\subsection{$\boldsymbol{{\cal C}_q\ge 0}$ in the soft approximation}
\label{subsection:Cqpos}

Since we neglect NMLLA corrections and the running of $\alpha_s$, we can make use 
of (\ref{eq:rapMLLA}) in order to derive the positivity constrain for the quark 
correlator. In the r.h.s. of (\ref{eq:rapMLLA}) we can indeed neglect the MLLA 
correction in the square brackets because it is numerically small
(for instance, for $\gamma_0\simeq0.5$ it is $\approx10^{-3}$). Therefore, ${\cal C}_q$
changes sign when 

$$
{\cal C}_g\ge1-\frac{C_F}{N_c}=\frac59\approx\frac12,
$$

(\ref{eq:elliyi}) gets therefore replaced by
$$
\sqrt{\frac{\ell_1}{y_1}}+\sqrt{\frac{\ell_2}{y_2}}>\frac45(a+2b)\gamma_0,
$$
which finally, following the same steps, gives
$$
\sqrt{\ell_1}+\sqrt{\ell_2}>\frac45\frac{a+2b}{\sqrt{\beta}}\simeq2.6.
$$
The last inequality is satisfied as soon as $\ell_1>1.6$ ($\ell_2>\ell_1$). This condition
slightly differs from that of the gluon correlator in \ref{subsection:signG}.

\subsection{The sign of $\boldsymbol{({\cal C}_q-1)}$}
\label{subsection:signQ}

From (\ref{eq:QMLLAap}), ${\cal C}_q-1$ changes sign for

\begin{equation}
{\cal C}_q-1\approx\frac{N_c}{C_F}\,\frac {
\displaystyle{\cal C}_g\Big[1 -a\big(\psi_{1,\ell}+\psi_{2,\ell}\big)
\Big]}
{2 + \Delta -a\big(\psi_{1,\ell}+\psi_{2,\ell}\big)}>0
\end{equation}

which gives the condition

$$
1=a\big(\psi_{1,\ell}+\psi_{2,\ell}\big).
$$

This gives a formula identical to (\ref{eq:pmest}) with the exchange
$b\rightarrow a$; $a$ being slightly larger than $b$, we find now a parameter
$M_q = 4a^2/\beta \simeq 4.66$. The corresponding $\ell_\pm$ at
which $({\cal C}_q-1)$ will change sign is slightly higher than for gluons;
for example at $Y=5.2$, $\ell_\pm \simeq 2.5\; (x_\pm \simeq 1/12)$, $Y=6.0$, 
$\ell_\pm \simeq 2.7\;  (x_\pm \simeq 1/13)$, $Y=7.5$, $\ell_\pm \simeq 2.9\;  (x_\pm \simeq 1/16)$. This is confirmed numerically in figures \ref{fig:3qbandsLEP}, \ref{fig:3qbandsTeV} and \ref{fig:3qbandsLHC}.

\section{NUMERICAL RESULTS}
\label{section:numer}

In order to lighten the core of the paper, only the main lines and ideas of
the calculations, and the results, are given here; the numerical analysis of
(MLLA and NMLLA) corrections occurring in (\ref{eq:CGfull}) and 
(\ref{eq:Qcorr}) is the object of appendix \ref{section:numcorr}, that we
summarize in subsection \ref{subsection:commE} below. We present our 
results as functions of $(\ell_1+\ell_2)$ and  $(\ell_1-\ell_2)$.

\subsection{The gluon jet correlator}
\label{subsection:gcorr}

In order to implement the iterative solution of the first line of
(\ref{eq:CGfull}), we define

\begin{equation}
\Upsilon_g = \ln\Bigg[1+\displaystyle\frac
{1-b(\psi_{1,\ell} + \psi_{2,\ell} -[\beta\gamma_0^2])}
{1+\Delta +[a\beta\gamma_0^2]}\Bigg]
\label{eq:upsg}
\end{equation}
as the starting point of the procedure. It represent the zeroth order of
the iteration for $\chi \equiv \ln{\cal C}_g$. The terms proportional to derivatives of 
$\chi$ in the numerator and denominator of (\ref{eq:CGfull}) are the objects
of the iteration and do not appear in (\ref{eq:upsg});  
the parameter $\Delta$ depends (see (\ref{eq:deltabis}))
only on the logarithmic 
derivatives $\psi_{\ell},\psi_{y}$ of the inclusive spectrum $G$ which are
 determined at each step, by the exact solution (\ref{eq:ifD})
(\ref{eq:calFdef}) for $G$ demonstrated in appendix \ref{section:ESEE}. The leading 
piece (DLA) of (\ref{eq:upsg}) 

$$
\Upsilon_g \stackrel{DLA}{=} \ln\Bigg[1+\displaystyle\frac
{1}{1+\Delta}\Bigg]
\label{eq:upsgDLA}
$$

is the one that should be used when reducing (\ref{eq:CGfull}) to MLLA.
We have instead consistently kept sub-leading (MLLA and NMLLA) corrections in 
(\ref{eq:upsg}) in order to 
follow the same logic that proved successful for the single inclusive spectrum.

\subsection{The quark jet correlator}
\label{subsection:qcorr}

We start now from (\ref{eq:Qcorr}) and define, like for
gluons

\begin{equation}
\Upsilon_q =\ln\left\{1+\displaystyle\frac{
     \frac {N_c}{C_F} {\cal C}_g
       \Big[ 1-\textstyle {\frac34}\Big(\psi_{1,\ell}
  +\psi_{2,\ell} +[\chi_{\ell} - \beta\gamma_0^2]\Big) \Big]
\frac{C_F}{N_c}\frac{G_1}{Q_1} \frac{C_F}{N_c}\frac{G_2}{Q_2}}
{\widetilde\Delta + \Big[1-\textstyle {\frac34}
  \big(\psi_{1,\ell}-[\beta\gamma_0^2]\big)\Big]\frac{C_F}{N_c}\frac{G_1}{Q_1}
+\Big[1-\textstyle {\frac34} \big(\psi_{2,\ell}-[\beta\gamma_0^2]\big)\Big]
\frac{C_F}{N_c}\frac{G_2}{Q_2}}\right\}
\label{eq:upsq}
\end{equation}

as the starting point of the iterative procedure, {\em i.e.} the zeroth
order of the iteration for $\sigma \equiv \ln{\cal C}_q$;
it again includes MLLA
(and some NMLLA) corrections. Since the iteration concerns ${\cal C}_q$,
the terms proportional to ${\cal C}_g$ and to its derivative $\chi_{\ell}$
must be present in (\ref{eq:upsq}).
All other functions are
determined, like above, by the exact solution of 
(\ref{eq:ifD}) and (\ref{eq:calFdef}) for $G$.

We have replaced in the denominator of (\ref{eq:upsq}) $\tilde\Delta$ with
$\Delta$, which amounts to neglecting ${\cal O}(\gamma_0^2)$ corrections,
 because the coefficient of $\gamma_0^{-2}(\tilde\Delta
-\Delta)$ is numerically very small; this occurs for two
combined reasons: it is proportional to $(a-3/4)$ which is small,
and the combination $(\psi_{1,\ell\,y}\psi_{2,\ell}
+\psi_{2,\ell,\ell}\psi_{1,y} +\psi_{2,\ell\,y}\psi_{1,\ell}
+\psi_{1,\ell\,\ell}\psi_{2,y} )$ that appears in (\ref{eq:Deltatilde})
is very small (see Fig.~\ref{fig:doublepsi}). Accordingly,

$$
\Upsilon_q \stackrel{DLA}{=} \ln\Bigg[1+\frac{N_c}{C_F}\displaystyle\frac
{1}{1+\Delta}\Bigg].
\label{eq:upsqDLA}
$$

We can use this simplified expression for the MLLA reduction of (\ref{eq:Qcorr}).

\subsection{The role of corrections; summary of appendix \ref{section:numcorr}}
\label{subsection:commE}

Analysis have been done separately for a gluon and a quark jet; their
conclusions are very similar.

That $\psi_{\ell}$ and $\psi_{y}$, which are ${\cal O}(\gamma_0)$ should
not exceed reasonable values (fixed arbitrarily to $1$) provides an
interval of reliability of our calculations; for example, at LEP-I
\begin{equation}
2.5 \leq \ell \leq  4.5\ \text{or}\ 5 \leq \ell_1 + \ell_2 \leq 9,\quad Y=5.2.
\label{eq:confintLEP}
\end{equation}
This interval is shifted upwards and gets larger when $Y$ increases.

$\Upsilon_g$ and $\Upsilon_q$ defined in (\ref{eq:upsg}) and (\ref{eq:upsq})
and their derivatives are shown to behave smoothly in the confidence
interval (\ref{eq:confintLEP}).

The roles of all corrections $\delta_1, \delta_2, \Delta$ for a gluon jet,
$\tilde\delta_1, \tilde\delta_2, \tilde\Delta$ for a quark jet,
 have been investigated individually. They stay under control in
(\ref{eq:confintLEP}).
While, in its center, their relative values coincide
with what is expected from subsection \ref{subsection:estimate}, NMLLA
corrections can become larger than MLLA close to the bounds; this could
make our approximations questionable. Two cases may occur which depend on
NMLLA corrections  not included in the present frame of
calculation;  either they largely cancel with the included ones and the sum
of all NMLLA corrections is (much) smaller than those of MLLA: then pQCD is
trustable at $Y=5.2$; or they do not, the confidence in our results at
this energy is weak, despite the fast convergence of the iterative
procedure which occurs thanks to the ``accidental'' observed cancellation
between MLLA and those of NMLLA which are included.
The steepest descent method \cite{RPR3}\cite{these}, in which a better
control is obtained of MLLA corrections alone, will shed some more light
on this question.
The global role of all corrections in the iterative process 
does not exceed $30 \%$ for $Y=5.2$ (OPAL) at the bounds
of (\ref{eq:confintLEP}); it is generally much smaller, though never
negligible. In particular, $\delta_1 + \delta_2 + a\Upsilon_{g,\ell}$ for
gluons (or $\tilde\delta_1 + \tilde\delta_2$ for quarks) sum up to ${\cal
O}(10^{-2})$ at LEP energy scale (they reach their maximum
${\cal O}(10^{-1})$ at the bound of the
interval corresponding to the $30\%$ evoked above). 

The role of corrections decreases when the total energy $Y$ of the jet
increases, which makes our calculations all the more reliable.

\subsection{Results for LEP-I}

\begin{figure}
\begin{center}
\epsfig{file=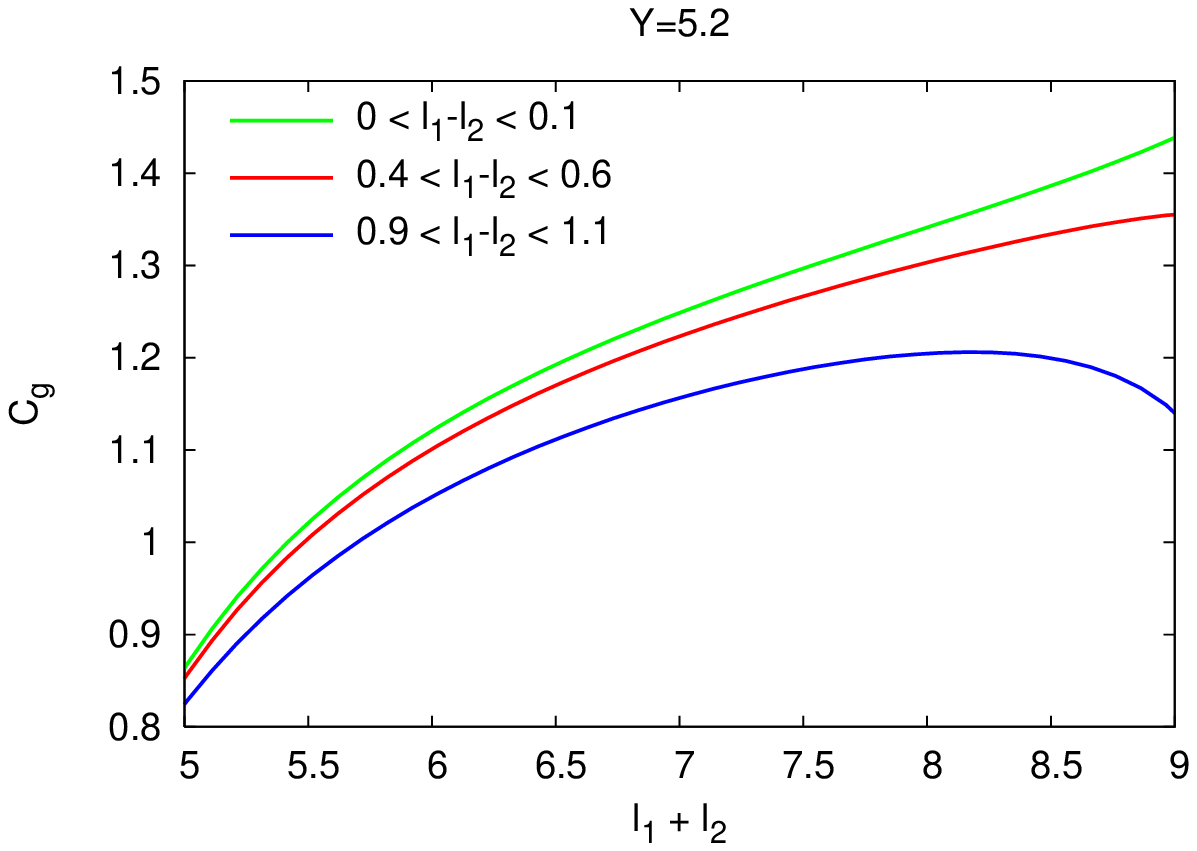, height=6truecm,width=0.47\tw}
\epsfig{file=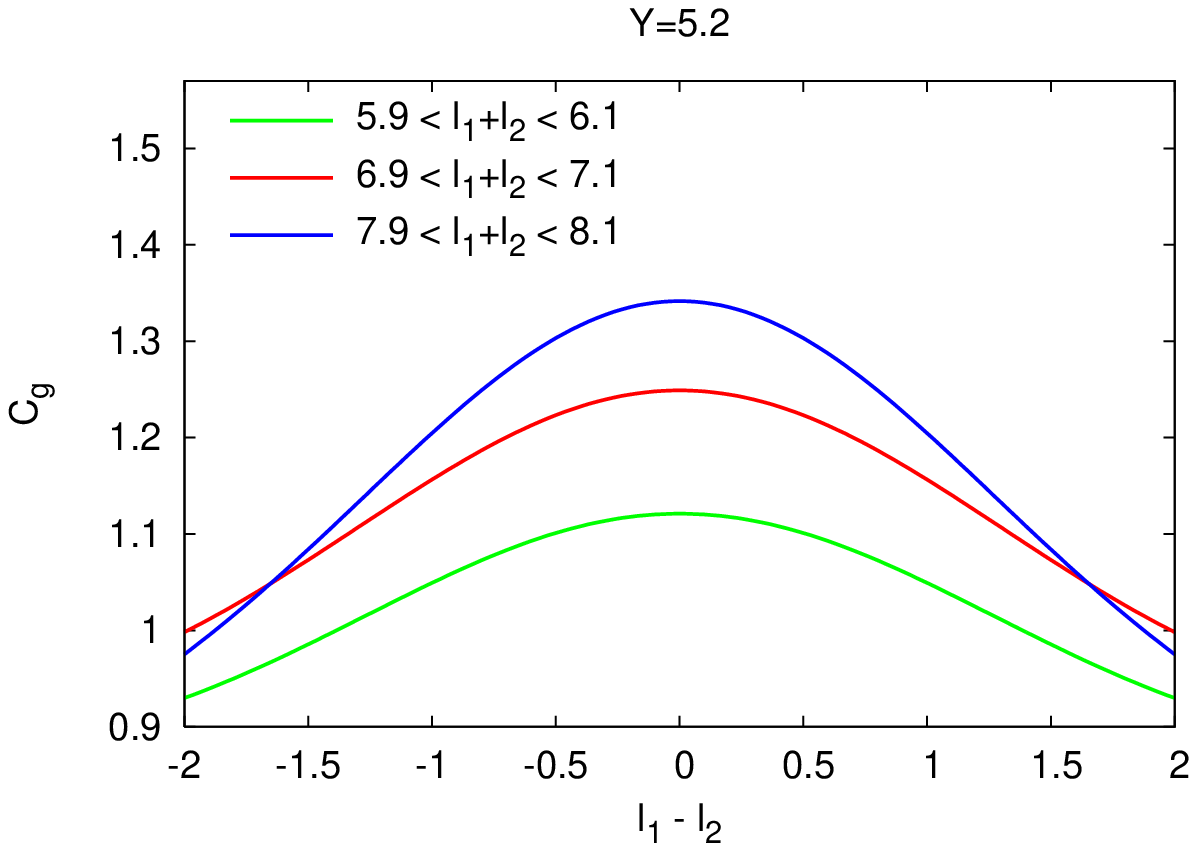, height=6truecm,width=0.47\tw}
\vskip .5cm
\caption{${\cal C}_g$ for the LEP-I ($Y=7.5$) inside a gluon jet as function of $\ell_1+\ell_2$ (left) and of $\ell_1-\ell_2$ (right)}
\label{fig:3gbandsLEP}
\end{center}
\end{figure}

\begin{figure}
\begin{center}
\epsfig{file=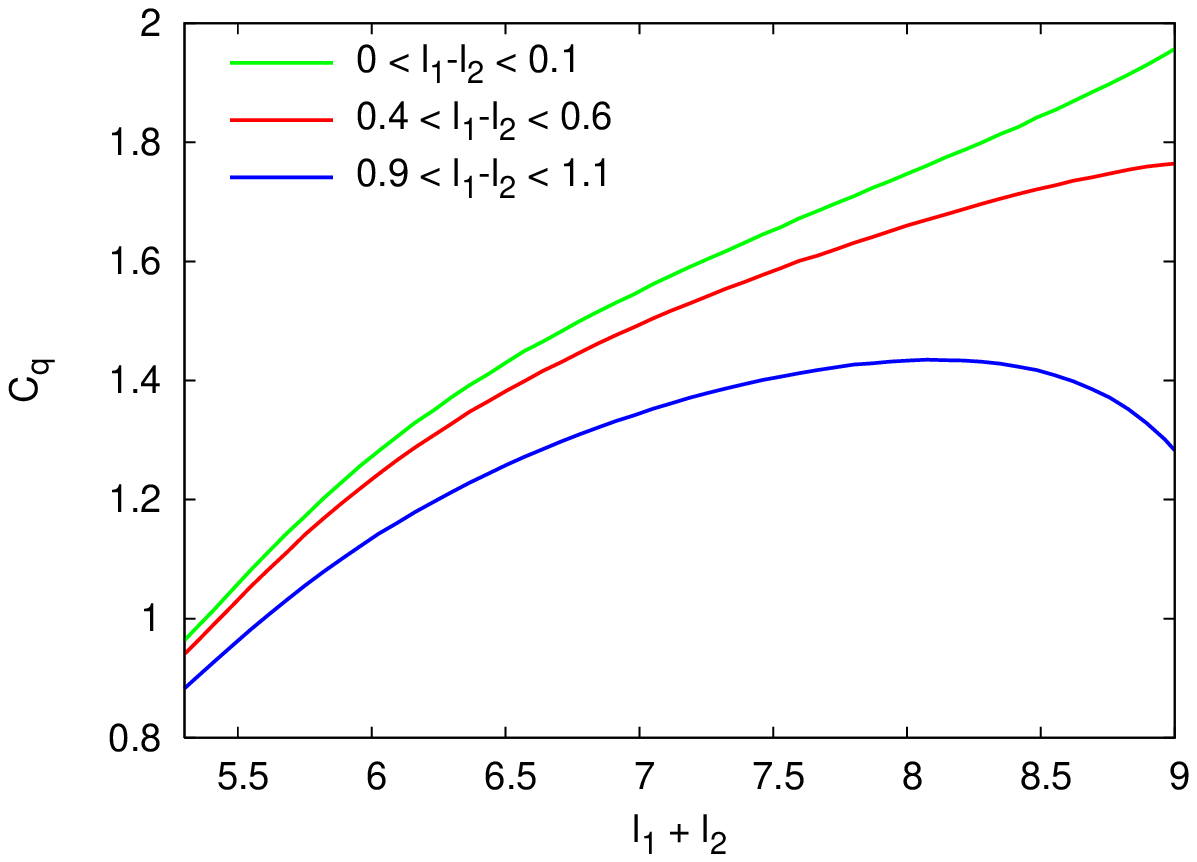, height=6truecm,width=0.47\tw}
\epsfig{file=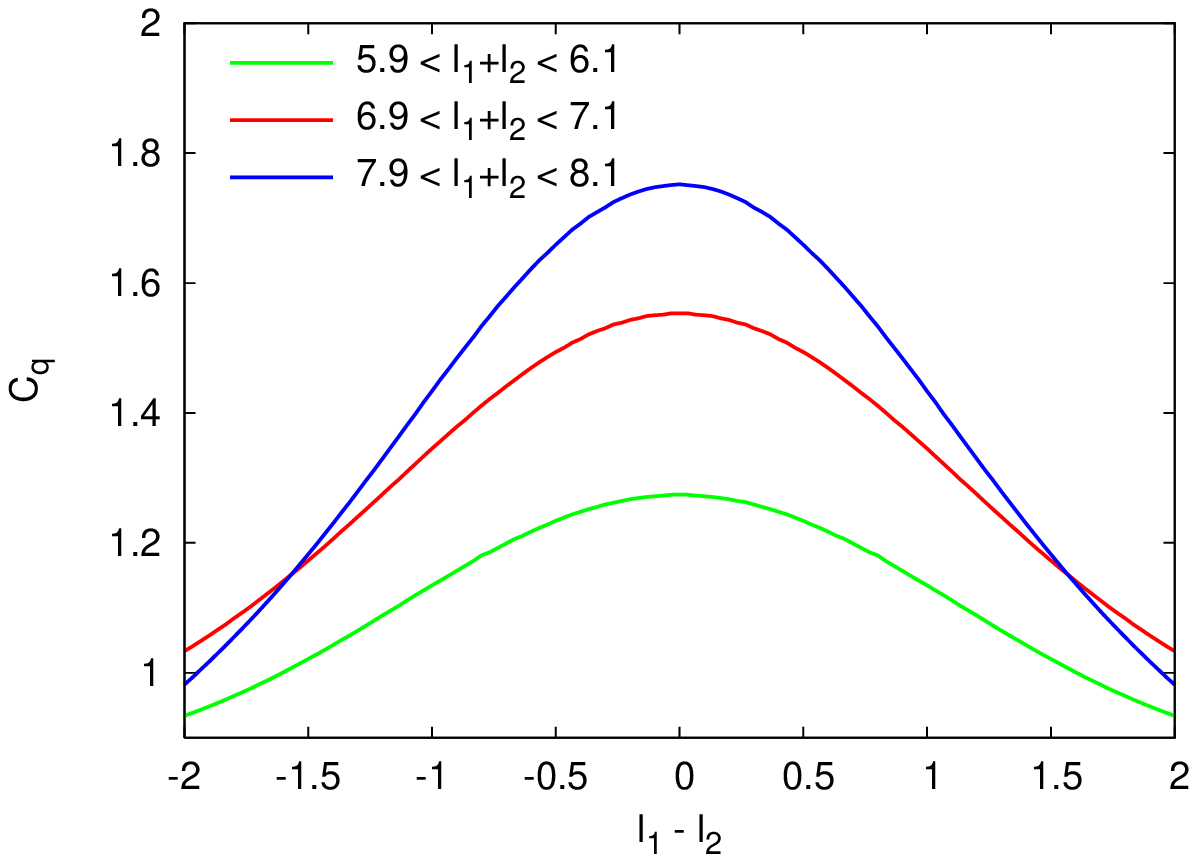, height=6truecm,width=0.47\tw}
\vskip .5cm
\caption{ ${\cal C}_q$ for the LEP-I ($Y=7.5$) inside  a quark jet as function of $\ell_1+\ell_2$ (left) and of $\ell_1-\ell_2$ (right)}
\label{fig:3qbandsLEP}
\end{center}
\end{figure}

In $e^+e^-\rightarrow q\bar{q}$ collisions 
at the $Z^0$ peak, $Q=91.2\,\text{GeV}$, $Y=5.2$, and $\gamma_0\simeq0.5$.
In Fig.~\ref{fig:3gbandsLEP} we give the results for  gluon jets  and
in Fig.~\ref{fig:3qbandsLEP} for quark jets.


\subsubsection{Comments}
\label{subsubsection:commentLEP}

Near the maximum of the single inclusive distribution 
($\ell_1\approx\ell_2\approx \frac{Y}2(1+a\gamma_0)$) our curves are linear functions
of $(\ell_1+\ell_2)$ and quadratic functions of $(\ell_1-\ell_2)$, in agreement with the Fong-Webber analysis \cite{FW}.

$({\cal C}_q-1)$ is roughly twice $({\cal C}_g-1)$ since
gluons cascade twice more than quarks ($\frac{N_c}{C_F}\approx2$). The difference 
is clearly observed from Fig.~\ref{fig:3gbandsLEP} and Fig.~\ref{fig:3qbandsLEP} (left) 
near the hump of the single inclusive distribution ($\ell_1+\ell_2\simeq7.6$), 
that is where most of the partonic multiplication takes place.

In both cases, $\cal C$ reaches its largest value for $\ell_1\approx \ell_2$ 
and steadily increases as a function of $(\ell_1+\ell_2)$ 
(Fig.~\ref{fig:3gbandsLEP}, left);
for $\ell_1\ne \ell_2$, it increases with $(\ell_1+\ell_2)$, then flattens off 
and decreases.

Both ${\cal C}$'s decrease as $|\ell_1-\ell_2|$ becomes large 
(Fig.~\ref{fig:3gbandsLEP} and \ref{fig:3qbandsLEP}, right).
The quark's tail is steeper than the gluon's;
for $5.9<\ell_1+\ell_2<6.1$, $({\cal C}-1)$ becomes negative when $\ell_1-\ell_2$
increases; ${\cal C}\geq1$ as soon as $\ell_1,\ell_2\geq2.75\,(x_1,x_2\leq0.06)$;
this bounds is close to $\ell\geq2.4$ found in subsection \ref{subsection:signG} or 
$\ell\geq2.5$ of (\ref{eq:confint1}).

One finds the limit 

\begin{equation}
{\cal C}_{g\,or\,q}\stackrel{\ell_1+\ell_2\to 2Y}{\longrightarrow}1.
\end{equation}

\begin{figure}
\begin{center}
\epsfig{file=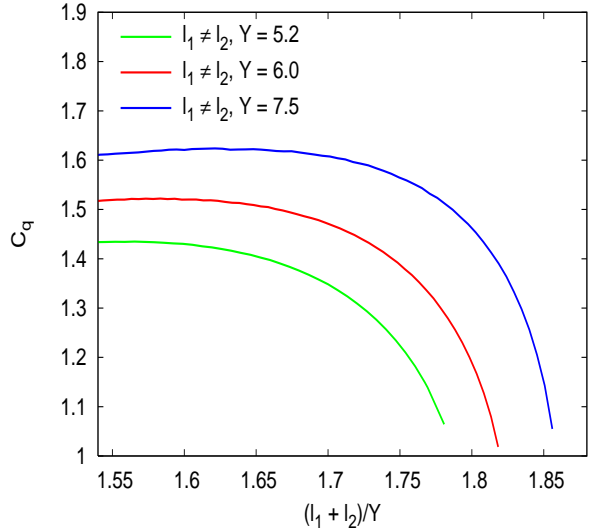, height=7truecm,width=0.48\tw}
\caption{Decrease of the correlation for $\ell_1\ne\ell_2$ at $Y=5.2$, $Y=6.0$ and $Y=7.5$}
\label{fig:Colliders}
\end{center}
\end{figure}

Actually, one observes on  Figs.~\ref{fig:3gbandsLEP}, \ref{fig:3qbandsLEP}
and \ref{fig:Colliders}
that a stronger statement holds. Namely, when 
 we take the limit $\ell_2\to Y$ for the softer particle,
the correlator goes to $1$.
This is the consequence of QCD coherence. The softer gluon is emitted at larger angles
by the total color charge of the jet and thus becomes de-correlated with the internal partonic structure of the jet.

The same phenomenon explains the flattening and the decrease of ${\cal C}$'s at $\ell_1\ne\ell_2$.

An interesting phenomenon is the seemingly continuous increase of ${\cal
C}_g$ and ${\cal C}_q$ at large $Y$ for $\ell_1 \approx \ell_2$ (green
curves in figs. \ref{fig:3gbandsLEP} and \ref{fig:3qbandsLEP} left). Like
we discussed in \cite{PerezMachet} concerning inclusive distributions, here we
reach a domain where a perturbative analysis cannot be trusted {\em
because of the divergence of $\alpha_s$}. Indeed, when $(\ell_1+\ell_2)$ gets close to
its limiting kinematical value ($2Y$), both $y_1$ and
$y_2$ get close to $0$, such that the corresponding
$\alpha_s(k_{1\perp}^2)$ and $\alpha_s(k_{2\perp}^2)$ cannot but become out
of control. Away from the $\ell_1 \approx \ell_2$ diagonal, taking $\ell_2\to Y$ 
($y_2\to0$), we have $y_1\to\eta>0$ and the emission of the harder parton still stays under control.

The two limitations of our approach already pointed at in
\cite{PerezMachet} are found again here:

$\ast$\ $x$ should be small enough such that our soft approximation stays valid;

$\ast$\ no running coupling constant should get too large such that pQCD stays reliable.

\subsection{Comparison with the data from LEP-I}

OPAL results are given in terms of

$$
 R\left(\ell_1, \ell_2, Y\right) \>=\>
 \frac1{2}+\frac1{2}{\cal C}_q\left(\ell_1, \ell_2, Y\right).
$$

\begin{figure}
\vbox{
\begin{center}
\epsfig{file=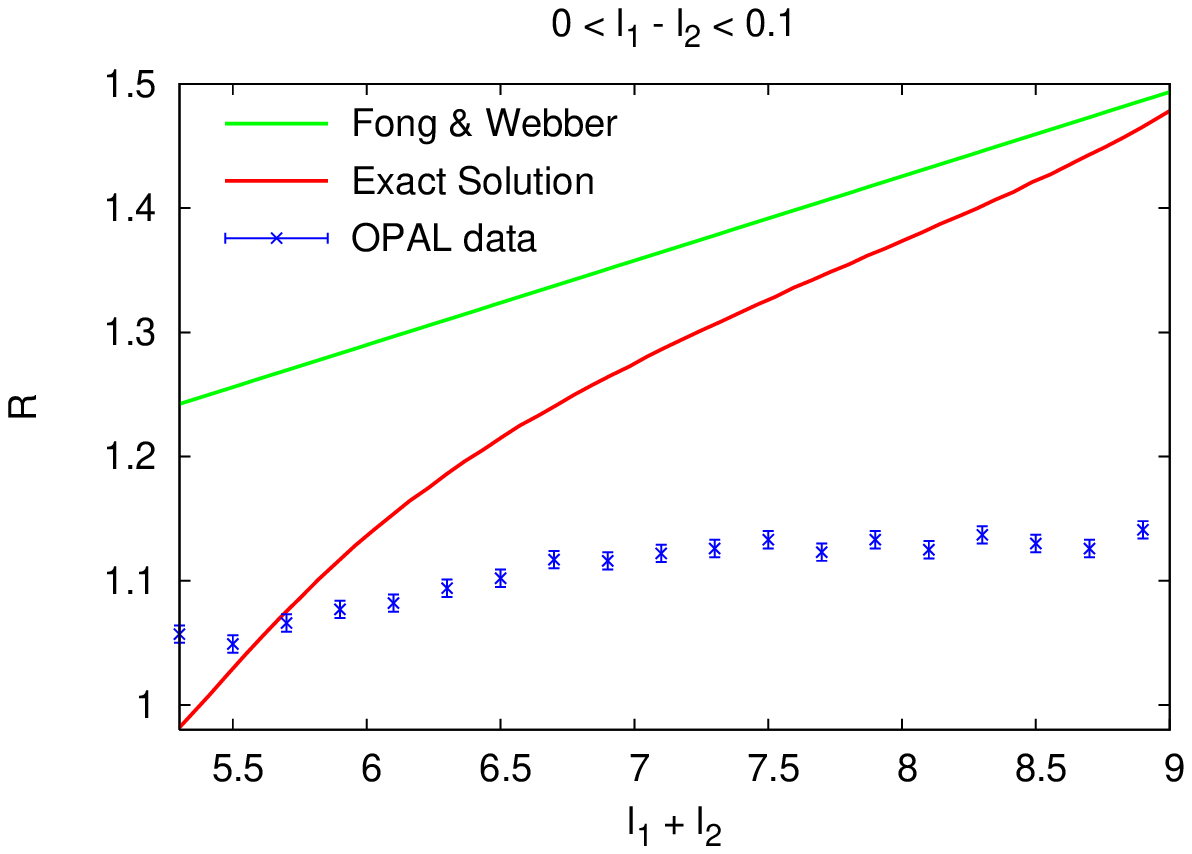, height=6truecm,width=0.48\tw}
\hfill
\epsfig{file=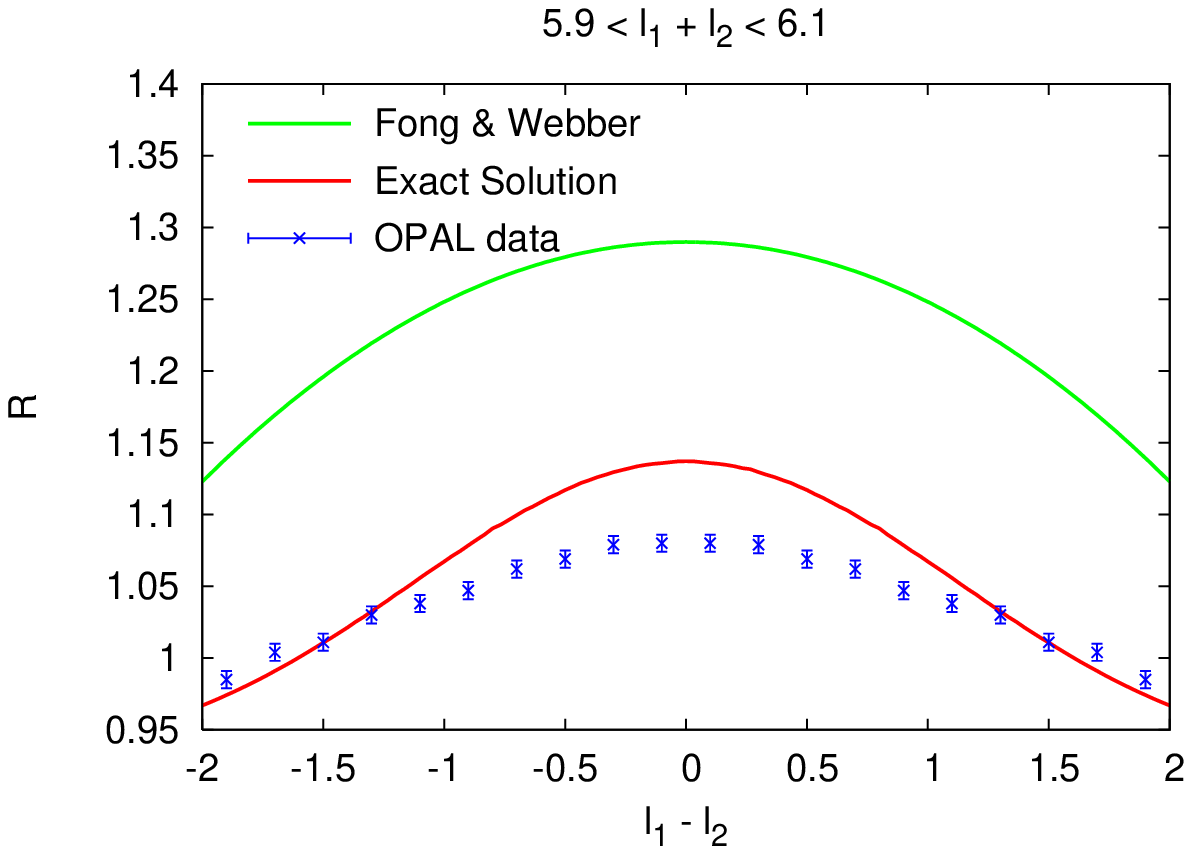, height=6truecm,width=0.48\tw}
\vskip .5cm
\epsfig{file=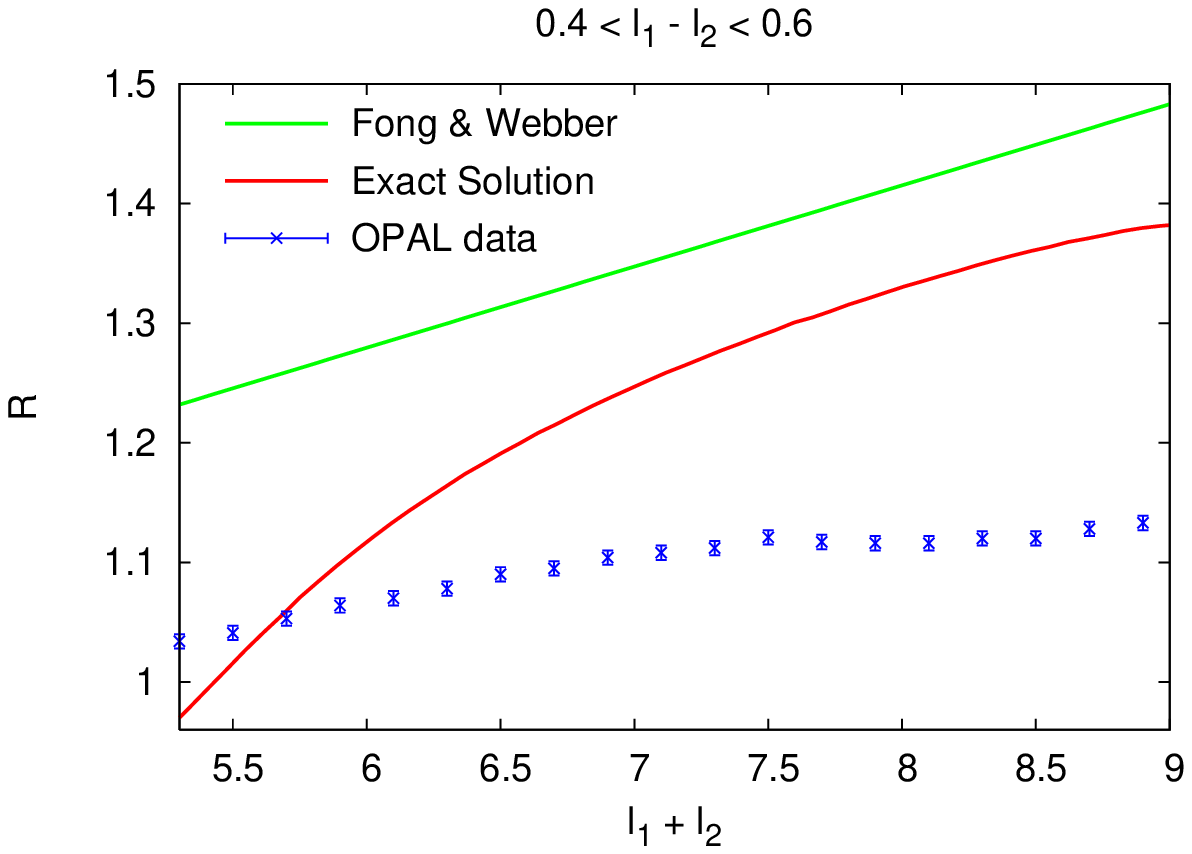, height=6truecm,width=0.48\tw}
\hfill
\epsfig{file=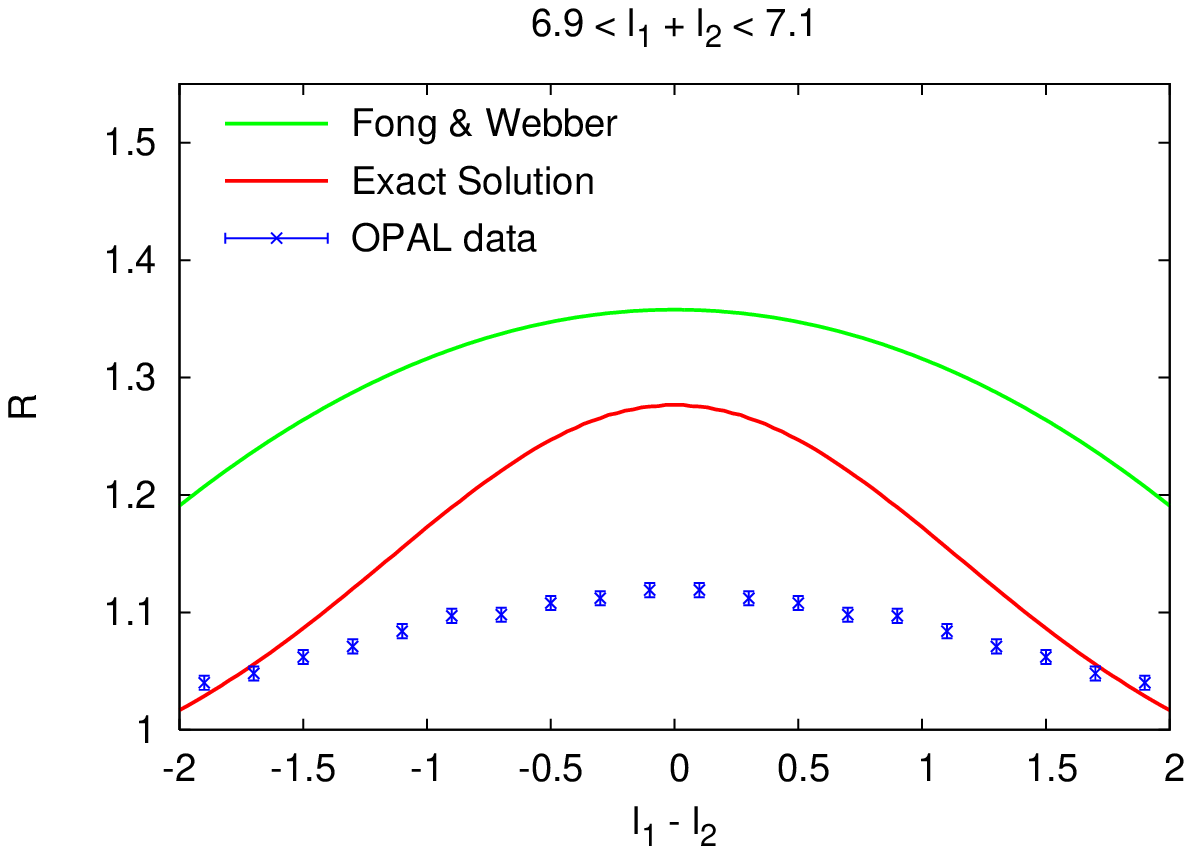, height=6truecm,width=0.48\tw}
\vskip .5cm
\epsfig{file=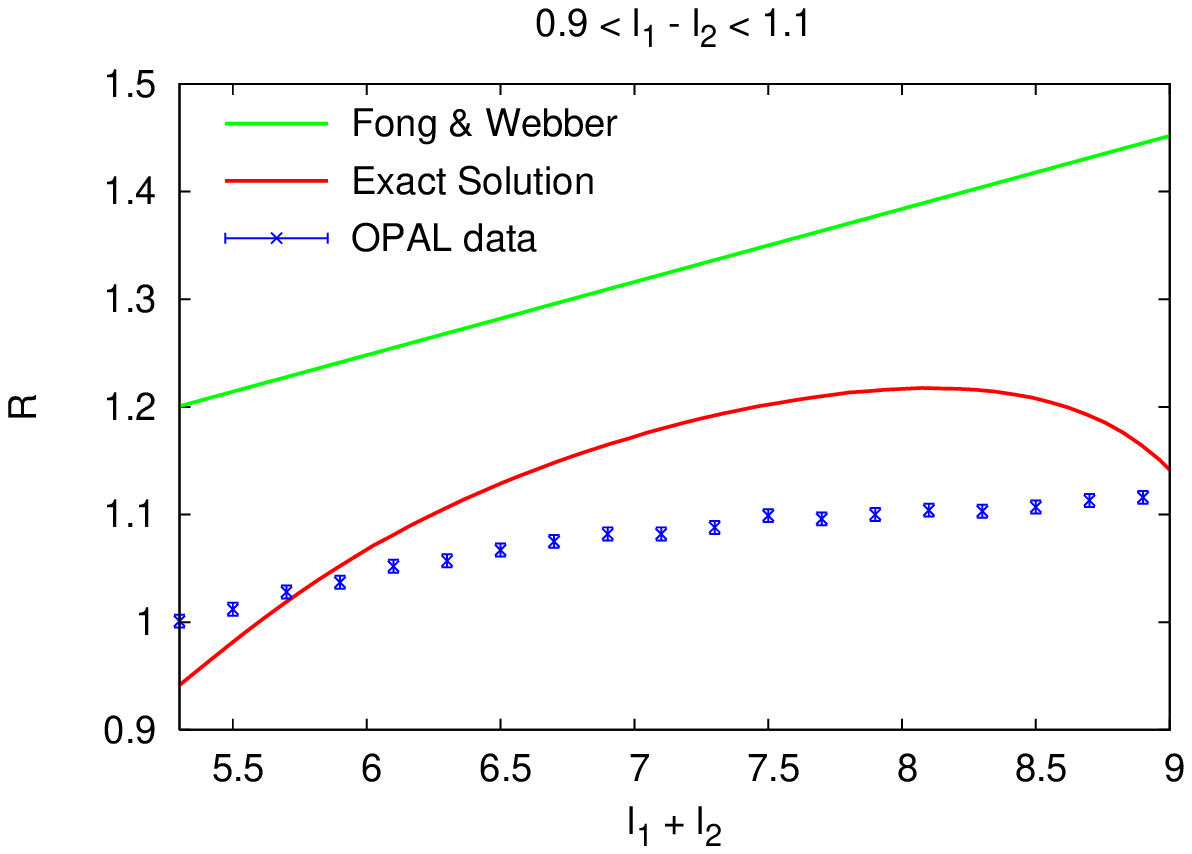, height=6truecm,width=0.48\tw}
\hfill
\epsfig{file=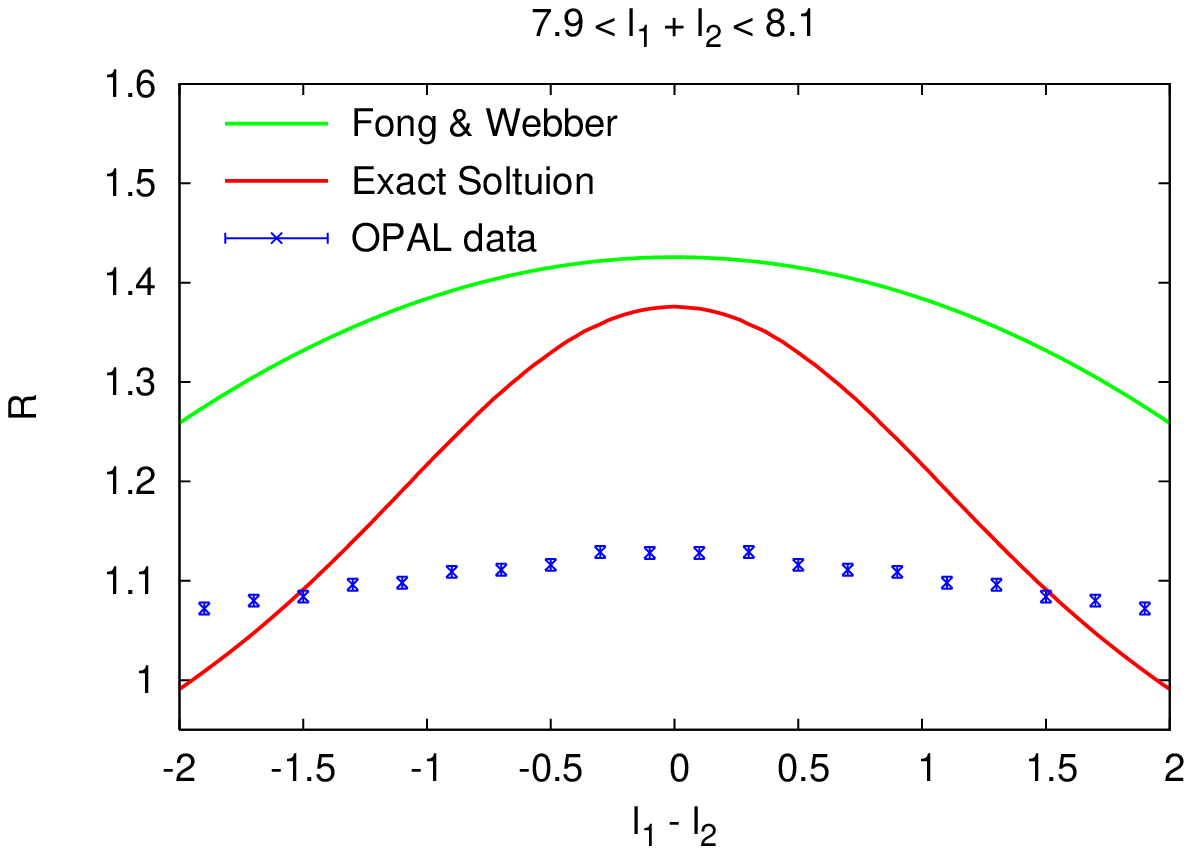, height=6truecm,width=0.48\tw}
\vskip .5cm
\caption{Correlations $R$ between two particles produced in $e^+e^-\rightarrow q\bar{q}$ 
compared with the OPAL data and the Fong-Webber approximation}
\label{fig:corrqqbar}
\end{center}
}
\end{figure}

In Fig.~\ref{fig:corrqqbar} we compare our prediction with the OPAL data 
\cite {OPAL} and the Fong-Webber curves (see subsection \ref{sub:FW}
and \cite{FW}).

\subsection{Comparing with the Fong-Webber approximation}
\label{sub:FW}

The only pQCD analysis of two-particle correlations in jets
beyond DLA was performed by Fong and Webber in 1990. In~\cite{FW}
the next-to-leading ${\cal O}({\gamma_0})$ correction,
${\cal C}_{g\, or\, q}=1+\sqrt{\alpha_s}+\cdots$, to the normalized two-particle
correlator was calculated. This expression was derived in the region 
$|\ell_1-\ell_2|/Y \ll 1$,
that is when the energies of the registered particles are 
close to each other (and to the maximum of the inclusive
distribution \cite{EvEq}\cite{KO}\cite{FW1}).
In this approximation the correlation function is
quadratic in $(\ell_1-\ell_2)$ and increases linearly
with $(\ell_1+\ell_2)$, see (\ref{eq:FWS}). 
For example, if one replaces the expression of the 
single inclusive distribution distorted gaussian \cite{FW1} 
(obtained in the region $\ell\approx \frac{Y}2(1+a\gamma_0)$) 
into (\ref{eq:CGMLLA}) the MLLA result for a gluon jet reads

\begin{equation}
{\cal C}_g(\ell_1,\ell_2,Y)\approx1+ \frac{1- \bigg(5b-3b
\displaystyle{\frac{\ell_1+\ell_2}Y}\bigg)\gamma_0+{\cal O}(\gamma_0^2)}
{3+9\bigg(\displaystyle
{\frac{\ell_1-\ell_2}Y}\bigg)^2 
 -\left(2\beta + 5a-3a\displaystyle{\frac{\ell_1+\ell_2}Y}\right)\gamma_0+
{\cal O}(\gamma_0^2)},
\label{eq:CGMLLAFW}
\end{equation}

where we have neglected the MLLA correction $\delta_1\simeq(\ell_1-\ell_2)^2\sqrt{\alpha_s}\simeq0$ near
the hump of the single inclusive distribution ($\ell_1\approx\ell_2\approx\frac{Y}2(1+a\gamma_0)$). The Fong-Webber answer is obtained by
expanding (\ref{eq:CGMLLAFW}) in $\gamma_0$ to get 
\cite{FW}

\begin{eqnarray}
{\cal C}_g^{(\mbox{\scriptsize FW})} \approx \frac43 - \left(\frac{\ell_1 - \ell_2}{Y}\right)^2 + \left[-\frac53\left(b-\frac13 a\right) +\frac{2}{9}\beta +\left(b-\frac13 a\right)
\left(\frac{\ell_1+\ell_2}{Y}\right)\right]\gamma_0 + {\cal O}(\gamma_0^2).
\label{eq:FWS}
\end{eqnarray}

\begin{figure}
\vbox{
\begin{center}
\epsfig{file=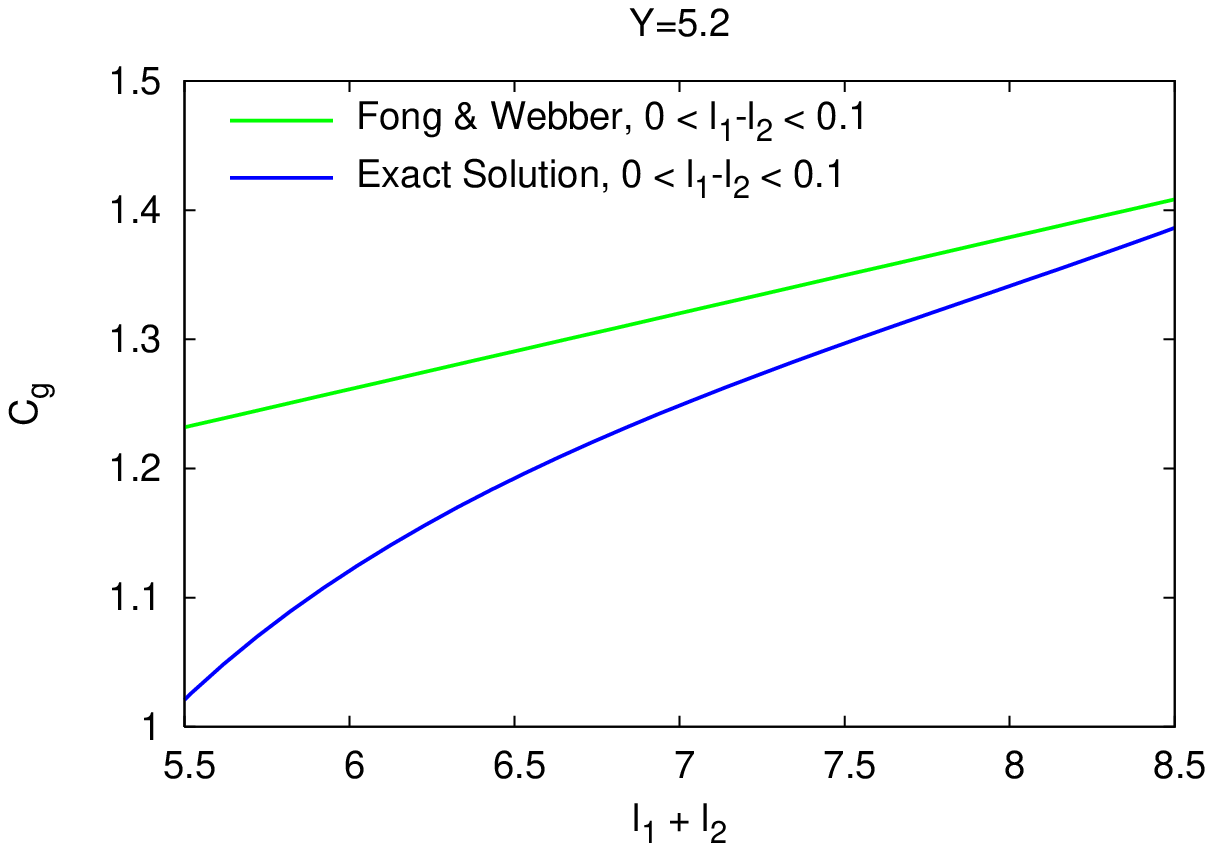, height=5truecm,width=0.48\tw}
\hfill
\epsfig{file=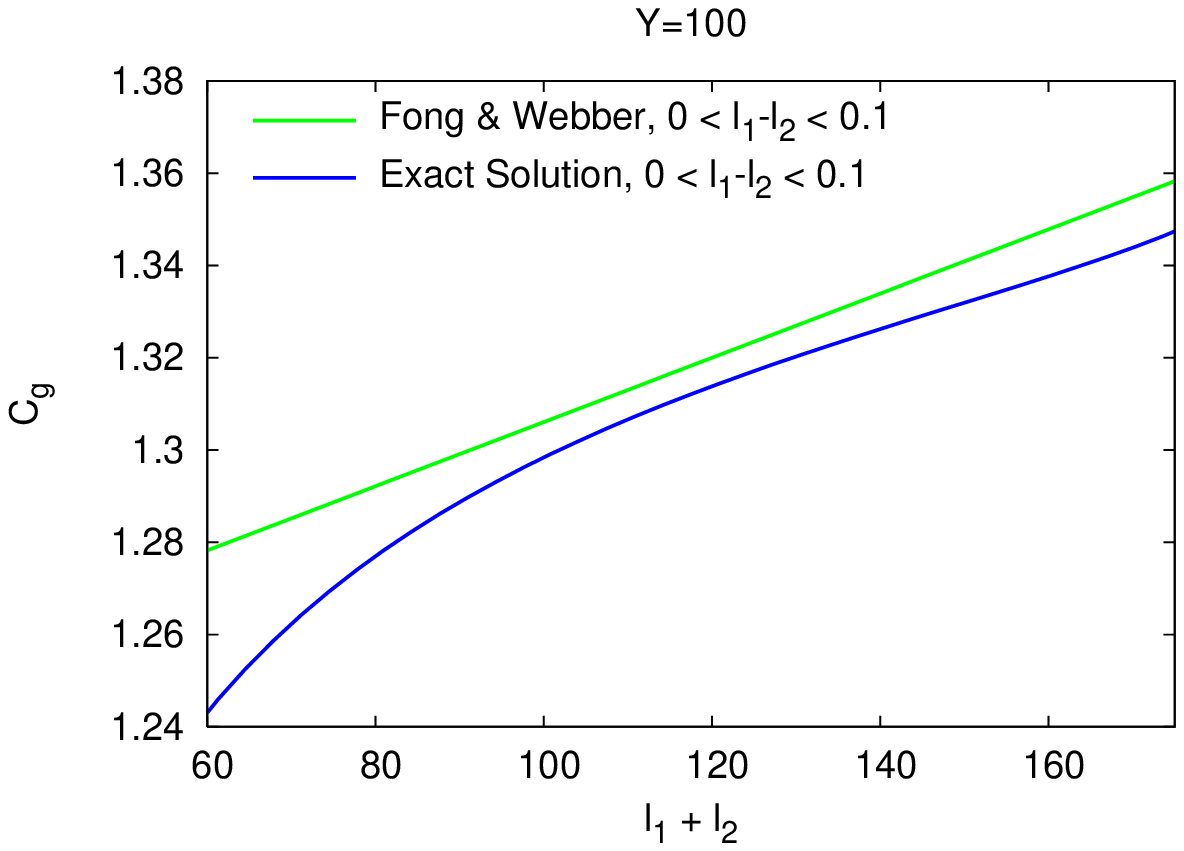, height=5truecm,width=0.48\tw}
\vskip .5cm
\caption{Exact ${\cal C}_g$ compared with Fong-Webber's at $Y=5.2$ (left) and 
$Y=100$ (right)}
\label{fig:FWES}
\end{center}
}
\end{figure}

In Fig.~\ref{fig:FWES} we compare, choosing for pedagogical reasons $Y=5.2$ and
$Y=100$, our exact solution of the evolution equation with
the Fong-Webber predictions \cite{FW} for two particle correlations. 
The mismatch in both cases is, as seen on (\ref{eq:FWS}),  
${\cal O}(\gamma_0^2)$, and decreases for
smaller values of the perturbative expansion parameter $\gamma_0$. 
In particular, at $Y=100$, ($\gamma_0^2\simeq0.01$) 
the exact solution 
(\ref{eq:CGfull}) gets close to (\ref{eq:FWS}).
This comparison is analogous in the case of a quark jet.

We do not perform in the present work such an expansion but keep instead the ratios
(\ref{eq:CGfull}) and  (\ref{eq:Qcorr}) as exact solutions of the evolution equations.

\subsection{Predictions for Tevatron and LHC}
\label{subsection:tevatronlhc}

In hadronic high energy colliders, the nature of the jet (quark or gluon) is not
determined, and one simply detects outgoing hadrons, which can originate
from  either type; one then introduces a ``mixing'' parameter $\omega$,
which is to be determined experimentally, such that, the expression for two particle
correlations can be written as a linear combination of ${\cal C}_g$ and ${\cal C}_q$

\begin{equation}
{\cal C}^{mixed}(\omega;\ell_1,\ell_2,Y)=A(\omega;\ell_1,\ell_2,Y)\,
{\cal C}_q(\ell_1,\ell_2,Y)+ B(\omega;\ell_1,\ell_2,Y)\,{\cal C}_g(\ell_1,\ell_2,Y),
\label{eq:corrmix}
\end{equation}

where

$$
A(\omega;\ell_1,\ell_2,Y)=
\frac{\omega\, \left[\displaystyle{\frac{Q(\ell_1,Y)}{G(\ell_1,Y)}}
\displaystyle{\frac{Q(\ell_2,Y)}{G(\ell_2,Y)}}\right]}
{\bigg[1+\omega\bigg(\displaystyle{\frac{Q(\ell_1,Y)}{G(\ell_1,Y)}}-1\bigg)\bigg]
\bigg[1+\omega\bigg(\displaystyle{\frac{Q(\ell_2,Y)}{G(\ell_2,Y)}}-1\bigg)\bigg]}
$$

and

$$ B(\omega;\ell_1,\ell_2,Y)=
\frac{(1-\omega)}{\bigg[1+\omega\bigg(\displaystyle{\frac{Q(\ell_1,Y)}{G(\ell_1,Y)}}
-1\bigg)\bigg]
\bigg[1+\omega\bigg(\displaystyle{\frac{Q(\ell_2,Y)}{G(\ell_2,Y)}}-1\bigg)\bigg]}.
$$

\begin{figure}
\begin{center}
\epsfig{file=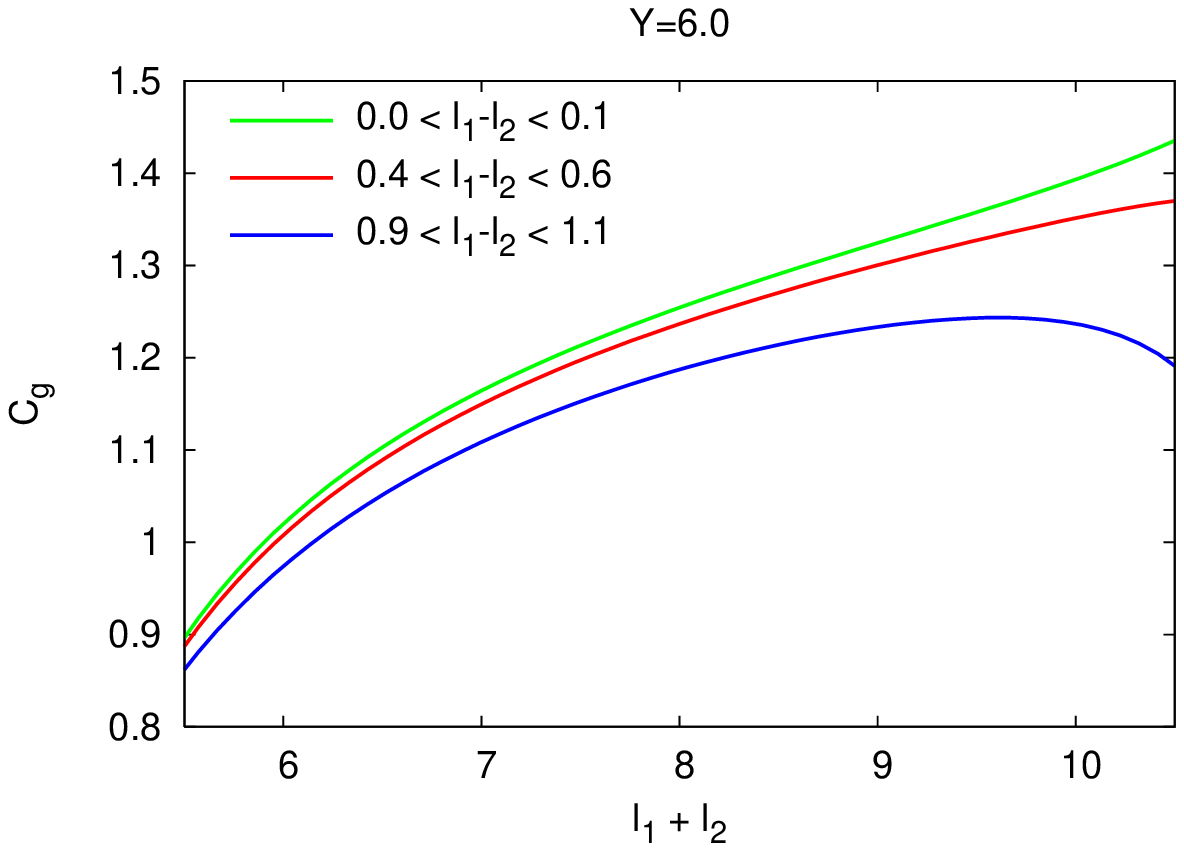, height=6truecm,width=0.47\tw}
\epsfig{file=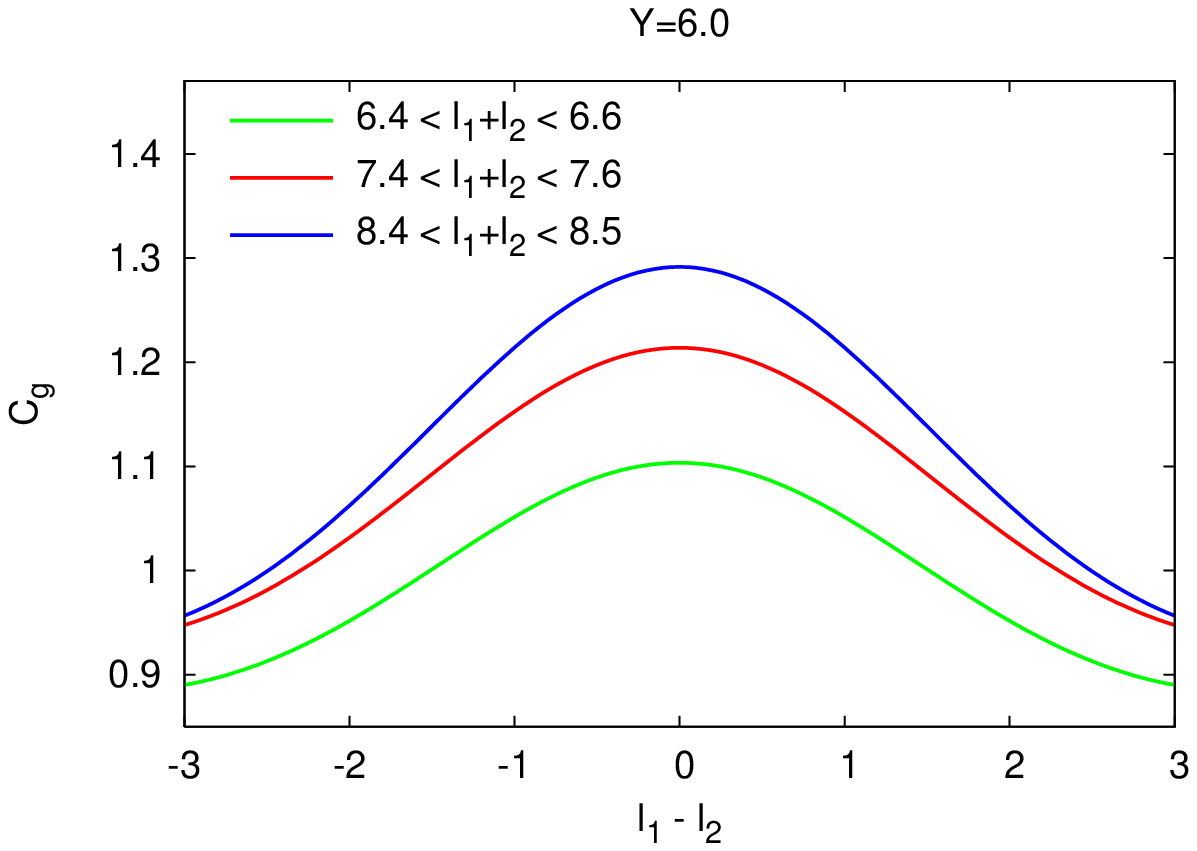, height=6truecm,width=0.47\tw}
\vskip .5cm
\caption{${\cal C}_g$ for the Tevatron ($Y=6.0$) as function of $\ell_1+\ell_2$ (left)
and of $\ell_1-\ell_2$ (right)}
\label{fig:3gbandsTeV}
\end{center}
\end{figure}

\begin{figure}
\begin{center}
\epsfig{file=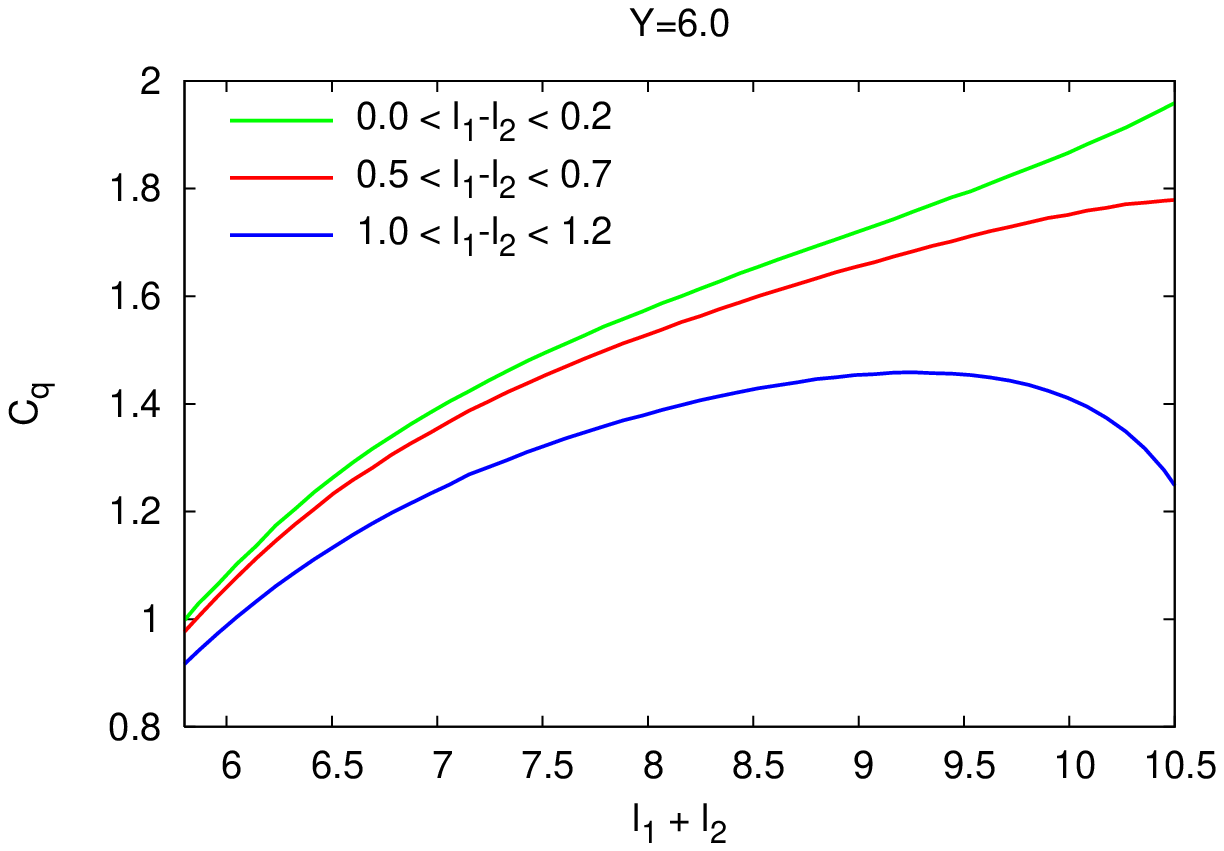, height=6truecm,width=0.47\tw}
\epsfig{file=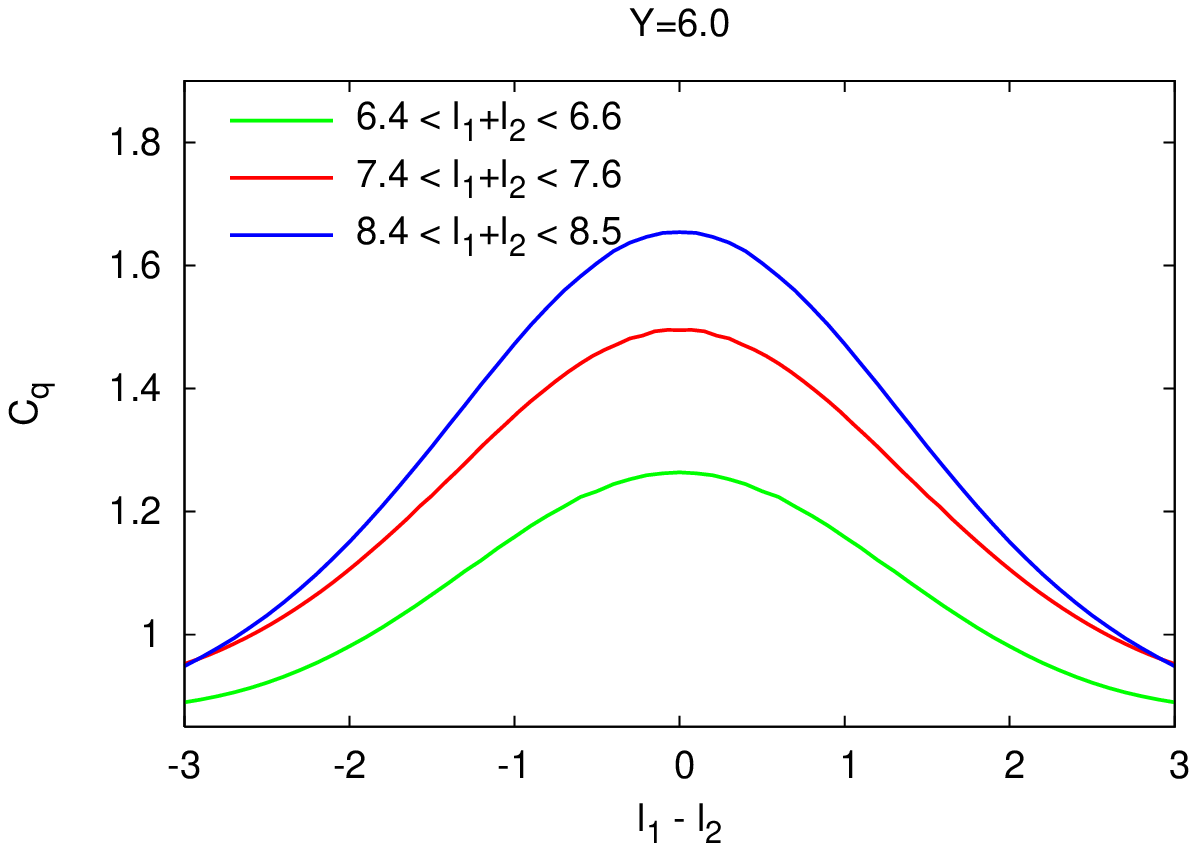, height=6truecm,width=0.47\tw}
\vskip .5cm
\caption{${\cal C}_q$ for the Tevatron ($Y=6.0$) as function of $\ell_1+\ell_2$ (left)
and of $\ell_1-\ell_2$ (right)}
\label{fig:3qbandsTeV}
\end{center}
\end{figure}

\begin{figure}
\begin{center}
\epsfig{file=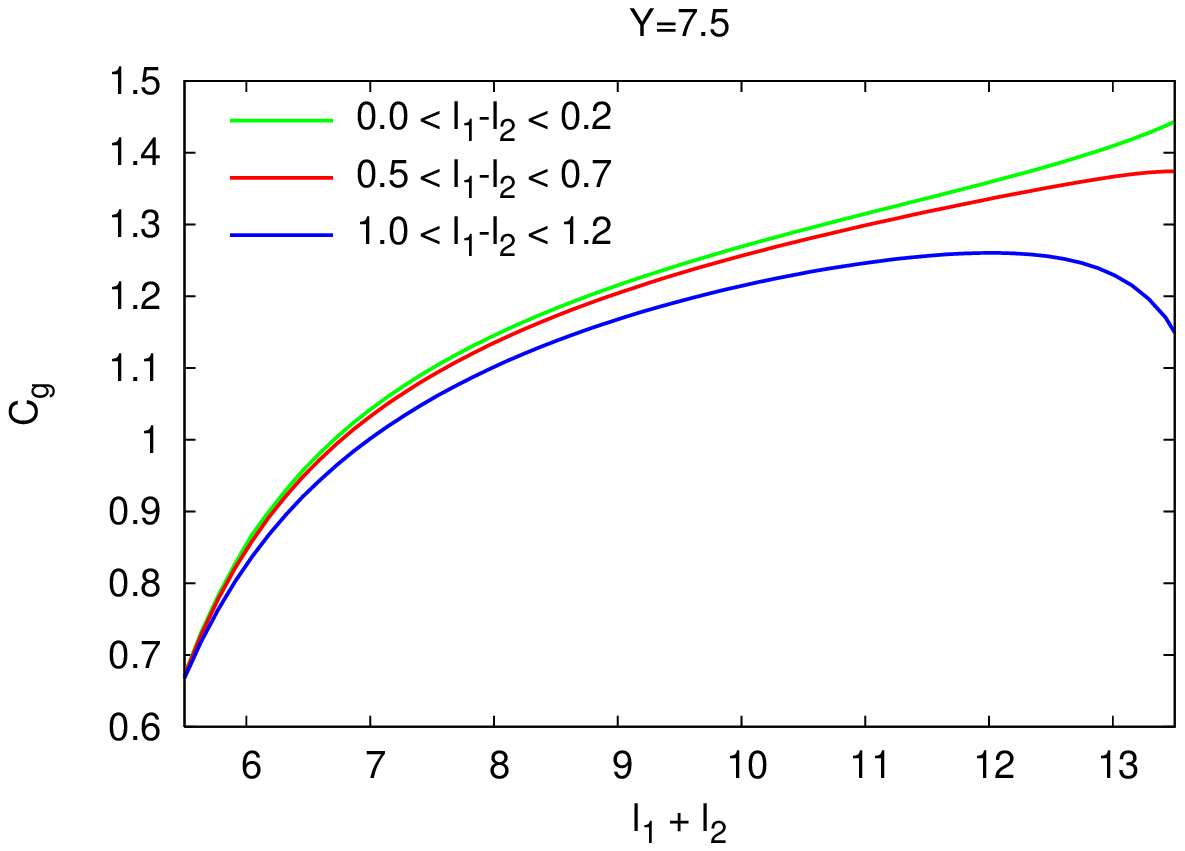, height=6truecm,width=0.47\tw}
\epsfig{file=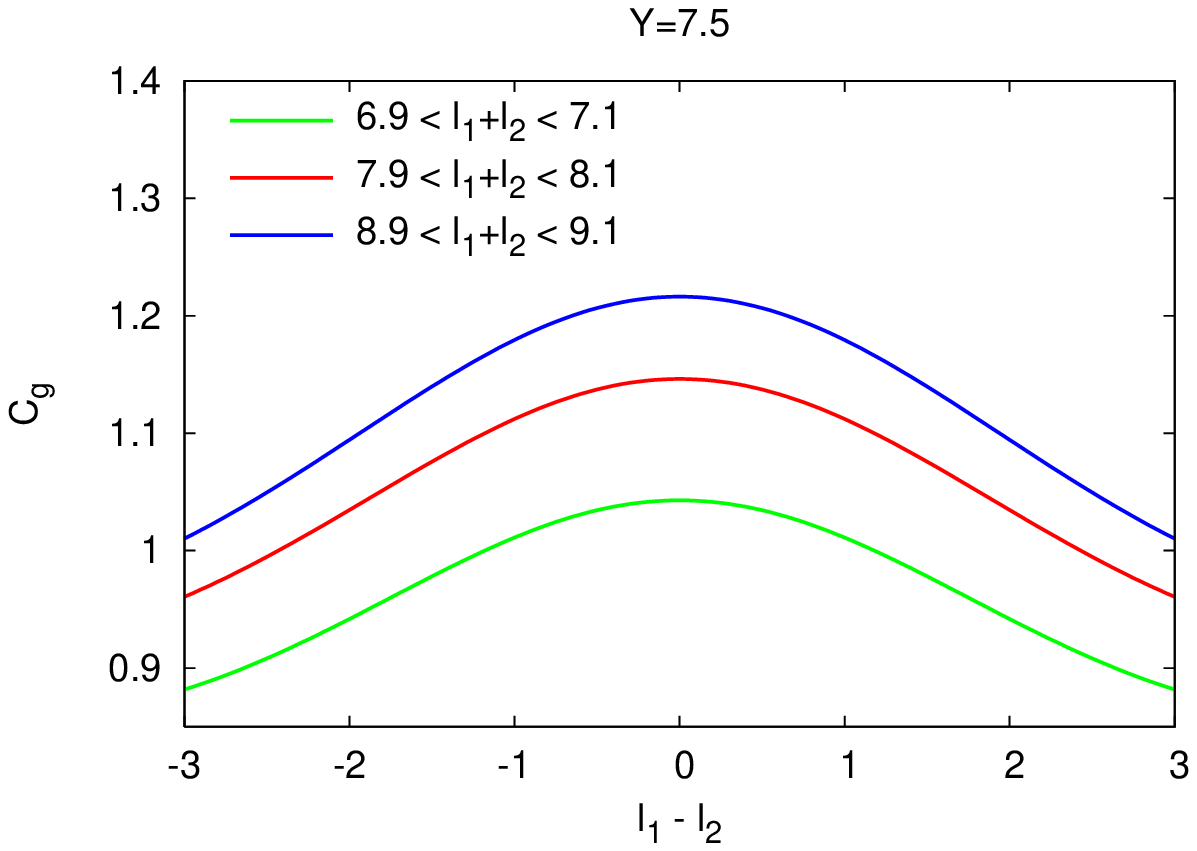, height=6truecm,width=0.47\tw}\vskip .5cm
\caption{ ${\cal C}_g$ for the LHC ($Y=7.5$) inside a gluon jet as function of $\ell_1+\ell_2$ (left) and of $\ell_1-\ell_2$ (right)}
\label{fig:3gbandsLHC}
\end{center}
\end{figure}

\begin{figure}
\begin{center}
\epsfig{file=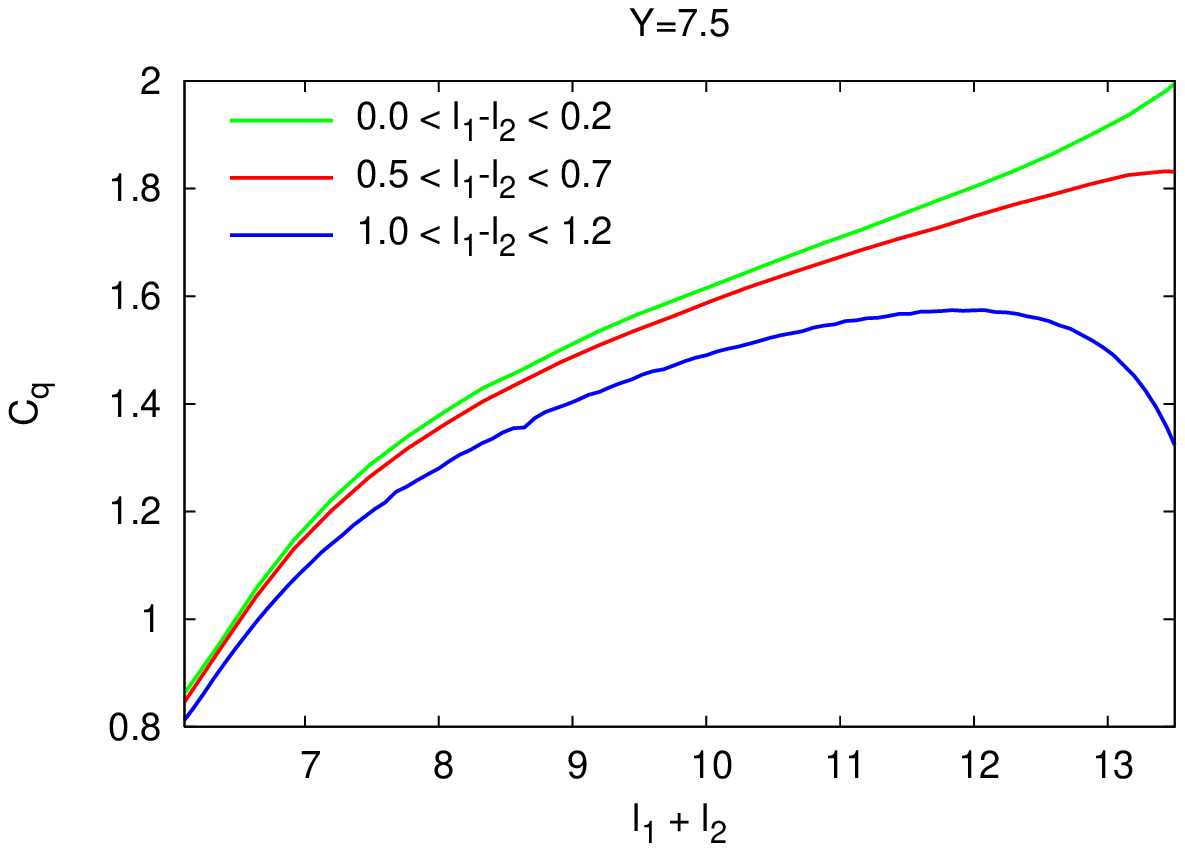, height=6truecm,width=0.47\tw}
\epsfig{file=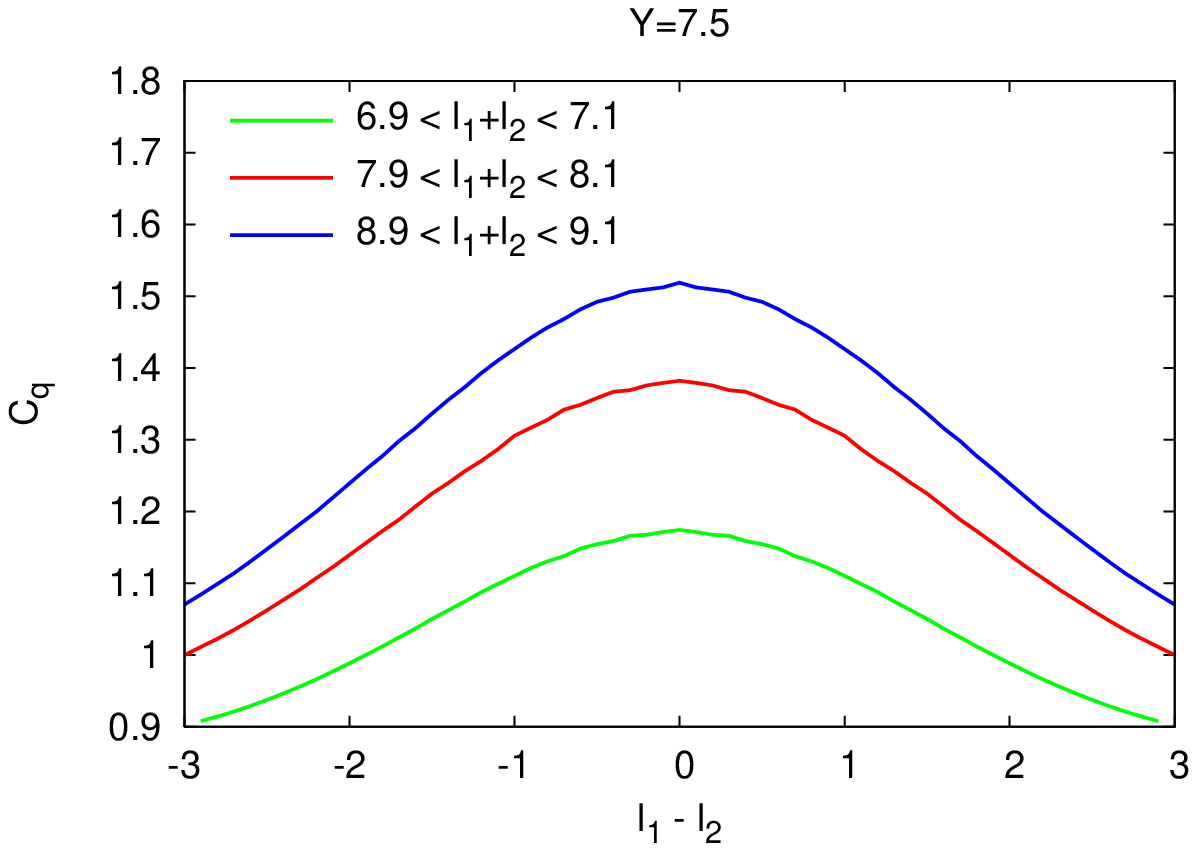, height=6truecm,width=0.47\tw}
\vskip .5cm
\caption{${\cal C}_q$ for the LHC ($Y=7.5$) inside a gluon jet as function of $\ell_1+\ell_2$ (left) and of $\ell_1-\ell_2$ (right)}
\label{fig:3qbandsLHC}
\end{center}
\end{figure}

We plug in respectively (\ref{eq:CGfull}) (\ref{eq:Qcorr}) for 
${\cal C}_g$ and ${\cal C}_q$; the predictions for the latter are given in
Figs.~\ref{fig:3gbandsTeV} and \ref{fig:3qbandsTeV} for the Tevatron,
Figs.~\ref{fig:3gbandsLHC} and \ref{fig:3qbandsLHC} for the LHC.



\subsubsection{Comments}

For both $Y=6.0$ (Tevatron) and $Y=7.5$ (LHC), the global behavior given in 
\ref{subsubsection:commentLEP} also holds. 
The interval corresponding to the condition ${\cal C}_{g\,or\,q}>1$ is shifted toward larger values of $\ell$ (smaller $x$) as compared with the $Y=5.2$ case, in agreement 
with the predictions of (\ref{subsection:signG}) and (\ref{subsection:signQ}).
Numerically, this is achieved for $\ell>2.9$ ($\ell>3.2$) at $Y=6.0$ ($Y=7.5$) in
a gluon jet at the Tevatron (LHC). For a quark jet, these values become
respectively $\ell>3.1$ ($\ell>3.3$) and one can check that they are close to
the approximated ones obtained in (\ref{subsection:signG}) and
(\ref{subsection:signQ}). 

One notices that correlations increase as the total energy (Y) increases (LHC $>$ TeV $>$ LEP-I).

\subsection{Asymptotic behavior of $\boldsymbol{{\cal C}_{g\,or\,q}}$}

\begin{figure}
\vbox{
\begin{center}
\epsfig{file=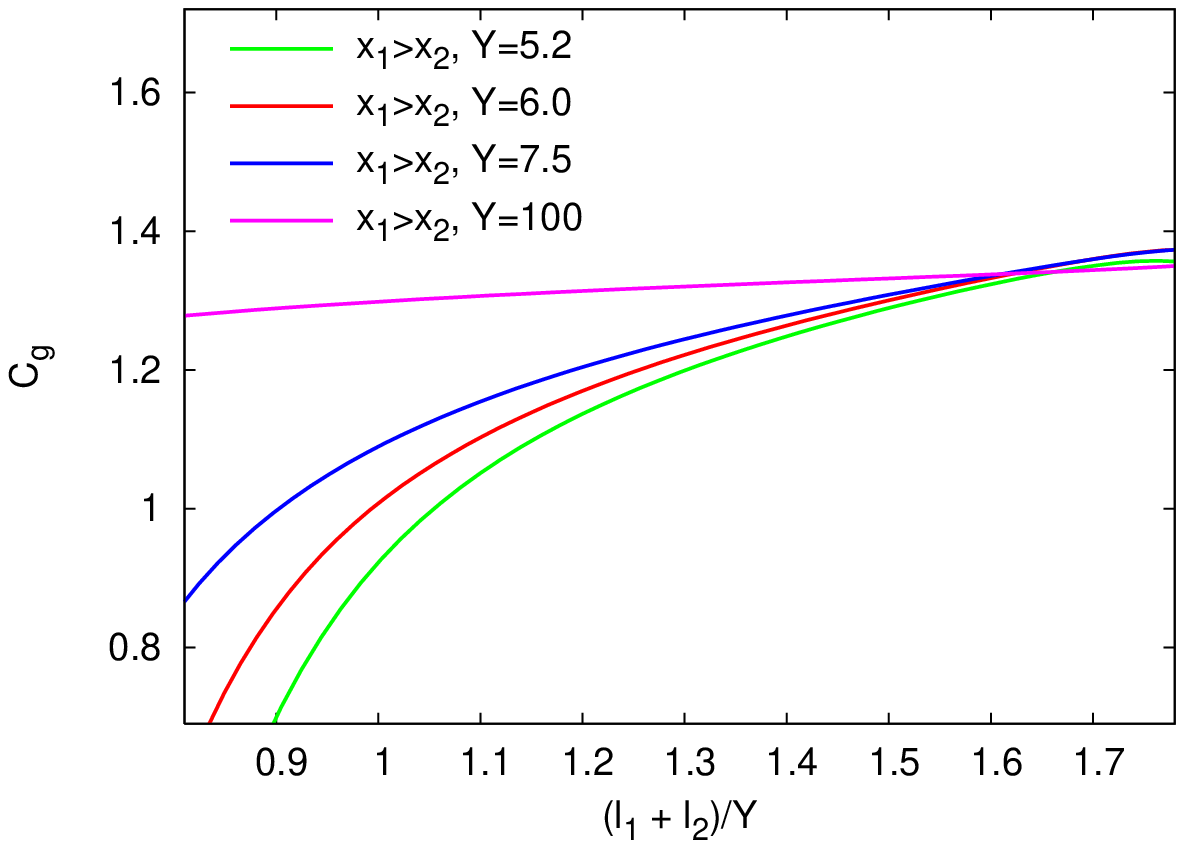, height=7truecm,width=0.48\tw}
\epsfig{file=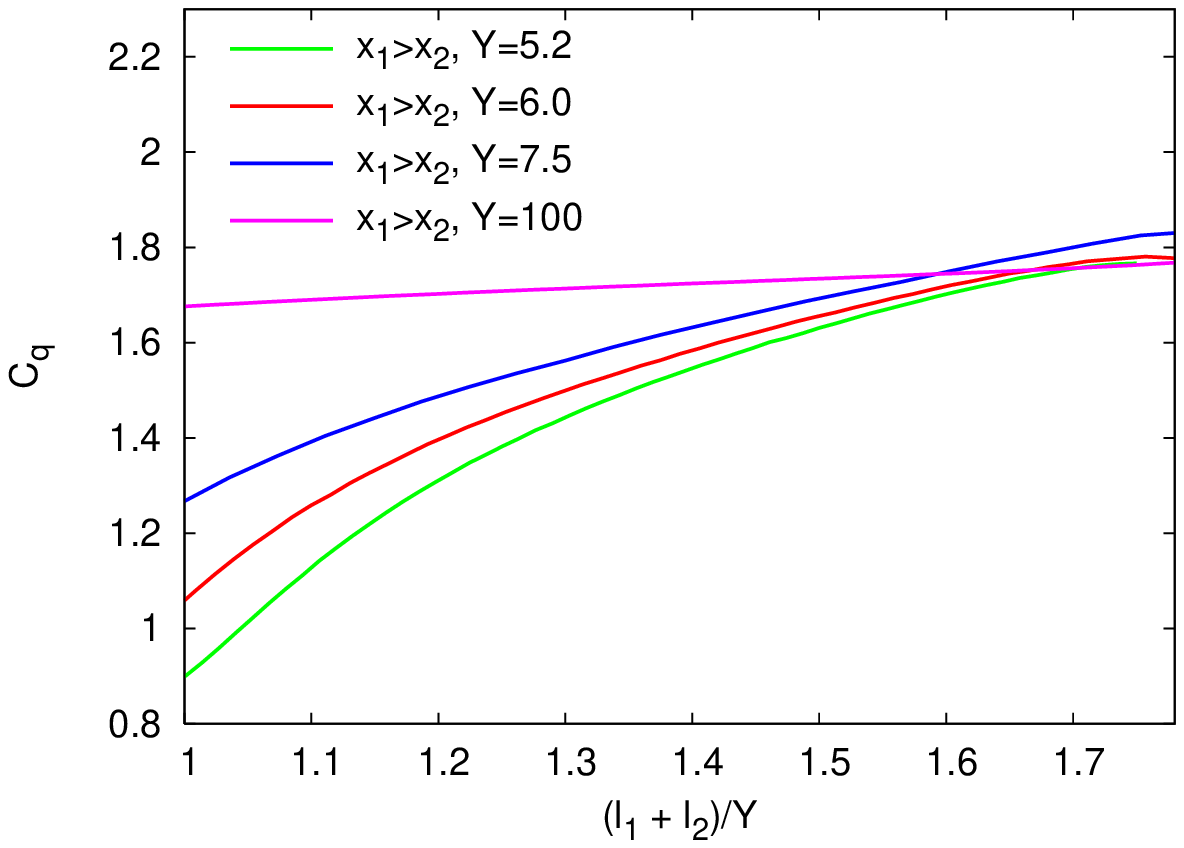, height=7truecm,width=0.48\tw}
\vskip .5cm
\caption{Asymptotic behavior of ${\cal C}_g$ and ${\cal C}_q$ when $Y$ increases}
\label{fig:corr0.1}
\end{center}
}
\end{figure}

We display in Fig.~\ref{fig:corr0.1}  the asymptotic behavior of
${\cal C}_g$ and ${\cal C}_q$ when $Y$ increases. 
$$
{\cal C}_g\stackrel{Y\to\infty}{\longrightarrow}\frac{<n(n-1)>_g}{<n>_g^2}\approx1+\frac13
\approx1.33,\quad 
{\cal C}_q\stackrel{Y\to\infty}{\longrightarrow}\frac{<n(n-1)>_q}{<n>_q^2}\approx1+\frac13
\frac{N_c}{C_F}=1.75,
$$
where $n$ is the multiplicity inside one jet. These limits coincide with
those of the DLA multiplicity correlator \cite{MW}\cite{DFK}.
It confirms the consistency of our approach.
%

\section{CONCLUSION}
\label{section:conclusion}
In this paper two particle correlations between soft partons in quark
and gluon jets were considered.

Corresponding evolution equations for parton correlators were derived in
the next to leading approximation of perturbative QCD, known as MLLA,
which accounts for QCD coherence (angular ordering) on soft gluon
multiplication, hard corrections to parton splittings and the running
coupling effects.

The MLLA equations for correlators were analyzed and solved
iteratively. This allowed us to generalize the result previously
obtained by Fong and Webber in \cite{FW} that was valid in the vicinity
of the maximum of the single inclusive parton energy distribution
("hump").

In particular, we have analyzed the regions of moderately small $x$
above which the correlation becomes "negative" (${\cal C}-1<0$). This happens
when suppression because of the limitation of the phase space takes over
the positive correlation due to gluon cascading.

Also, the correlation vanishes (${\cal C}\to1$) when one of the partons
becomes very soft ($\ell = \ln1/x \to Y=\ln E\Theta/Q_0$). The reason
for that is dynamical rather than kinematical: radiation of a soft
gluon occurs at {\em large angles}\/ which makes the radiation
coherent and thus insensitive to the internal parton structure of the
jet ensemble.

Qualitatively, our MLLA result agrees better with available OPAL data
than the Fong--Webber prediction. There remains however a significant
discrepancy, markedly at very small $x$. In this region
non-perturbative effects are likely to be more pronounced. They may
undermine the applicability {\em to particle correlations}\/ of the local
parton--hadron duality considerations that were successful in
translating parton level predictions to hadronic observations in the
case of more inclusive {\em single particle energy spectra}.

Forthcoming data from Tevatron as well as future studies at LHC should
help to elucidate the problem.

\vskip 1cm

{\underline{\em Acknowledgments}}: It is a great pleasure to thank Yuri Dokshitzer and Bruno Machet for their guidance and encouragements. I thank Fran\c cois Arl\'eo, Bruno Durin for many discussions and Gavin Salam for his expert help in numerical calculations.


\newpage
\appendix

\section{DERIVATION OF THE GLUON CORRELATOR $\boldsymbol{{\cal C}_g}$ IN
(\ref{eq:CGfull})}
\label{section:Gcorr}

One differentiates $G^{(2)} -G_1G_2 \equiv G_1G_2\big({\cal C}_g-1\big)$ 
 with respect to $\ell_1$ and $y_2$  and use the
evolution equations (\ref{eq:solg}) and (\ref{eq:eveeqglu}).
 
By explicit differentiation and using the definitions
(ref{eq:nota4bis})-(\ref{eq:nota4}) one gets
\begin{eqnarray}
\Big[ G_1G_2\big({\cal C}_g-1\big)\Big]_{\ell\,y}\!\!&\!\!=\!\!&\!\!
G_1G_2\Big[{\cal C}_{g,\ell\,y}+{\cal
C}_{g,\ell}\big(\psi_{1,y}+\psi_{2,y}\big) +{\cal
C}_{g,y}\big(\psi_{1,\ell}+\psi_{2,\ell}\big)\Big]\cr
 \!\!&\!\!+\!\!&\!\!({\cal
C}_g-1)\Big[G_1G_2\big(\psi_{1,\ell}\psi_{2,y}+\psi_{2,\ell}\psi_{1,y}\big)
+G_1G_{2,\ell\,y} +G_2G_{1,\ell\,y}\Big];
\label{eq:gc2}
\end{eqnarray}
the definition (\ref{eq:nota4bis}) of $\chi$ entails
${\cal C}_{g,\ell}=\chi_{\ell}{\cal C}_g$,
${\cal C}_{g,y}=\chi_{y}{\cal C}_g$,
${\cal C}_{g,\ell\,y} = {\cal C}_g\big(\chi_{\ell\,y} +
\chi_{\ell}\chi_{y}\big)$, such that (\ref{eq:gc2}) rewrites
\begin{eqnarray}
\Big[G^{(2)} -G_1G_2\Big]_{\ell\,y} \!\!&\!\!=\!\!&\!\!
 {\cal C}_g\;G_1G_2 \Big[
\big(\chi_{\ell\,y} + \chi_{\ell}\chi_{y}\big) 
+ \chi_{\ell}\big(\psi_{1,y}+\psi_{2,y}\big)
+  \chi_{y}\big(\psi_{1,\ell}+\psi_{2,\ell}\big)
\Big]\cr
\!\!&\!\!+\!\!&\!\!({\cal C}_g-1)\Big[
 G_1G_2 \big( \psi_{1,\ell}\psi_{2,y}+\psi_{1,y}\psi_{2,\ell} \big)
 + G_{1} G_{2,\ell\,y}+ G_2G_{1,\ell\,y}
 \Big].
\label{eq:gc3}
\end{eqnarray}
By differentiating the evolution equation for the inclusive spectra
(\ref{eq:solg}) with respect to $y$ and $\ell$ one gets 
\begin{equation}
  G_{k,\ell\,y} =  \gamma_0^2
  \Big(1-a\big(\psi_{k,\ell}- {\beta}\gamma_0^2\big)\Big) G_k,
\label{eq:gc4}
\end{equation}
where one has used the definition (\ref{eq:psi1})(\ref{eq:psi2}) of
$\psi_{k,\ell}$ to replace 
$\frac{dG_k}{d\ell}$ with $G_k \psi_{k,\ell}$, and (\ref{eq:gammabeta})
to evaluate $\frac{d}{d\ell}\gamma_0^2 = -\beta \gamma_0^4$.
Substituting into (\ref{eq:gc3}) yields
\begin{eqnarray}
\frac{l.h.s. (\ref{eq:eveeqglu})\big\vert_{\ell\,y}} {\gamma_0^2G_1G_2 }=
({\cal C}_g-1)\bigg( 2- a\left(\psi_{1,\ell}+\psi_{2,\ell}\right) +
\frac{ \psi_{1,\ell}\psi_{2,y}+\psi_{1,y}\psi_{2,\ell}}
{\gamma_0^{2}} + 2a {\beta}\gamma_0^2 \bigg)
+  {\cal C}_g\big(\delta_1+\delta_2\big), 
\label{eq:gc5}
\end{eqnarray}
where $\delta_1$ and $\delta_2$ are defined in (\ref{eq:delta1})
(\ref{eq:nota4}).

Differentiating now the r.h.s. of (\ref{eq:eveeqglu}) with respect to $y_2$
and $\ell_1$, one gets
\begin{eqnarray}
\frac{r.h.s. (\ref{eq:eveeqglu})\Big\vert_{\ell\,y}}
{\gamma_0^2\,G_1G_2 }=
{\cal C}_g \big( 1-a\left(\psi_{1,\ell} +\psi_{2,\ell} 
    -{\beta}\gamma_0^2\right)\Big)
- {\cal C}_g a\chi_{\ell}
 + (a-b)\left(\psi_{1,\ell} +\psi_{2,\ell}- {\beta}\gamma_0^2 \right).  
\label{eq:rhsdiff2} 
\end{eqnarray}
Equating the expressions (\ref{eq:gc5}) and (\ref{eq:rhsdiff2})
for the correlation function we derive 
\begin{eqnarray} 
({\cal C}_g-1)
\Big(1+\Delta + \delta_1
+ a\left(\chi_{\ell} + {\beta\gamma_0^2}\right) + \delta_2\Big)
= 1-b\big(\psi_{1,\ell} +\psi_{2,\ell} - \beta\gamma_0^2 \big)
-\delta_1 - (a\chi_{\ell}+ \delta_2),
\end{eqnarray}
which gives (\ref{eq:CGfull}).

\section{DERIVATION OF THE QUARK CORRELATOR $\boldsymbol{{\cal C}_q}$ IN
(\ref{eq:Qcorr})}
\label{section:Qcorr}

\subsection{Derivation of (\ref{eq:Qcorr})}
\label{subsection:deriveq}

The method is the same as in appendix \ref{section:Gcorr}:
one evaluates now $\big[Q^{(2)} -{Q_1}{Q_2}\big]_{\ell\,y}
\equiv \Big[{\cal C}_q-1)Q_1Q_2\Big]_{\ell\,y}$.

First, by differentiating the evolution equation (\ref{eq:eveeqq}), one gets
\begin{equation}
\big[Q^{(2)} -{Q_1}{Q_2}\big]_{\ell\,y} =
\frac{C_F}{N_c}\gamma_0^2 {\cal C}_g G_1G_2 
\Big(1-\frac34
\big(\psi_{1,\ell}+\psi_{2,\ell}+\chi_{\ell}-\beta\gamma_0^2\big)\Big);
\label{eq:rhsQ}
\end{equation}
then, one explicitly differentiates $\Big[({\cal C}_q-1)Q_1Q_2\Big]$ and
makes use of
\begin{equation}
Q_{k,\ell\,y}=\frac{C_F}{N_c}\gamma_0^2 G_k\Big(1-\frac34(\psi_{k,\ell}
-\beta\gamma_0^2)\Big),
\label{eq:relQ}
\end{equation}
which comes directly from differentiating the r.h.s of (\ref{eq:solq})
with respect to $\ell$ and $y$; this yields
\begin{eqnarray}
\big[Q^{(2)} -{Q_1}{Q_2}\big]_{\ell\,y}\!\!&\!\!=\!\!&\!\!
{\cal C}_q Q_1Q_2\Big[\sigma_{\ell}\big(\varphi_{1,y} + \varphi_{2,y}\big)
+\sigma_{\ell}\big(\varphi_{1,\ell}+\varphi_{2,\ell}\big) +\sigma_{\ell\,y}
+\sigma_{\ell}\sigma_{y}\Big]\cr
&&+ \big({\cal C}_q-1\big)Q_1Q_2\gamma_0^2
\Big[\varphi_{1,\ell}\varphi_{2,y}+\varphi_{1,y}\varphi_{2,\ell}\Big]\cr
&&\hskip -2cm +\big({\cal C}_q-1\big)\gamma_0^2\frac{C_F}{N_c}\Big[
\big(G_1Q_2+Q_1G_2\big)
-\frac34 G_1Q_2\big(\psi_{1,\ell}-\beta\gamma_0^2\big)
-\frac34 Q_1G_2\big(\psi_{2,\ell}-\beta\gamma_0^2\big)\Big];\cr
&&
\label{eq:explidif}
\end{eqnarray}
equating (\ref{eq:rhsQ}) and (\ref{eq:explidif}) gives
\begin{eqnarray}
{\cal C}_q-1 = \frac{\frac{N_c}{C_F}{\cal C}_g
\bigg[ 1-\frac34\Big(
\psi_{1,\ell}+\psi_{2,\ell}+\chi_{\ell}-\beta\gamma_0^2 \Big)\bigg]
\frac{C_F}{N_c}\frac{G_1}{Q_1} \frac{C_F}{N_c}\frac{G_2}{Q_2}
-{\cal C}_q\big(\tilde\delta_1 +\tilde\delta_2\big)}
{ \widetilde\Delta +\Big[1-\frac34\big(\psi_{1,\ell}-\beta\gamma_0^2\big)\Big]
\frac{C_F}{N_c}\frac{G_1}{Q_1}
+\Big[1-\frac34\big(\psi_{2,\ell}-\beta\gamma_0^2\big)\Big]
\frac{C_F}{N_c}\frac{G_2}{Q_2}},
\label{eq:Qrel}
\end{eqnarray}
which leads (\ref{eq:Qcorr}).

\subsection{Expressing $\boldsymbol{\widetilde\Delta}$,
$\boldsymbol{\tilde\delta_1}$ and
$\boldsymbol{\tilde\delta_2}$ in terms of gluon-related quantities}
\label{subsection:corrections}

All the intricacies of (\ref{eq:Qrel}) lie in $\widetilde\Delta$,
$\tilde\delta_1$ and $\tilde\delta_2$ defined in (\ref{eq:nota5}),
which involve
the quark related quantities $\sigma$ and $\varphi$ (\ref{eq:phisigma}).
In what follows, we will express them in terms of the gluon related
quantities $\chi$ and $\psi$
(\ref{eq:nota4bis})(\ref{eq:psi1})(\ref{eq:psi2}).

\subsubsection{Expression for $\boldsymbol{\tilde\Delta}$}

Differentiating (\ref{eq:ratio}) with respect to $\ell$
yields
\begin{equation}
Q_{k,\ell}=\frac{C_F}{N_c}G_{k,\ell}\bigg[1+\Big(a-\frac34\Big)\psi_{k,\ell}\bigg]+
\frac{C_F}{N_c}G_k\bigg(a-\frac34\bigg)\psi_{k,\ell\,\ell} + {\cal
O}(\gamma_0^4);
\end{equation}
then  
\begin{equation}
\varphi_{\ell}=\frac{Q_{k,\ell}}{Q_k}=\left\{\frac{C_F}{N_c}G_{k,\ell}
\Big[1+\Big(a-\frac34\Big)\psi_{k,\ell}\Big]+
\frac{C_F}{N_c}G_k\Big(a-\frac34\Big)\psi_{k,\ell\,\ell}\right\}\Big[G_{k}^{-1}-
\Big(a-\frac34\Big)\psi_{k,\ell}G_{k}^{-1}\Big]
\end{equation}
yields
\begin{equation}
\varphi_{k,\ell} = \psi_{k,\ell} + \Big(a-\frac34\Big)\psi_{k,\ell\,\ell}
+{\cal O}(\gamma_0^4).
\label{eq:varphil}
\end{equation}

Differentiating (\ref{eq:ratio}) with respect to $y$ yields
\begin{equation}
Q_{k,y}=\frac{C_F}{N_c}G_{k,y}\bigg[1+\Big(a-\frac34\Big)\psi_{k,\ell}\bigg]+
\frac{C_F}{N_c}G_k\bigg(a-\frac34\bigg)\psi_{k,\ell\,y} + {\cal
O}(\gamma_0^4),
\end{equation}
and, finally,
\begin{equation}
\varphi_{k,y} = \psi_{k,y} +\Big( a-\frac34\Big) \psi_{k,\ell\,y}
+ {\cal O}(\gamma_0^4).
\label{eq:varphiy2}
\end{equation}
Using (\ref{eq:varphil}) and (\ref{eq:varphiy2}) in
$\widetilde{\Delta}$ given by (\ref{eq:nota5}) gives

\begin{equation}
\widetilde\Delta \approx
\Delta +\Big(a-\frac34\Big)
\Big(\psi_{1,\ell\,y}\psi_{2,\ell}
+\psi_{2,\ell\,\ell}\psi_{1,y} +\psi_{2,\ell\,y}\psi_{1,\ell}
+\psi_{1,\ell\,\ell}\psi_{2,y} \Big)\gamma_0^{-2},
\label{eq:Deltatilde}
\end{equation}
which shows in particular, that
$\widetilde\Delta \approx \Delta +{\cal O}(\gamma_0^2)$.

\subsubsection{Expression for $\boldsymbol{\tilde\delta_1,
\tilde\delta_2}$}

(\ref{eq:nota5}) entails
${\cal C}_q \gamma_0^2 \tilde\delta_1 = {\cal
C}_{q,\ell}\big(\varphi_{1,y}+\varphi_{2,y}\big) + {\cal
C}_{q,y}\big(\varphi_{1,\ell}+\varphi_{2,\ell}\big)$; since ${\cal
C}_{q,\ell}$ and ${\cal C}_{q,y}$ are ${\cal O}(\gamma_0^2)$ and
considering (\ref{eq:varphiy2}) and (\ref{eq:varphil}), we can
approximate
\begin{equation}
{\cal C}_q \gamma_0^2 \tilde\delta_1 = {\cal
C}_{q,\ell}\big(\psi_{1,y}+\psi_{2,y}\big) + {\cal
C}_{q,y}\big(\psi_{1,\ell}+\psi_{2,\ell}\big) +{\cal O}(\gamma_0^5),
\label{eq:delta1ap}
\end{equation}
which needs evaluating ${\cal C}_{q,\ell}$ and ${\cal C}_{q,y}$ in terms of
gluonic quantities.
Actually, since ${\cal C}_q\tilde\delta_1$ and ${\cal C}_q\tilde\delta_2$
occur as MLLA and NMLLA corrections in (\ref{eq:Qrel}),
it is enough to take the leading (DLA) term of ${\cal C}_q$ to estimate them
\begin{equation}
{\cal C}_q^{DLA}=1+\frac{N_c}{C_F}\frac1{1+\Delta}=1-\frac{N_c}{C_F}+\frac{N_c}{C_F}
\Big(1+\frac{1}{1+\Delta}\Big);
\end{equation}
differentiating then over $\ell$ and $y$ yields
\begin{eqnarray}\label{eq:cqell}
{\cal C}_{q,\ell}^{DLA}&=&-\frac{N_c}{C_F}
\frac{\Delta_{\ell}}{\Big(1+\Delta\Big)^2}=\frac{N_c}{C_F}{\cal
C}_{g,\ell}^{DLA},\\
\label{eq:cqy}
{\cal C}_{q,y}^{DLA}&=&-\frac{N_c}{C_F}
\frac{\Delta_{y}}{\Big(1+\Delta\Big)^2}=\frac{N_c}{C_F}{\cal C}_{g,y}^{DLA}.
\end{eqnarray}
Substituting (\ref{eq:cqell}), (\ref{eq:cqy}) into
(\ref{eq:delta1ap})
one gets
\begin{equation}
{\cal C}_q\tilde\delta_1={\cal C}_g\delta_1+{\cal {O}}(\gamma_0^3).
\end{equation}
Likewise, calculating $\gamma_0^2 {\cal C}_q \tilde\delta_2$ needs
evaluating ${\cal C}_{q,\ell\,y}^{DLA}$ in terms of gluonic quantities. Using
(\ref{eq:cqell}) one gets
\begin{equation}
{\cal C}_q\tilde\delta_2={\cal C}_g\delta_2+{\cal {O}}(\gamma_0^4).
\end{equation}
Accordingly, ${\cal C}_q(\tilde\delta_1+\tilde\delta_2)$ can be replaced by 
${\cal C}_g(\delta_1+\delta_2)$ to get the solution (\ref{eq:Qrel}). This 
approximation is used to get the MLLA solution (\ref{eq:QMLLAap}) of 
(\ref{eq:Qrel}).

\section{DLA INSPIRED  SOLUTION OF THE MLLA EVOLUTION EQUATIONS
FOR THE INCLUSIVE SPECTRUM}
\label{section:inspiredDLA}

This appendix completes subsection \ref{subsection:estimate}.
For pedagogical reasons we will estimate the solution of 
(\ref{eq:solg}) when neglecting the running of 
$\alpha_s$ (constant-$\gamma_0^2$) (see \cite{EvEq}\cite{KO} and references therein). 
We perform a Mellin's transformation of
$G(\ell,y)$

\begin{equation}
G\left(\ell,y\right)=\iint_{C}\frac{d\omega\, d\nu}
{\left(2\pi i\right)^2}\,e^{\omega \ell}
\,e^{\nu y}\,{\cal G}\left(\omega,\nu\right).
\label{eq:GG}
\end{equation}

The contour C lies to the right of all singularities. In (\ref{eq:solg})
one set the lower
bounds for $\ell$ and $y$ to $-\infty$ since these integrals are
vanishing when one closes the C-contour to the right. Using the Mellin's representation
for $\delta(\ell)$

\begin{equation}
\delta(\ell)=\iint_C\frac{d\omega d\nu}
{\left(2\pi i\right)^2}\,e^{\omega \ell}\,e^{\nu y}\frac{1}{\nu},
\label{eq:deltaell}
\end{equation}

one gets

\begin{equation}\label{eq:GSMellin}
{\cal {G}}\left(\omega,\nu\right)=\displaystyle{\frac{1}
{\nu-\gamma_0^2\big(1/\omega-a\big)}}.
\end{equation}

Substituting (\ref{eq:GSMellin}) into (\ref{eq:GG}) and extracting the pole 
($\nu_0=\gamma_0^2\big(1/\omega-a\big)$)
from the denominator of (\ref{eq:GSMellin}) one gets rid of the integration over $\nu$
and obtains the following representation
\footnote{by making use of Cauchy's theorem.}

\begin{equation}
G\left(\ell,y\right)=\int_{C}\frac{d\omega}{2\pi i}\exp{\Big[\omega\ell+\gamma_0^2\big(1/\omega-a\big) y\Big]};
\label{eq:naiveG}
\end{equation}

finally treating $\ell$ as a large variable (soft approximation $x\ll1$)
allows us to have an estimate of (\ref{eq:naiveG}) by performing the steepest
descent method; one then has

\begin{equation}
G(\ell,y)\stackrel{x\ll1}{\approx}
\frac12\sqrt{\frac{\gamma_0\,y^{1/2}}{{\pi\,\ell^{3/2}}}}
\exp{\left(2\gamma_0\sqrt{\ell\,y}-a\gamma_0^2\,y\right)}.
\label{eq:naivesol}
\end{equation}

However, since we are interested in getting logarithmic derivatives;
in this approximation we can drop the normalization factor of (\ref{eq:naivesol})
which leads to sub-leading corrections that we do not take into account here;
we can use instead

\begin{equation}
G(\ell,y)\stackrel{x\ll1}{\simeq}
\exp{\left(2\gamma_0\sqrt{\ell\,y}-a\gamma_0^2\,y\right)},
\label{eq:naivesolbis}
\end{equation}

which is (\ref{eq:Gmod}).

\section{EXACT SOLUTION OF THE MLLA EVOLUTION EQUATION FOR THE
INCLUSIVE SPECTRUM}
\label{section:ESEE}

We  solve  (\ref{eq:solg}) by performing a
Mellin's transformation of the following function ($\gamma_0^2$ 
, $\beta$ and $\lambda$ are defined in (\ref{eq:gammabeta}), (\ref{eq:beta})):

\begin{equation*}
F\left(\ell,y\right)=\gamma_0^2(\ell+y)G\left(\ell,y\right),
\end{equation*}

that is,

\begin{equation}\label{eq:MellRep}
F\left(\ell,y\right)=\iint_{C}\frac{d\omega d\nu}{\left(2\pi
    i\right)^2}\,e^{\omega \ell}\,
 e^{\nu y}\,{\cal F}\left(\omega,\nu\right).
\end{equation}

Substituting (\ref{eq:MellRep}) into (\ref{eq:solg}) we 
obtain:
\begin{eqnarray*}
{\beta}\left(\ell+y+\lambda\right)\iint\frac{d\omega d\nu}{\left(2\pi i\right)^2}\,e^{\omega \ell}\,e^{\nu y}\,{\cal F}\left(\omega,\nu\right)\!\!&\!\!=\!\!&\!\!\iint\frac{d\omega d\nu}{\left(2\pi i\right)^2}\,e^{\omega \ell}\,e^{\nu y}\left[\frac{1}{\nu}+\frac{{\cal F}\left(\omega,\nu\right)}{\omega\nu}\right]\\\nonumber\\
\!\!&\!\!-\!\!&\!\!a\!\!\iint\frac{d\omega d\nu}{\left(2\pi i\right)^2}\,e^{\omega \ell}\,e^{\nu y}\frac{{\cal F}(\omega,\nu)}{\nu},
\end{eqnarray*}

where we have again replaced $\delta(\ell)$ by its Mellin's representation (\ref{eq:deltaell}).
Then using the equivalence $\ell\leftrightarrow\frac{\partial}{\partial\omega},\, y\leftrightarrow\frac{\partial}{\partial\nu}$, we integrate the l.h.s. by parts and obtain:

\begin{eqnarray*}
&&\beta\!\!\iint\frac{d\omega d\nu}{\left(2\pi i\right)^2}\left[\left(\frac{\partial}{\partial\omega}\!+\!\frac{\partial}
{\partial\nu}\!+\!\lambda\right)e^{\omega\ell+\nu y}\right]{\cal F}\left(\omega,\nu\right)
=\beta\!\!\iint\frac{d\omega d\nu}{\left(2\pi i\right)^2}
\left(\lambda {\cal F}\!-\!\frac{\partial {\cal F}}
{\partial\omega}\!-\!\frac{\partial {\cal F}}
{\partial\nu}\right)\,e^{\omega\ell+\nu y}.
\end{eqnarray*}

We are finally left with the following inhomogeneous differential equation:

\begin{equation}\label{eq:diffeq}
\beta\left(\lambda {\cal F}\!-\!\frac{\partial {\cal F}}{\partial\omega}\!-\!\frac{\partial {\cal F}}{\partial\nu}\right)=\frac{1}{\nu}\!+\!\frac{{\cal F}}{\omega\nu}-a\frac{{\cal F}}{\nu}.
\end{equation}

The variables $\omega$ and $\nu$ can be changed conveniently to

$$
\omega'=\frac{\omega+\nu}{2},\qquad \nu'=\frac{\omega-\nu}{2},
$$

such that (\ref{eq:diffeq}) is now decoupled and can be 
easily solved:

\begin{equation*}
\beta\left(\lambda {\cal F}-\frac{d{\cal F}}{d\omega'}\right)=\frac{1}{\omega'-\nu'}+\frac{{\cal F}}{\omega'^2-\nu'^2}
-a\frac{{\cal F}}{\omega'-\nu'}.
\end{equation*}
The solution of the corresponding homogeneous equation, written as a function of
$\omega$ and $\nu$, is the following:

\begin{equation*}
{\cal F}^h\left(\omega,\nu\right)=\frac1{\beta}\int_{0}^{\infty}
\frac{ds}{\nu+s}\left(\frac{\omega\left(\nu+s\right)}
{\left(\omega+s\right)\nu}\right)^{1/\beta\left(\omega-\nu\right)}
\left(\frac{\nu}{\nu+s}\right)^{a/\beta}.
\end{equation*}

We finally obtain the exact solution of (\ref{eq:solg}) given by the following 
Mellin's representation:

\begin{equation}\label{eq:MLLAalphasrun}
G\left(\ell,y\right)=\left(\ell\!+\!y\!+\!\lambda\right)\!\!\iint\frac{d\omega\, d\nu}
{\left(2\pi i\right)^2}e^{\omega\ell+\nu y}
\!\!\int_{0}^{\infty}\frac{ds}{\nu+s}\!\!
\left(\frac{\omega\left(\nu+s\right)}
{\left(\omega+s\right)\nu}\right)^{1/\beta\left(\omega-\nu\right)}\!\!
\left(\frac{\nu}{\nu+s}\right)^{a/\beta}\,e^{-\lambda s}.
\end{equation}

(\ref{eq:MLLAalphasrun}) will be estimated using 
the steepest descent method in a forthcoming work that will treat two particles
correlations at $Q_0\geq\Lambda_{QCD}$
($\lambda=\ln(Q_0/\Lambda_{QCD})\ne0$) \cite{RPR3}\cite{these}.
Substituting (\ref{eq:MLLAalphasrun}) into (\ref{eq:eveeqglu}) one has the Mellin's 
representation inside a quark jet

\begin{eqnarray*}
Q(\ell,y)\!\!=\!\!(\ell\!+\!y\!+\!\lambda)\!\!\iint\frac{d\omega\, d\nu}
{\left(2\pi i\right)^2}e^{\omega\ell+\nu y}
\left(\frac{\gamma_0^2}{\omega\nu}-\frac34\frac{\gamma_0^2}{\nu}\right)
\!\!\int_{0}^{\infty}\frac{ds}{\nu+s}\!\!
\left(\!\frac{\omega\left(\nu+s\right)}
{\left(\omega+s\right)\nu}\!\right)^{1/\beta\left(\omega-\nu\right)}\!\!
\left(\frac{\nu}{\nu+s}\right)^{a/\beta}\!\!e^{-\lambda s};
\end{eqnarray*}
where $\gamma_0^2/\omega\nu={\cal O}(1)$ and the second term is the MLLA 
correction $\gamma_0^2/\nu={\cal O}(\gamma_0)$.

\subsection{Limiting Spectrum, $\boldsymbol{\lambda=0}$}

We set $\lambda=0$ (that is $Q_0=\Lambda_{QCD}$) in (\ref{eq:MLLAalphasrun})
and change variables as follows

$$
\bar{\omega}=\omega-\nu,\quad 
s+\bar{\omega}t=\bar{\omega}/u,\quad A\equiv A(\bar{\omega})=\frac1{\beta\bar{\omega}},\quad
B=a/\beta
$$
to get ($\ell+y=Y$ is used as a variable)

\begin{equation}
    G\left(\ell,Y\right)=\int_{\epsilon_1-i\infty}^{\epsilon_1+i\infty}
    \frac{d\bar{\omega}}{2\pi i}
    x^{-\bar{\omega}}\bar{\omega}Y\int_{\epsilon_2-i\infty}^{\epsilon_2+i\infty}
    \frac{dt}{2\pi i}\,e^{\bar{\omega}Yt}\left(\frac{t}{1+t}\right)^{-A}t^{B}
    \int_{0}^{t^{-1}}\,du\,u^{B-1}\left(1+u\right)^{-A};
\label{eq:red12}
\end{equation}
the last integral of (\ref{eq:red12}) is the representation of the
hypergeometric functions of the second kind (see ~\cite{GR})
\begin{equation*}
 \int_{0}^{t^{-1}}\,du\,u^{B-1}\left(1+u\right)^{-A}
 =\frac{t^{-B}}{B}\,_2F_1\left(A,B;B+1;-t^{-1}\right);
\end{equation*}

for $\Re B>0$, we also have

\begin{equation*}
_2F_1\left(a,b;c;x\right)=\sum_{n=0}^{\infty}
 \frac{\left(a\right)_n\left(b\right)_nx^n}{\left(c\right)_n n!},
\end{equation*}
where for example
$$
\left(a\right)_n=\frac{\Gamma\left(a+n\right)}{\Gamma\left(a\right)}=a\left(a+1\right)
...\left(a+n-1\right).
$$
Therefore (\ref{eq:red12}) can be rewritten in the form:

\begin{eqnarray}\label{eq:red13}
G\left(\ell,Y\right)=
\frac{Y}{B}\int_{\epsilon_1-i\infty}^{\epsilon_1+i\infty}
\frac{d\bar{\omega}}{2\pi  i}
x^{-\bar{\omega}}\bar{\omega}\int_{\epsilon_2-i\infty}^
{\epsilon_2+i\infty}\frac{dt}{2\pi i}\,
 e^{\bar{\omega}Yt}\left(\frac{t}{1+t}\right)^{-A}\,
 _2F_1\left(A,B;B+1;-t^{-1}\right).
\end{eqnarray}

\vskip 0.5cm

By making use of the identity \cite{SDP}:
\begin{eqnarray}
 \left(1+t^{-1}\right)\,_2F_1\left(-A+B+1,1;B+1;-t^{-1}\right)
=\left(\frac{t}{1+t}\right)^{-A}\,_2F_1\left(A,B;B+1;-t^{-1}\right),\notag
\end{eqnarray}
we split (\ref{eq:red13}) into two integrals. The solution of the
second one is given by the hypergeometric function of the first kind
~\cite{SDP}:

\begin{eqnarray}
 \int_{\epsilon_2-i\infty}^{\epsilon_2+i\infty}\frac{dt}{2\pi i}
 \,e^{\bar{\omega}Yt}\,
 t^{-1}\,_2F_1\left(-A+B+1,1;B+1;-t^{-1}\right)
\!=\!_1F_1\left(-A+B+1;B+1;-\bar{\omega}Y\right).
\label{eq:red14}
\end{eqnarray}

Taking the derivative of (\ref{eq:red14}) over $(\bar{\omega}Y)$ we
obtain:

\begin{eqnarray*}
\int_{\epsilon_2-i\infty}^{\epsilon_2+i\infty}
 \frac{dt}{2\pi i}\,e^{\bar{\omega}Yt}\,
 _2F_1\left(-A+B+1,1;B+1;-t^{-1}\right)
\!=\!-\frac{d}{d\left(-\bar{\omega}Y\right)}
 \,_1F_1\left(-A+B+1;B+1;-\bar{\omega}Y\right),
\end{eqnarray*}

where,

\begin{equation*}
 _1F_1\left(a;b;x\right)\equiv\Phi\left(a;b;x\right)=\sum_{n=0}^{\infty}
 \frac{\left(a\right)_nx^n}{\left(b\right)_n n!}.
\end{equation*}

We finally make use of the identity ~\cite{SDP}:

\begin{eqnarray*}
_1F_1\left(-A+B+1;B+2;-\bar{\omega}Y\right)\!\!&\!\!=\!\!&\!\!\frac{B+1}{A}
 \Big[\,_1F_1\left(-A+B+1;B+1;-\bar{\omega}Y\right)\Big.\\
\!\!&\!\!-\!\!&\!\!\Big.\frac{d}{d\left(-\bar{\omega}Y\right)}\,
 _1F_1\left(-A+B+1;B+1;-\bar{\omega}Y\right)\Big]
\end{eqnarray*}

to get ($_1F_1\equiv\Phi$):

\begin{equation}\label{eq:hyprep}
G\left(\ell,Y\right)=\frac{Y}{\beta B\left(B+1\right)}
 \int_{\epsilon-i\infty}^{\epsilon+i\infty}\frac{d\bar{\omega}}{2\pi  i}
 x^{-\bar{\omega}}\Phi\left(-A+B+1,B+2,-\bar{\omega} Y\right);
\end{equation}

we can rename $\bar{\omega}\rightarrow\omega$ and set $Y=\ell+y$, 
which yields

\begin{eqnarray}\nonumber
G\left(\ell,y\right)\!\!&\!\!=\!\!&\!\!\frac{\ell+y}{\beta B\left(B+1\right)}
 \int_{\epsilon-i\infty}^{\epsilon+i\infty}\frac{d{\omega}}{2\pi  i}
 x^{-{\omega}}\Phi\left(-A+B+1,B+2,-{\omega} (\ell+y)\right)\\\nonumber\\
\!\!&\!\!=\!\!&\!\!\frac{\ell+y}{\beta B\left(B+1\right)}
 \int_{\epsilon-i\infty}^{\epsilon+i\infty}\frac{d{\omega}}{2\pi  i}
 e^{-{\omega}y}\Phi\left(A+1,B+2,{\omega} (\ell+y)\right).\label{eq:hyprep1}
\end{eqnarray}

We thus demonstrated that the integral representation (\ref{eq:MLLAalphasrun}) 
is equivalent to (\ref{eq:hyprep}) in the limit $\lambda=0$. In this problem
all functions are derived 
using  (\ref{eq:hyprep1}), and one fixes
the value of $Y=\ln(Q/Q_0)$ (that is fixing the hardness of the process under 
consideration), such that each result is presented as a function of the 
energy fraction in the logarithmic scale $\ell=\ln(1/x)$. As demonstrated in \cite{EvEq}
\cite{PerezMachet}, the inclusive spectrum can be obtained using (\ref{eq:hyprep}) and
the result is

\begin{equation}
G(\ell,Y) = 2\frac{\Gamma(B)}{\beta}
\Re\left( \int_0^\frac{\pi}{2}
  \frac{d\tau}{\pi}\, e^{-B\alpha}\  {\cal F}_B(\tau,y,\ell)\right),
\label{eq:ifD}
\end{equation}
where the integration is performed with respect to $\tau$ defined by
$\displaystyle \alpha = \frac{1}{2}\ln\left(\frac{Y}{\ell}-1\right)  + i\tau$,
\begin{eqnarray}
{\cal F}_B(\tau,\ell,Y) \!\!&\!\!=\!\!&\!\! \left[ \frac{\cosh\alpha
-\displaystyle{\left(1-\frac{2\ell}{Y}\right)}
\sinh\alpha} 
 {\displaystyle \frac{
Y}{\beta}\,\frac{\alpha}{\sinh\alpha}} \right]^{B/2}
  I_B\Big(2\sqrt{Z(\tau,\ell,Y)}\Big), \cr
&& \cr
&& \cr
 Z(\tau,\ell,Y) \!\!&\!\!=\!\!&\!\!
\frac{Y}{\beta}\,
\frac{\alpha}{\sinh\alpha}\,
 \left[\cosh\alpha
-\left(1-\frac{2\ell}{Y}\right)
\sinh\alpha\right]; 
\label{eq:calFdef}
\end{eqnarray}
$I_B$ is the modified Bessel function of the first kind.

\subsection{Logarithmic derivatives of the spectrum, $\boldsymbol{\lambda=0}$}
\label{subsection:Logder}

Using the expressions derived in \cite{PerezMachet} and fixing the sum $\ell+y=Y$,
one gets
\begin{equation}
\frac{d}{d\ell}  G\left(\ell,Y\right)
 = 2\frac{\Gamma(B)}{\beta} \int_0^{\frac{\pi}2}\frac{d\tau}{\pi}\,
 e^{-B\alpha}
 \left[\frac1{Y}\left(1+2e^{\alpha}
\sinh{\alpha}\right){\cal{F}}_B
+\frac1{\beta}e^{\alpha}{\cal{F}}_{B+1}\right];
\label{eq:derivl}
\end{equation}
and
\begin{equation}
\frac{d}{dy} G\left(\ell,y\right)
\!=\! 2 \frac{\Gamma(B)}{\beta}\! \int_0^{\frac{\pi}2}
\frac{d\tau}{\pi}\,  e^{-B\alpha}
 \left[\frac1{Y}
\left(1+2e^{\alpha}\sinh{\alpha}\right)
 {\cal{F}}_B
 +\frac1{\beta}
 e^{\alpha}{\cal{F}}_{B+1}\right.
\left.-\frac{2\sinh\alpha}{Y}{\cal{F}}_{B-1}\right].
\label{eq:derivy}
\end{equation}

Logarithmic derivatives $\psi_{\ell}$ and $\psi_{y}$ are then constructed
according to their definition
 (\ref{eq:psi1})(\ref{eq:psi2}) by dividing (\ref{eq:derivl}) and 
(\ref{eq:derivy}) by the inclusive spectrum (\ref{eq:ifD}).
 
Using the expression of Bessel's series, one gets

$\bullet$\quad for $\ell\rightarrow 0$;
\begin{eqnarray}
\psi_{\ell}&\stackrel{\ell \to 0}{\simeq}&
\frac{a}{\beta\ell} + c_1\ln\left(\frac{Y}{\ell}-1\right)
\to \infty,\cr
c_1 &=& \frac{2^{a/\beta+2}}{\pi(a+2\beta)}\int_{0}^{\pi/2}d\tau\,
(\cos\tau)^{a/\beta+2}
\left[\cos\left(\frac{a}{\beta}\tau\right)-\tan\tau\sin\left(\frac{a}
{\beta}\tau\right)\right]=0.4097>0,\cr
\psi_{y}&\stackrel{\ell\to 0}{\simeq}&-a\gamma_0^2+c_1\frac{\ell}{y}
\to -a\gamma_0^2.
\label{eq:psilzero}
\end{eqnarray}

$\bullet$\quad for $\ell\rightarrow Y \Leftrightarrow y\rightarrow0$;
\begin{eqnarray}
\psi_{\ell}&\stackrel{y\to 0}{\simeq}& c_2\left(\frac{Y}{\ell}-1\right)
\to 0,\cr
c_2 &=&\frac{2^{a/\beta+2}}{\pi(a+2\beta)}\int_{0}^{\pi/2}d\tau\,
(\cos\tau)^{a/\beta+2}
\left[\cos\left(\frac{a}{\beta}\tau\right)+\tan\tau\sin\left(\frac{a}
{\beta}\tau\right)\right]=0.9218>0;\cr
\psi_{y}&\stackrel{y\to 0}{\simeq}&-c_2\ln\left(\frac{Y}\ell-1\right)
\to \infty.
\label{eq:psiyzero}
\end{eqnarray}

\vskip 0.5cm

They are represented in Fig.~\ref{fig:psily} as functions of $\ell$
for two different values of $Y$ ($=5.2, 15$).

\subsection{Double derivatives}
\label{subsection:doublederiv}

In the core of this paper we also need the expression for $\psi_{,\ell,\ell}$

\begin{equation}
\psi_{\ell\,\ell}=\frac1G G_{\ell\,\ell}-(\psi_{\ell})^2.
\end{equation}

By differentiating twice (\ref{eq:hyprep1}) with respect to $\ell$, one gets

\begin{eqnarray}
G_{\ell\,\ell}(\ell,y)&\!\!=\!\!&\frac2{\ell+y}\left(G_{\ell}(\ell,y)-
\frac1{\ell+y}G(\ell,y)\right)\cr
&&\cr
&&+\frac{(\ell+y)\Gamma(B)}{\beta \Gamma(B+3)}
 \int_{\epsilon-i\infty}^{\epsilon+i\infty}\frac{d\omega}{2\pi  i}
 e^{-\omega y}\omega^2
 \left(A^2+3A+2\right)
\Phi\left(A+3,B+4;\omega (\ell+y)\right).\cr
&&
\end{eqnarray}
Using the procedure of \cite{PerezMachet} (appendix A.2) and setting
$y=Y-\ell$, the result for $G_{\ell\,\ell}$ reads

\begin{eqnarray}
G_{\ell\,\ell}(\ell,Y)&\!\!=\!\!&\frac2{Y}\left(G_{\ell}(\ell,Y)-
\frac1{Y}G(\ell,Y)\right)\cr
&&\cr
&& + 2\frac{\Gamma(B)}{\beta} 
\int_{0}^{\frac{\pi}2} \frac{d\alpha}{\pi}\, e^{-(B-2)\alpha}\left[\frac1{\beta^2}\,{\cal{F}}_{B+2}+\frac6{\beta Y}\sinh\alpha\,
{\cal{F}}_{B+1}+\frac8{Y^2}\sinh^2\alpha\,{\cal{F}}_{B}\right].\cr
&&
\end{eqnarray}

Likewise, for
\begin{equation}
\psi_{y\,y}=\frac1G G_{y\,y}-(\psi_{y})^2,
\end{equation}
where

\begin{eqnarray}
G_{y\,y}(\ell,y)&\!\!=\!\!&\gamma_0^2 G(\ell,y) + \frac1{Y}\bigg(G_{y}(\ell,y)-G_{\ell}
(\ell,y)\bigg)\cr
&&\cr
&& +
\frac1{\beta}\frac{(\ell+y)\Gamma(B)}{\Gamma(B+2)}\int_{\epsilon-i\infty}^
{\epsilon+i\infty}\frac{d\omega}{2\pi  i} e^{-\omega y}
\left(\omega^2-\frac{\omega}{\beta}\right)
\Phi\left(A+1,B+3;\omega (\ell+y)\right),
\end{eqnarray}
one gets

\vbox{
\begin{eqnarray}
G_{y\,y}(\ell,Y)&\!\!=\!\!&\gamma_0^2 G(\ell,Y)
+ \frac1{Y}\bigg(G_{y}(\ell,Y)-G_{\ell}(\ell,Y)\bigg)\cr
&&\cr
&& + 4\frac{\Gamma(B)}{\beta}\int_{0}^{\frac{\pi}2} \frac{d\alpha}{\pi}\, 
e^{-(B+1)\alpha}
\left[2(B+1)\frac{\sinh^2\alpha}{Y^2}\,{\cal{F}}_{B-1}
-\frac1{\beta}\frac{\sinh\alpha}{Y}\,{\cal{F}}_{B}\right].
\end{eqnarray}
}

Finally, 
\begin{eqnarray}\nonumber
\psi_{\ell\, y} = \psi_{y\,\ell} = \gamma_0^2\left[1-a\left(\psi_{\ell}-\beta\gamma_0^2\right)\right]-\psi_{\ell}\psi_{y}. 
\end{eqnarray}
$\psi_{\ell\,\ell}$, $\psi_{y\,y}$ and $\psi_{\ell\, y}$ are drawn 
in Fig.~\ref{fig:doublepsi} of appendix \ref{subsub:psinum} as 
functions of $\ell$ for fixed $Y$. They are all ${\cal {O}}(\gamma_0^3)$.

\section{NUMERICAL ANALYSIS OF CORRECTIONS}
\label{section:numcorr}

In this section, we present plots for the derivatives of $\psi$,
and $\varphi$ (see (\ref{eq:psi1})(\ref{eq:psi2}) and (\ref{eq:phisigma})),
for $\Upsilon$ and its derivatives  (see (\ref{eq:upsg})(\ref{eq:upsq})),
for $\Delta$, $\delta_1$, $\delta_2$ (see
(\ref{eq:nota4bis})-(\ref{eq:nota4}))
and the combination $\delta_c\equiv \delta_1 + \delta_2 + a\Upsilon_{\ell}$,
$\tilde\delta_c\equiv \tilde\delta_1 + \tilde\delta_2$.

\subsection{Gluon jet}
\label{subsection:glucorr}

\subsubsection{$\boldsymbol\psi$ and its derivatives}
\label{subsub:psinum}

This subsection is associated with appendices \ref{subsection:Logder}
and \ref{subsection:doublederiv} . It enables in particular
to visualize the behaviors of $\psi_{\ell}$ and $\psi_{y}$ when $\ell
\to 0$ or $y\to 0$, as described in (\ref{eq:psilzero}) and
(\ref{eq:psiyzero}), and to set the $\ell$ interval within which our calculation
can be trusted.

\begin{figure}
\vbox{
\begin{center}
\epsfig{file=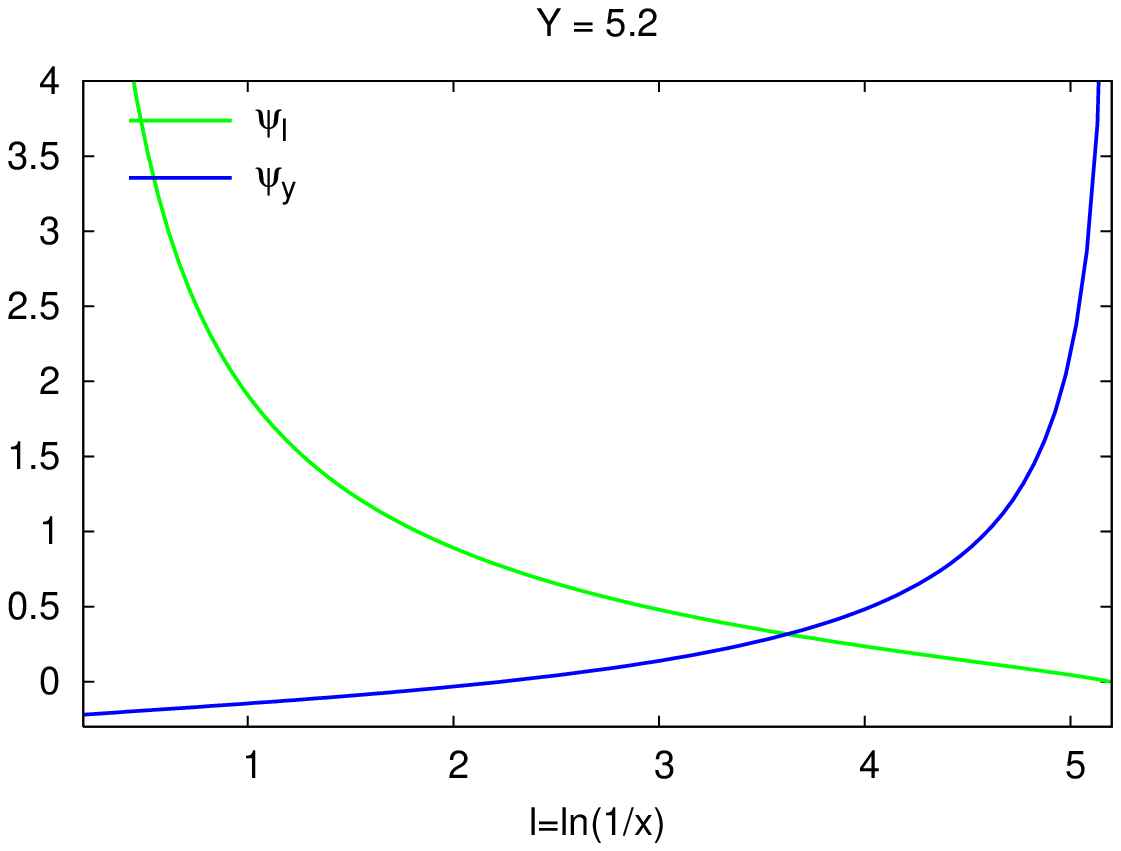, height=6truecm,width=0.45\tw}
\hfill
\epsfig{file=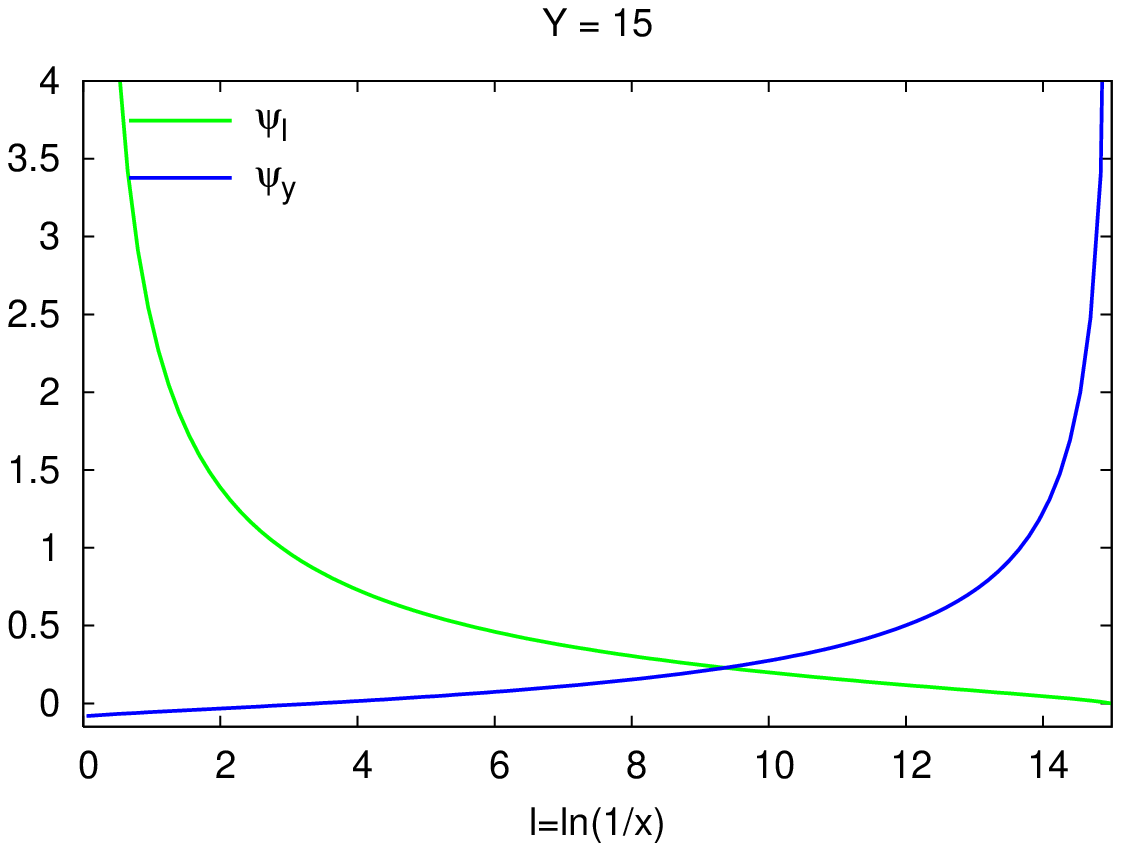, height=6truecm,width=0.45\tw}
\vskip .5cm
\caption{Derivatives $\psi_{\ell}$ and $\psi_{y}$ as functions of $\ell$ at fixed 
$Y=5.2$ (left) and $Y=15$ (right)}
\label{fig:psily}
\end{center}
}
\end{figure}

\begin{figure}
\vbox{
\begin{center}
\epsfig{file=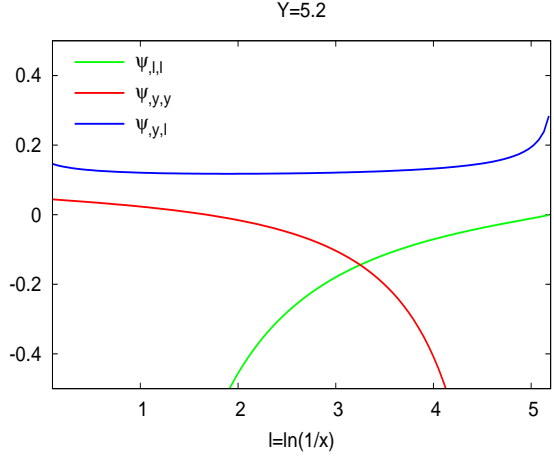, height=6truecm,width=0.45\tw}
\vskip .5cm
\caption{Double derivatives $\psi_{\ell\,\ell}$, $\psi_{\ell\,y}$ and
$\psi_{y\,y}$ as functions of $\ell$ at fixed $Y$}
\label{fig:doublepsi}
\end{center}
}
\end{figure}

In Fig.\ref{fig:psily} are drawn $\psi_{\ell}$ and $\psi_{y}$ as
functions of $\ell$ for two values $Y=5.2$ corresponding to LEP
working conditions, and $Y=15$ corresponding to an unrealistic ``high
energy limit''.

$\psi_{\ell}$ and ($\psi_{y}$) being both ${\cal {O}}(\gamma_0)$,
they should not exceed a ``reasonable value''; setting this value to
$1$, $|\psi_{\ell}|<1$ and  $|\psi_{y}|<1$ set, for $Y=5.2$, a confidence
interval 
\begin{equation}
2.5 \leq \ell \leq 4.5.
\label{eq:confint1}
\end{equation}

In the high energy limit $Y=15$, this interval becomes,  $4.5 \leq \ell
\leq 13$, in agreement with \ref{subsection:signG}.

\subsubsection{$\boldsymbol{\Delta(\ell_1,\ell_2,Y)}$}
\label{subsub:Deltanum}

$\Delta$ has been defined in (\ref{eq:deltabis}), in which $\psi_{1,\ell}$ and
$\psi_{1,y}$ are functions of $\ell_1$ and $Y$, $\psi_{2,\ell}$ and
$\psi_{2,y}$ are functions of $\ell_2$ and $Y$.

Studying the limits $\ell \to0$ and $\ell \to Y$ of subsection 
\ref{subsection:Logder}:

\begin{itemize}
\item for $\ell_1,\,\ell_2\rightarrow Y$ one gets (using the results of
\ref{subsection:Logder})
\begin{equation}
\Delta\simeq-c_2^{\,2}\left(\frac{Y-\ell_1}{\ell_1}\ln\frac{Y-\ell_2}{\ell_2}+
\frac{Y-\ell_2}{\ell_2}\ln\frac{Y-\ell_1}{\ell_1}\right),
\end{equation}
such that
\begin{equation}
\Delta \stackrel{\ell_1-\ell_2\to 0}{\longrightarrow}0,\quad \Delta
\stackrel{\ell_1-\ell_2\to\infty}{\longrightarrow}+\infty.
\end{equation}

\item for $\ell_1,\,\ell_2\rightarrow 0$ one gets (according to
\ref{subsection:Logder}):
\begin{equation}
\Delta\simeq-a\gamma_0^2\left[\frac{a}\beta\left( \frac1\ell_1+\frac1\ell_2 \right)+
c_1\left(\ln\frac{Y-\ell_1}{\ell_1}+\ln\frac{Y-\ell_2}{\ell_2}\right)\right]
\to -\infty.
\end{equation}

\end{itemize}

\begin{figure}
\vbox{
\begin{center}
\epsfig{file=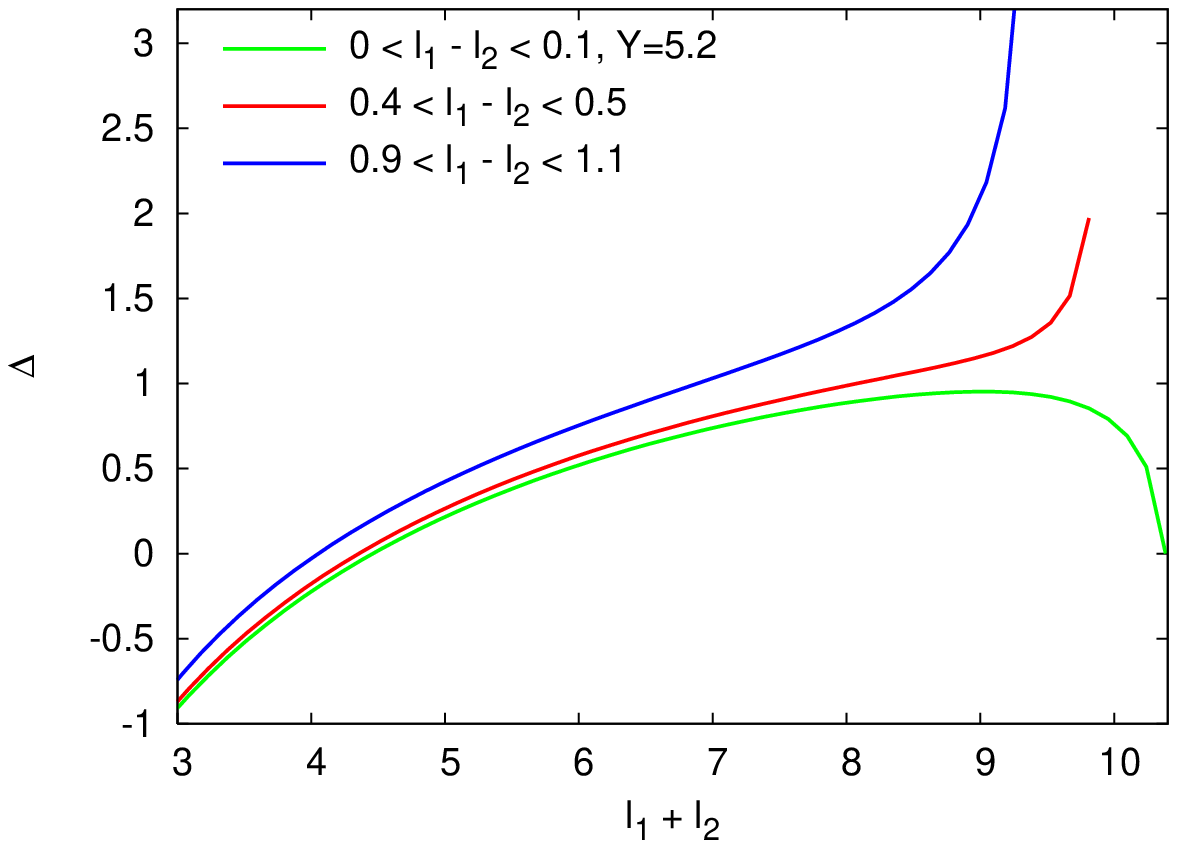, height=6truecm,width=0.47\tw}
\hfill
\epsfig{file=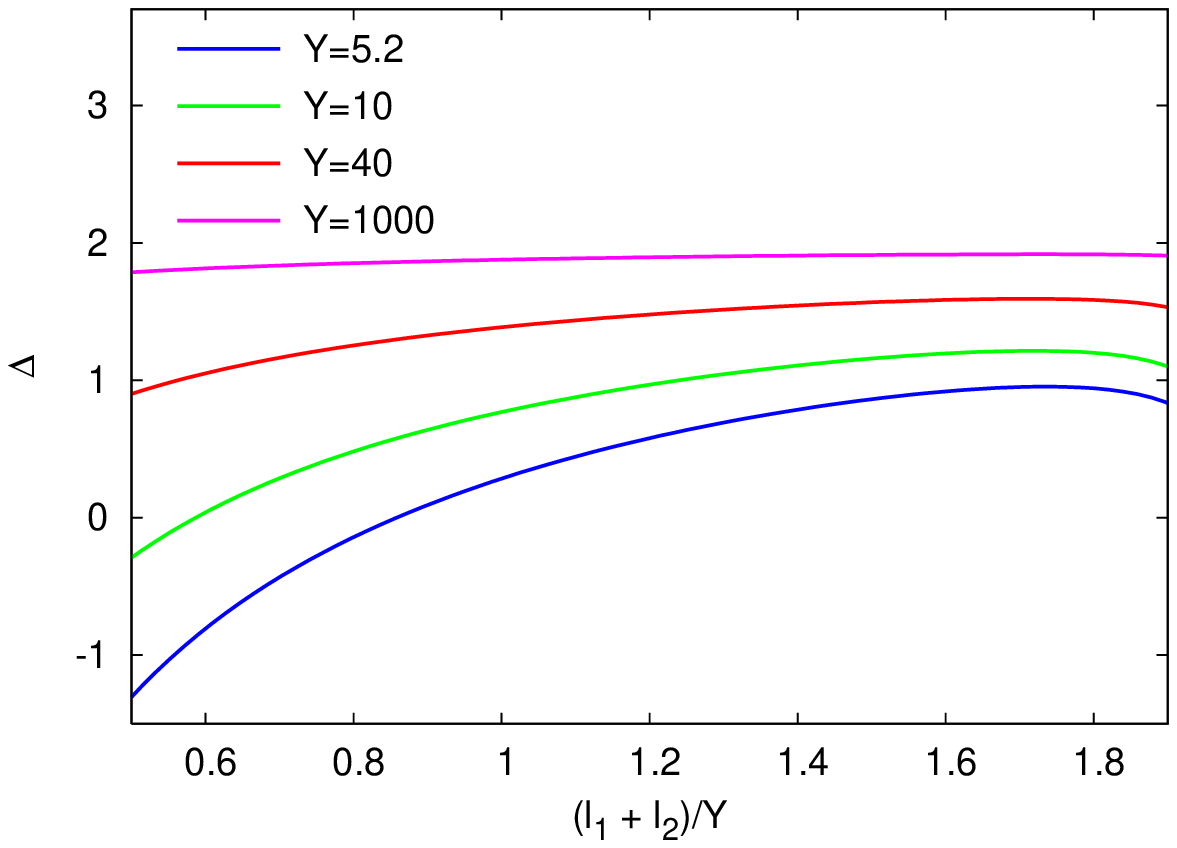, height=6truecm,width=0.47\tw}
\vskip .5cm
\caption{$\Delta$ as a function of $\ell_1+\ell_2$ for $Y=5.2$ (left) 
and its asymptotic behavior (right, $\ell_1-\ell_2=0.1$)}
\label{fig:Delta}
\end{center}
}
\end{figure}

In Fig.~\ref{fig:Delta} (left)  $\Delta$ is plotted as a function
of $\ell_1+\ell_2$ for three different values of $\ell_1-\ell_2$ $(0.1,\,0.5,\,1.0)$;
the condition (\ref{eq:confint1}) translates into
\begin{equation}
5.0\leq\ell_1+\ell_2\leq9.0;
\label{eq:confint2}
\end{equation}
on Fig.~\ref{fig:Delta} (right) the asymptotic limit $\Delta\rightarrow2$ 
for very large $Y$ clearly appears (we have taken $\ell_1-\ell_2=0.1$);
it is actually its DLA
value \cite{EvEq}; this is not surprising since, in the high energy
limit $\gamma_0$ becomes very small and sub-leading
corrections (hard corrections and running coupling effects) get suppressed.

\subsubsection{$\boldsymbol{\Upsilon_g}$ and its derivatives}
\label{subsub:upsgnum}

\begin{figure}
\vbox{
\begin{center}
\epsfig{file=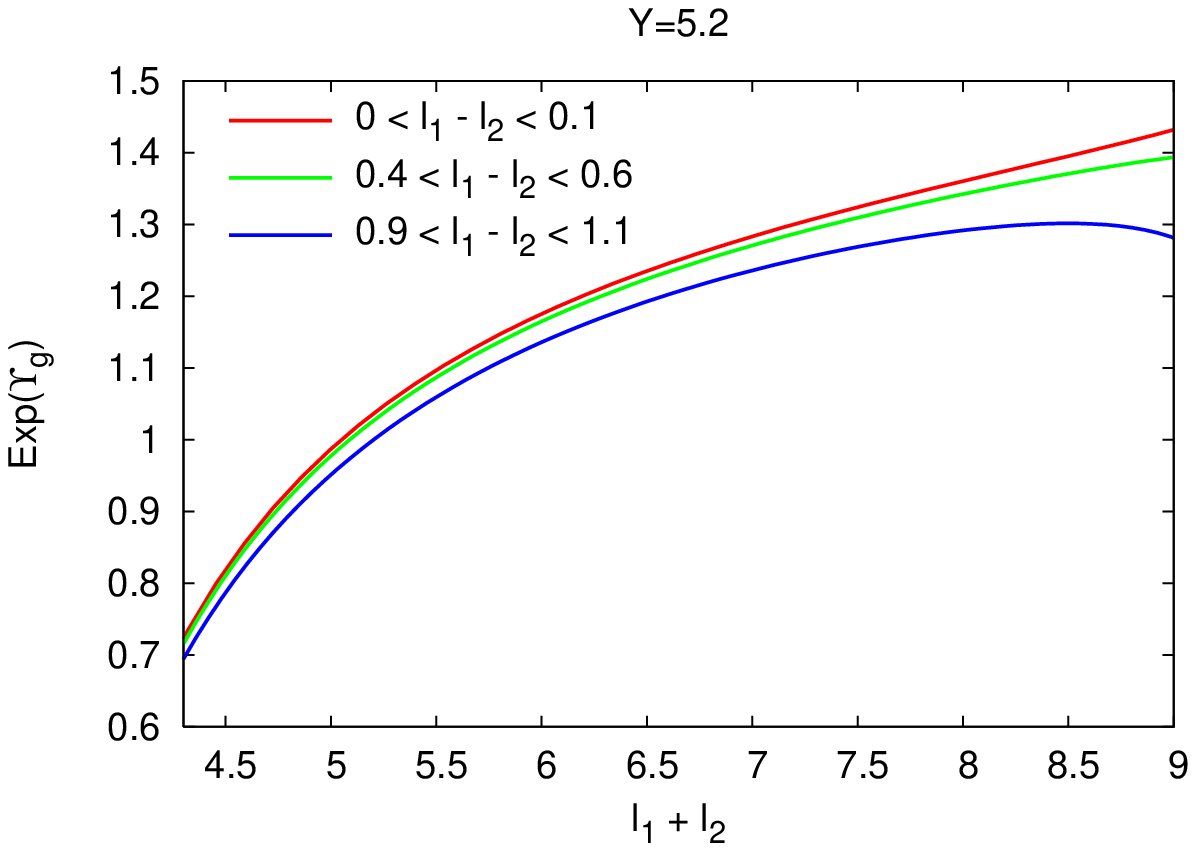, height=6truecm,width=0.47\tw}
\epsfig{file=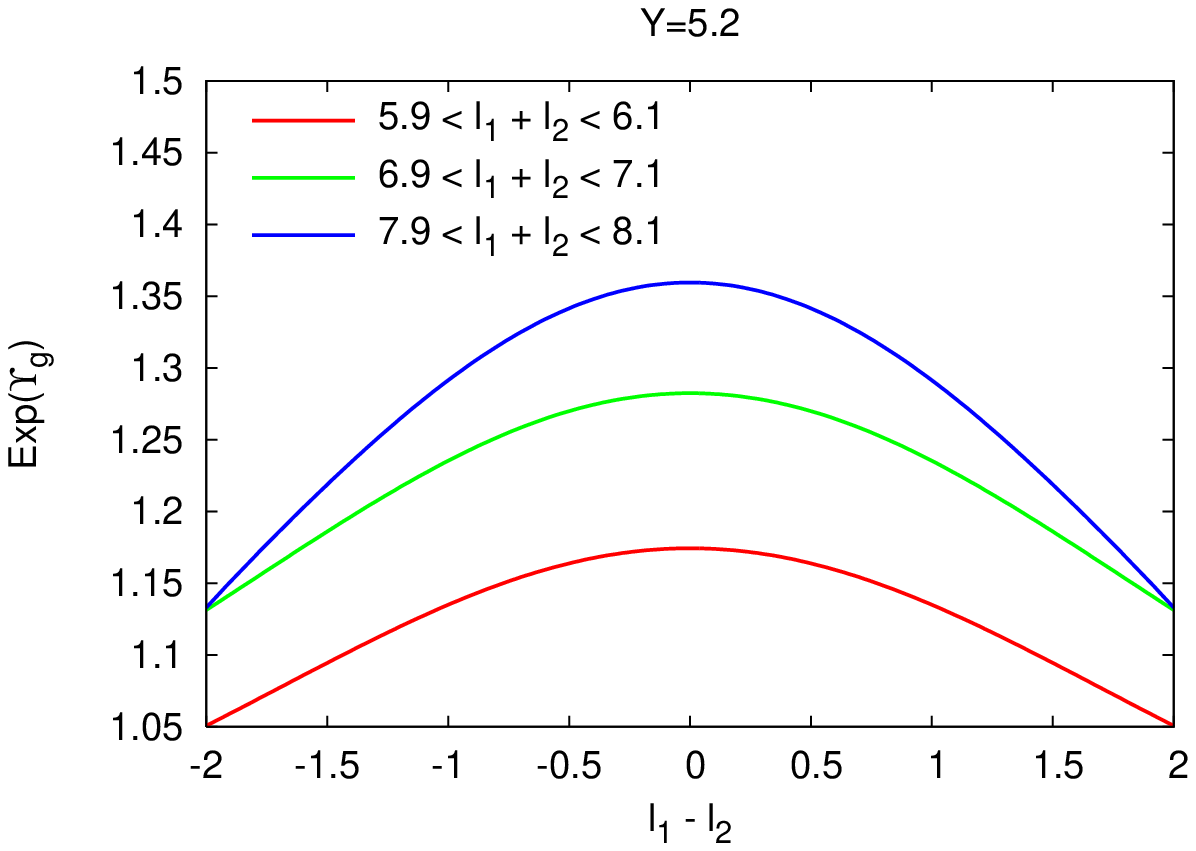, height=6truecm,width=0.47\tw}
\vskip .5cm
\caption{$\exp{(\Upsilon_g)}$ as a function of $\ell_1+\ell_2$ (left) and, 
$\ell_1-\ell_2$ (right) for $Y=5.2$}
\label{fig:Upsilong}
\end{center}
}
\end{figure}

Fig.~\ref{fig:Upsilong} exhibits the smooth behavior of
$\exp{(\Upsilon_g)}$ as a function of $(\ell_1+\ell_2)$ 
in the whole range of applicability of our approximation (we have chosen
the same values of $(\ell_1-\ell_2)$ as for Fig.~\ref{fig:Delta}),
and as a function of
$(\ell_1-\ell_2)$ for three values of $(\ell_1+\ell_2)$ 
($6.0, 7.0, 8.0$). So, the iterative procedure is 
safe and corrections stay under control.

\begin{figure}
\vbox{
\begin{center}
\epsfig{file=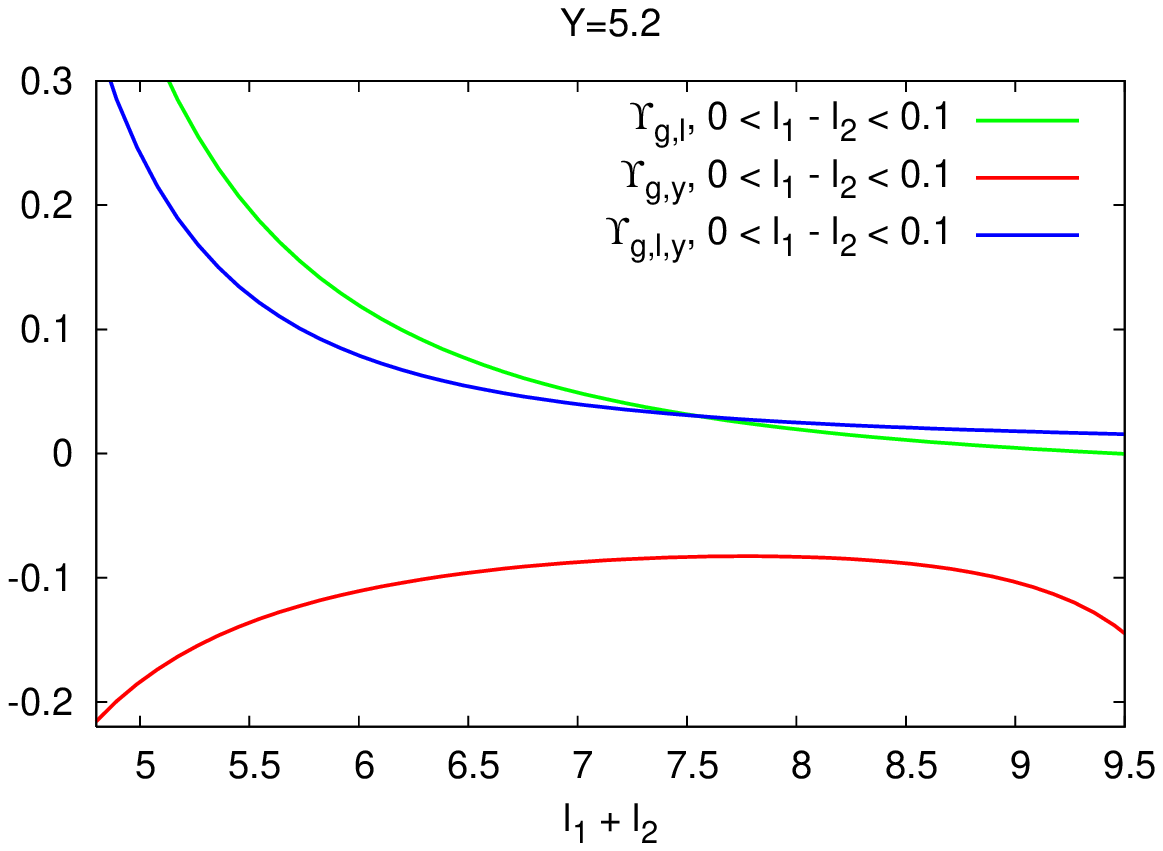, height=6truecm,width=0.47\tw}
\hfill
\epsfig{file=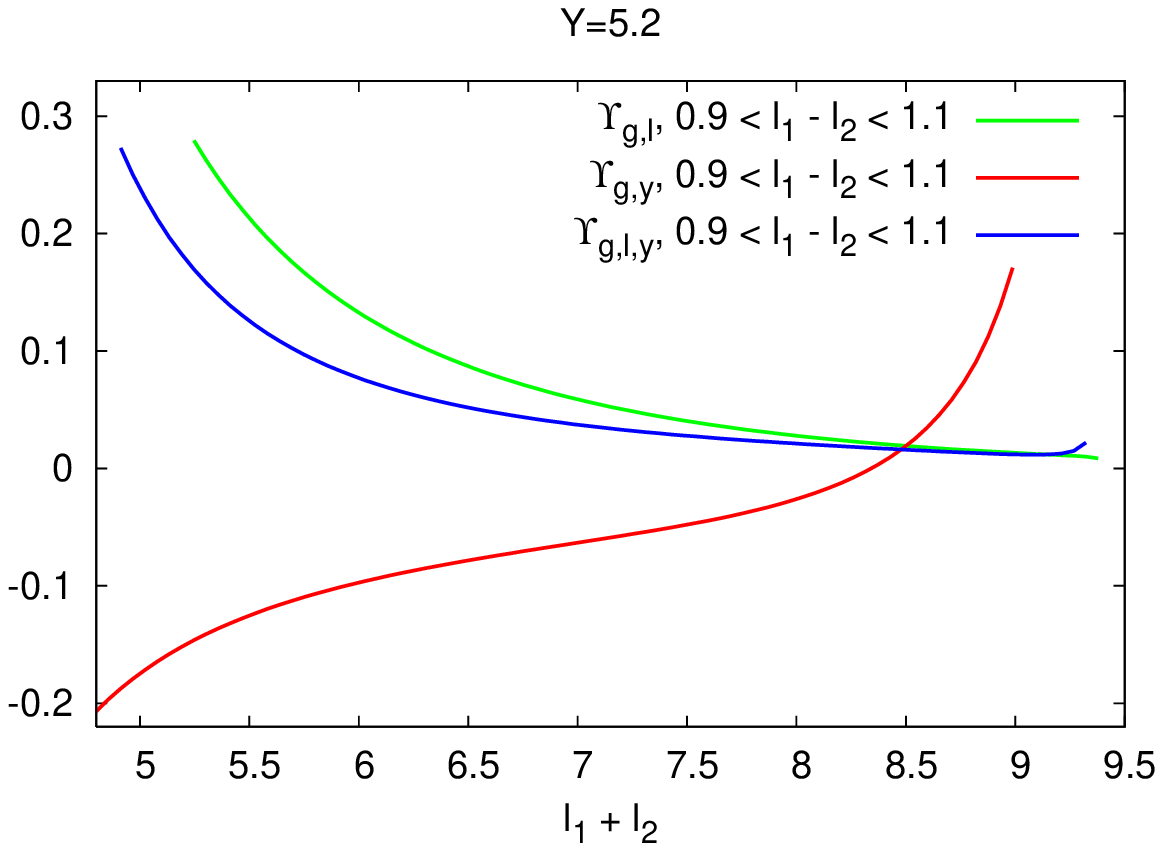, height=6truecm,width=0.47\tw}
\vskip .5cm
\caption{$\Upsilon_{g,\ell}$, $\Upsilon_{g,y}$ and $\Upsilon_{g,\ell\,y}$
as functions of $\ell_1+\ell_2$ for $Y=5.2$, $\ell_1-\ell_2=0.1$ (left) and 
$\ell_1-\ell_2=1.0$ (right)}
\label{fig:DUpsilon}
\end{center}
}
\end{figure}

Fig.~\ref{fig:DUpsilon} displays the derivatives of $\Upsilon_g$.
(\ref{eq:DUpsilonl}), (\ref{eq:DUpsilony}) and (\ref{eq:DUpsilonly}) 
have been plotted at $Y=5.2$, for  $(\ell_1-\ell_2)=0.1$ (left) and
$(\ell_1-\ell_2)=1.0$ (right).
The size and shape of these corrections agree with our expectations 
($\Upsilon_{g,\ell}=\Upsilon_{g,y}={\cal {O}}(\gamma_0^2)$, 
$\Upsilon_{g,\ell\,y}={\cal {O}}(\gamma_0^4)$).

For explicit calculations, we have used 
\begin{eqnarray}\label{eq:DUpsilonl}
\Upsilon_{g,\ell}\!\!\!&\!\!\!=\!\!\!&\!\!\! -\frac{\left[1\!-\!b\left(\psi_{1,\ell}
\!+\!\psi_{2,\ell}\!-\!
{\beta}\gamma_0^2 \right)\right]\left(\Delta_{\ell}\!-\!a\beta^2\gamma_0^4
\right)}{\left(1\!+\!\Delta\!+\!a\beta\gamma_0^2\right)\left[2\!+\!\Delta
  \!-\!b\left(\psi_{1,\ell}
\!+\!\psi_{2,\ell}\!-\!
  \,{\beta}\gamma_0^2 \right)\right]}
 \!-\!\frac{b\left(\psi_{1,\ell\,\ell}
\!+\!\psi_{2,\ell\,\ell}\!+\!
  {\beta^2}\gamma_0^4 \right)}
  {2\!+\!\Delta
  \!-\!b\left(\psi_{1,\ell}
\!+\!\psi_{2,\ell}\!-\!
  {\beta}\gamma_0^2 \right)},\\\nonumber\\
\label{eq:DUpsilony}
\Upsilon_{g,y}\!\!\!&\!\!\!=\!\!\!&\!\!\! -\frac{\left[1\!-\!b\left(\psi_{1,\ell}
\!+\!\psi_{2,\ell}\!-\!
 {\beta}\gamma_0^2\right)\right]\left(\Delta_{y}\!-\!a\beta^2\gamma_0^4
\right)}{\left(1\!+\!\Delta\!+\!a\beta\gamma_0^2\right)\left[2\!+\!\Delta
  \!-\!b\left(\psi_{1,\ell}
\!+\!\psi_{2,\ell}\!-\!
 {\beta}\gamma_0^2 \right)\right]}
 \!-\!\frac{b\left(\psi_{1,\ell\, y}
\!+\!\psi_{2,\ell\, y}\!+\!
 {\beta^2}\gamma_0^4 \right)}
  {2\!+\!\Delta
  \!-\!b\left(\psi_{1,\ell}
\!+\!\psi_{2,\ell}\!-\!
  {\beta}\gamma_0^2 \right)},\\\nonumber\\ 
\label{eq:DUpsilonly}
\Upsilon_{g,\ell\,y}\!\!\!&\!\!\!=\!\!\!&\!\!\!\frac{\partial\Upsilon_{g,y}}{\partial\ell},
\end{eqnarray}
where
\begin{eqnarray}
\Delta_{\ell}\!\!&\!\!=\!\!&\!\!\gamma_0^{-2}\left[\psi_{1,\ell\,\ell}\psi_{2,y}+
\psi_{1,\ell}\psi_{2,y\,\ell}+\psi_{2,\ell\,\ell}\psi_{1,y}+\psi_{2,\ell}
\psi_{1,y\,\ell}\right]+\beta\gamma_0^2\Delta,\cr
\Delta_{y}\!\!&\!\!=\!\!&\!\!\gamma_0^{-2}\left[\psi_{1,\ell\, y}\psi_{2,y}+
\psi_{1,\ell}\psi_{2,y\,y}+\psi_{2,\ell y}\psi_{1,y}+\psi_{2,\ell}
\psi_{1,y\,y}\right]+\beta\gamma_0^2\Delta.
\end{eqnarray}
For the expressions of $\psi_{\ell\,\ell}$, $\psi_{\ell\,y}=\psi_{y\,\ell}$ and 
$\psi_{y\,y}$, the reader is directed to  \ref{subsection:doublederiv}.
(\ref{eq:DUpsilonly})  has been computed numerically (its analytical
expression is too heavy to be easily manipulated).

\subsubsection{$\boldsymbol{\delta_1}$, $\boldsymbol{\delta_2}$,
$\boldsymbol{\delta_c}$}
\label{subsub:deltanum}

$\delta_1$ and $\delta_2$ are defined in
(\ref{eq:nota4bis})(\ref{eq:nota4}). We also define
\begin{equation}
\delta_c=\delta_1+\delta_2+a\Upsilon_{\ell},
\label{eq:deltac}
\end{equation}
which appears in the numerator of the first line of (\ref{eq:CGfull}).

\begin{figure}
\vbox{
\begin{center}
\epsfig{file=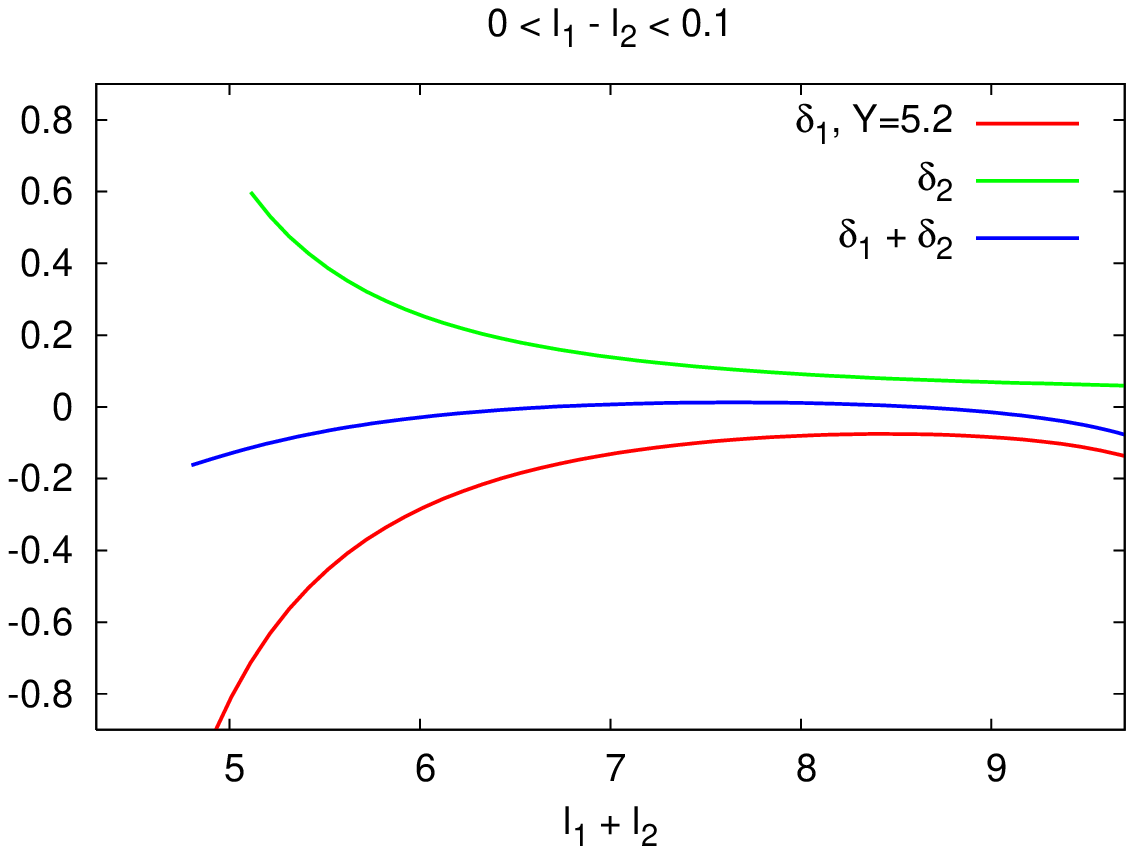, height=6truecm,width=0.48\tw}
\hfill
\epsfig{file=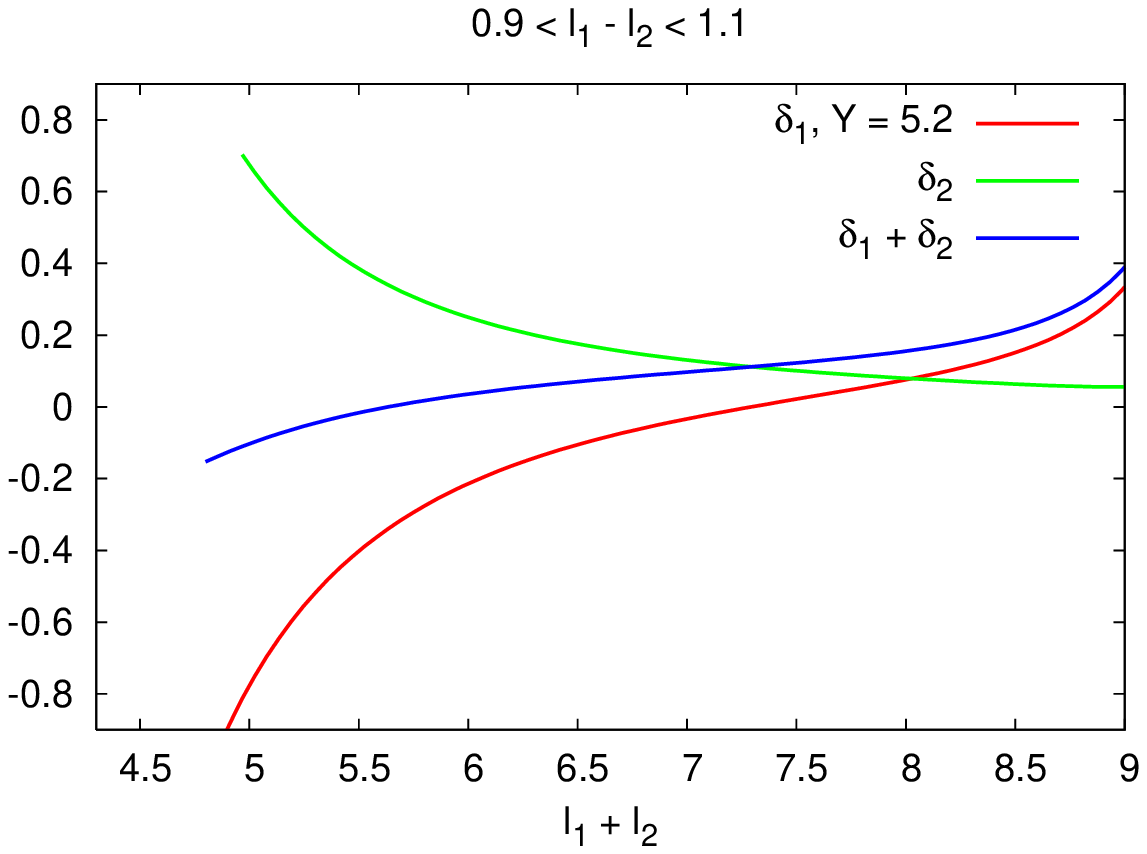, height=6truecm,width=0.48\tw}
\vskip .5cm
\caption{$\delta_1$, $\delta_2$ and $\delta_1+\delta_2$ as functions of
$\ell_1+\ell_2$ for $\ell_1-\ell_2=0.1$ (left) and $\ell_1-\ell_2=1.0$ (right)}
\label{fig:delta12}
\end{center}
}
\end{figure}

Fig.~\ref{fig:delta12} displays the behavior of $\delta_1$, $\delta_2$ and
$\delta_1+\delta_2$ at $Y=5.2$ for $\ell_1-\ell_2 =0.1$ and $\ell_1-\ell_2 = 1.0$.
We recall that these curves can only be reasonably trusted in the interval
(\ref{eq:confint2}).

Though $|\delta_1|={\cal {O}}(\gamma_0)$ (MLLA) should be numerically larger
than $|\delta_2|={\cal {O}}(\gamma_0^2)$ (NMLLA), it turns out that for 
relatively large $\gamma_0\sim0.5$ (Y=5.2), $|\delta_1|\sim|\delta_2|$, and 
 that strong
cancellations occur in their sum. As $\gamma_0$ decreases (or $Y$ increases) $|\delta_1|\gg|\delta_2|$,
in agreement with the perturbative expansion conditions.

\begin{figure}
\vbox{
\begin{center}
\epsfig{file=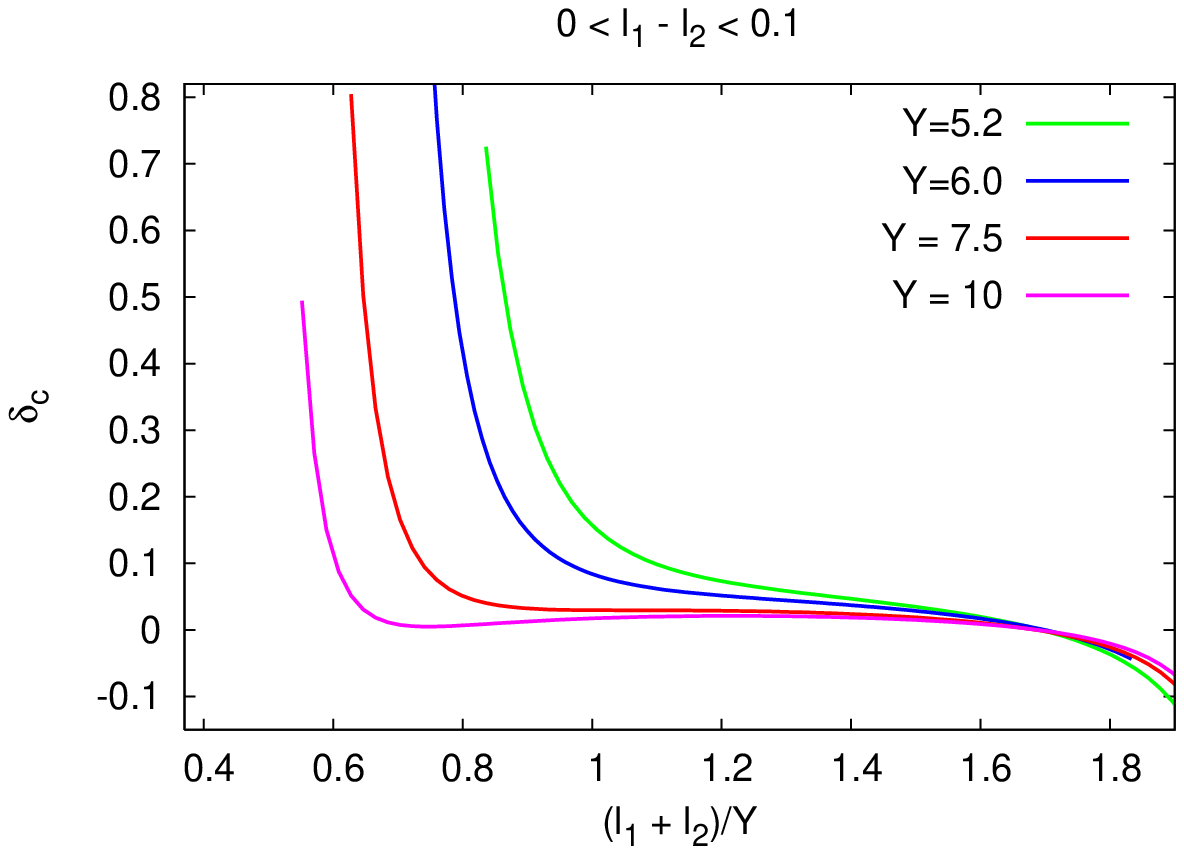, height=6truecm,width=0.48\tw}
\hfill
\epsfig{file=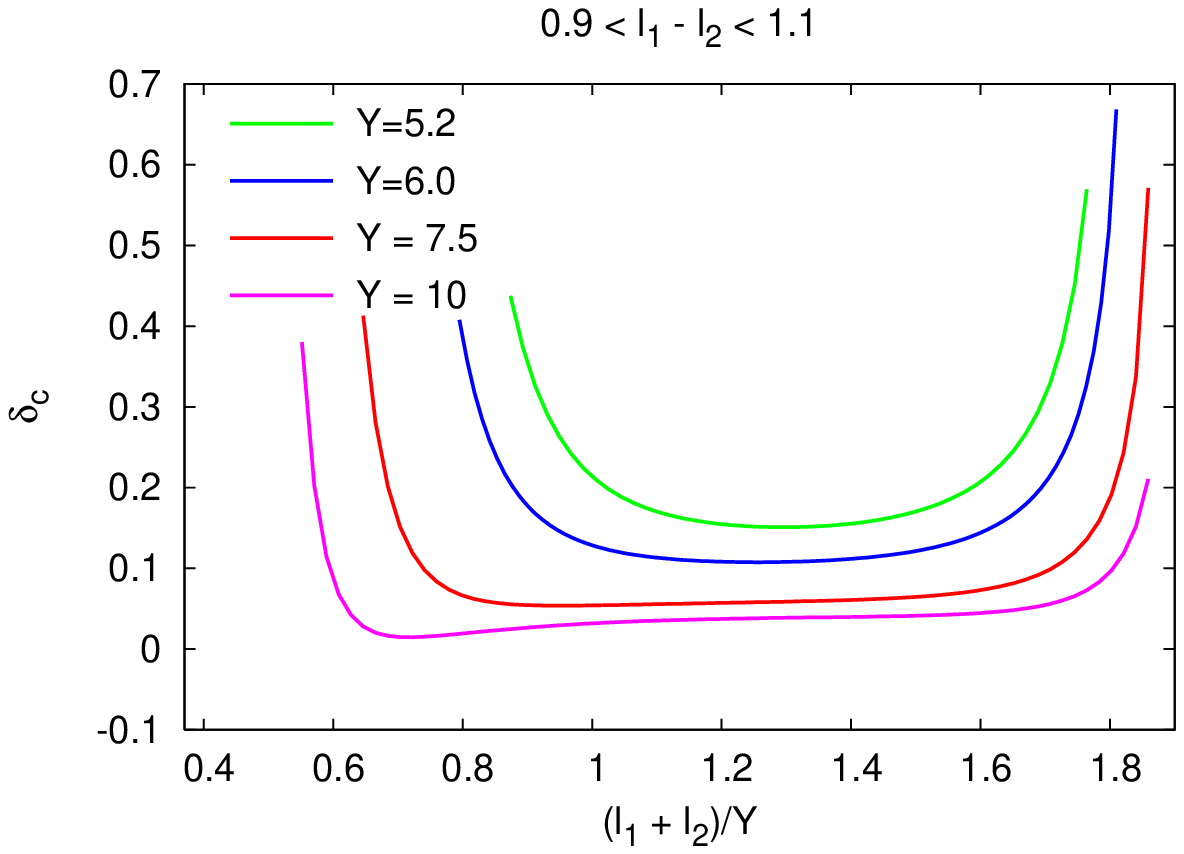, height=6truecm,width=0.48\tw}
\vskip .5cm
\caption{$\delta_c$ as a function of
$(\ell_1+\ell_2)/Y$ for $\ell_1-\ell_2=0.1$ (left) and $\ell_1-\ell_2=1.0$ (right)}
\label{fig:delta12chi}
\end{center}
}
\end{figure}

In Fig.~\ref{fig:delta12chi} we represent $\delta_c$ for different
values of $Y$;  it shows how the sum of corrections (MLLA and NMLLA) stay under
control in the confidence interval (\ref{eq:confint2}).
For $Y=5.2$ one reaches a regime where it
becomes slightly larger than $0.1$ away from the region $x_1\approx x_2$ 
(see upper curve on the right of Fig.~\ref{fig:delta12chi}) but still, since $1$ (which
is the leading term in the numerator of (\ref{eq:CGfull})) $\gg0.1$, our 
approximation can be trusted.

\subsubsection{The global role of corrections in the iterative procedure}
\label{subsub:allgcor}

\begin{figure}
\vbox{
\begin{center}
\epsfig{file=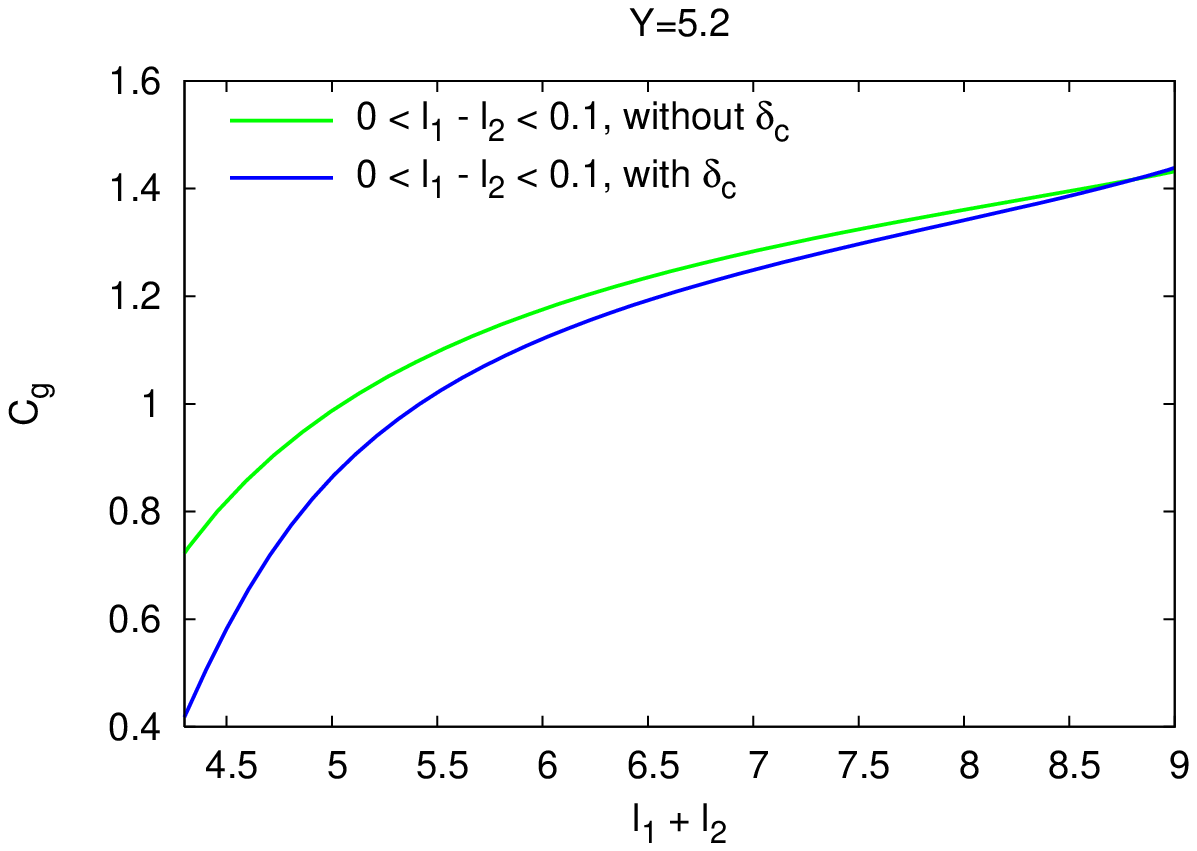, height=6truecm,width=0.48\tw}
\hfill
\epsfig{file=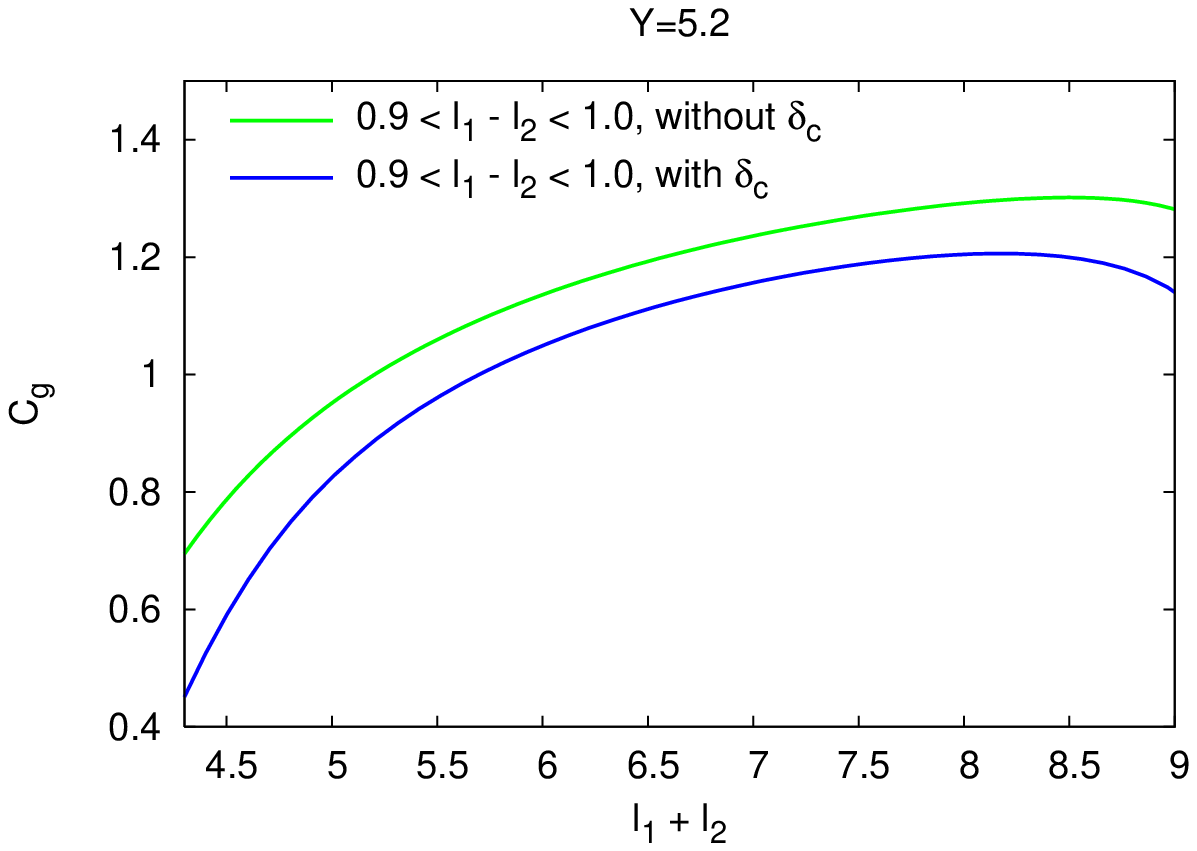, height=6truecm,width=0.48\tw}
\vskip 0.5cm
\epsfig{file=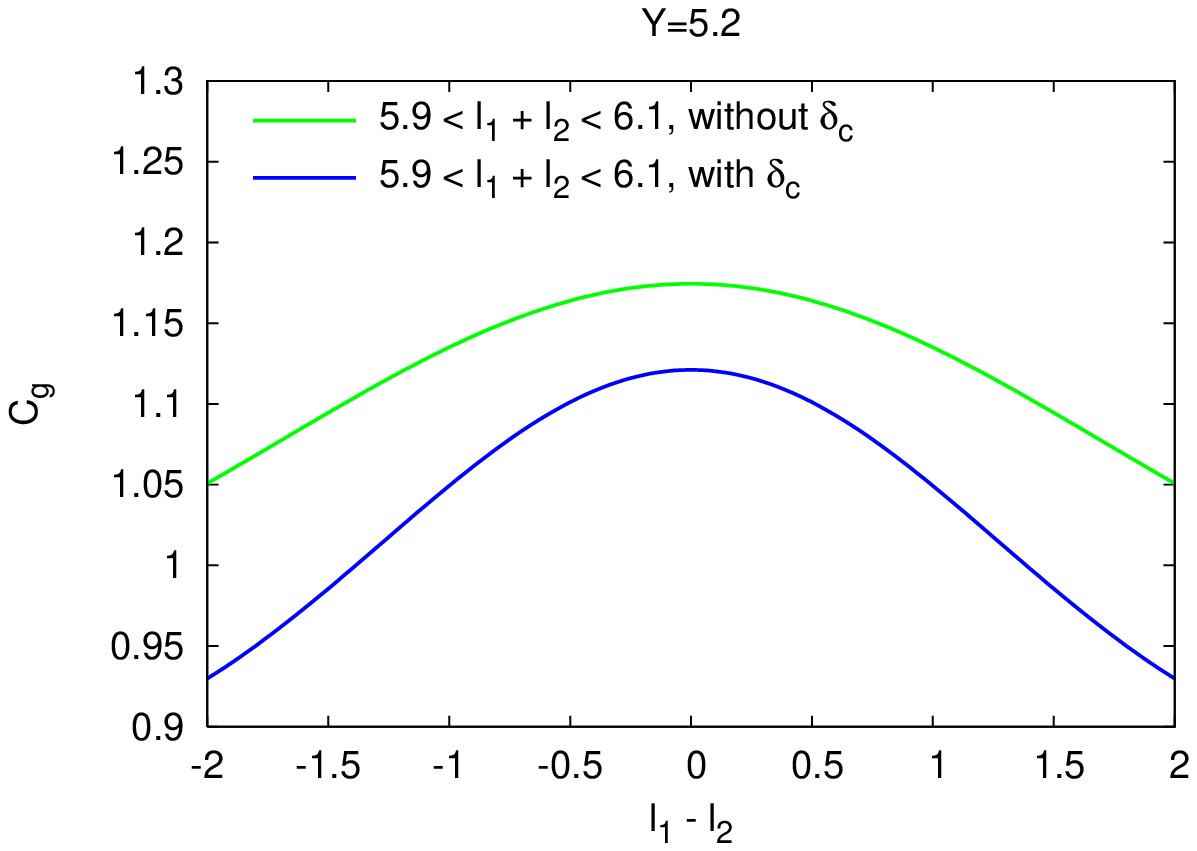, height=6truecm,width=0.48\tw}
\hfill
\epsfig{file=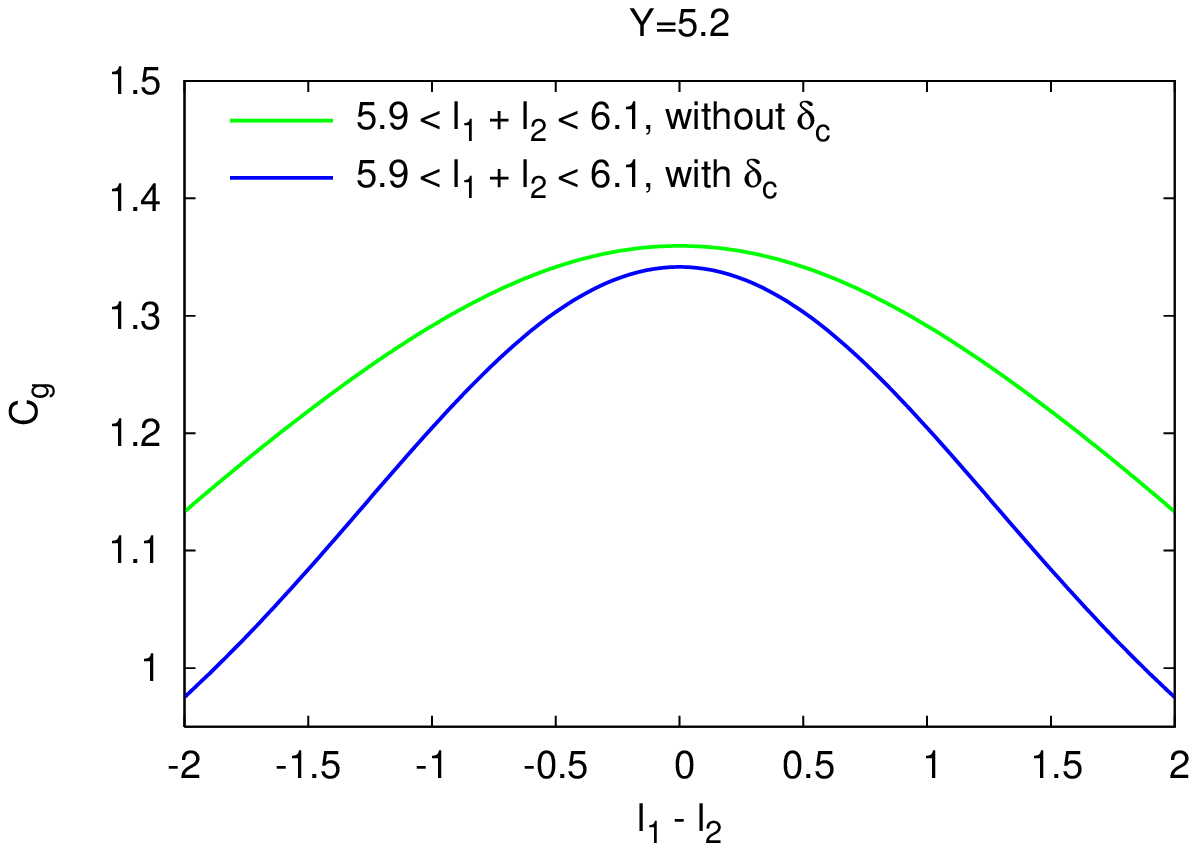, height=6truecm,width=0.48\tw}
\vskip .5cm
\caption{${\cal C}_g$ (blue) compared with $\exp\Upsilon_g$
(green)}
\label{fig:corrUpsilon}
\end{center}
}
\end{figure}

Fig.~\ref{fig:corrUpsilon} shows the role of $\delta_c$ on the 
correlation function: we represent the bare 
function $\exp{\Upsilon_g}$ (see \ref{eq:upsg}) as in Fig.~\ref{fig:Upsilong}, 
together with (\ref{eq:CGfull}). For $(\ell_1-\ell_2)=0.1$
($\ell_1\approx \ell_2$) and $(\ell_1-\ell_2)=1.0$, it is shown how
$\delta_c$ modifies the shape and size of $\exp{\Upsilon_g}$.
When $\ell_1\ne \ell_2$ ($(\ell_1-\ell_2)=1.0$),
$\delta_c$ decreases the correlations. 
They are also represented as a function of $(\ell_1-\ell_2)$
when $(\ell_1+\ell_2)$ is fixed ( to $6.0$ and $7.0$). The 
increase of $\delta_c$ 
as one goes away from the diagonal
$\ell_1\!\approx\! \ell_2$ (see Fig.~\ref{fig:delta12chi} for
$(\ell_1-\ell_2)=1.0$)
explain the difference between the green and blue curves; this substantially modifies
the tail of the correlations. 

When $Y$ gets larger, the role of $\delta_c$
decreases: at $Y=7.5$ (LHC conditions) the difference between the two curves becomes negligible.

\subsection{Quark jet}
\label{subsection:quacorr}

\subsubsection{$\boldsymbol\varphi$ and its derivatives}
\label{subsub:phinum}

\begin{figure}
\vbox{
\begin{center}
\epsfig{file=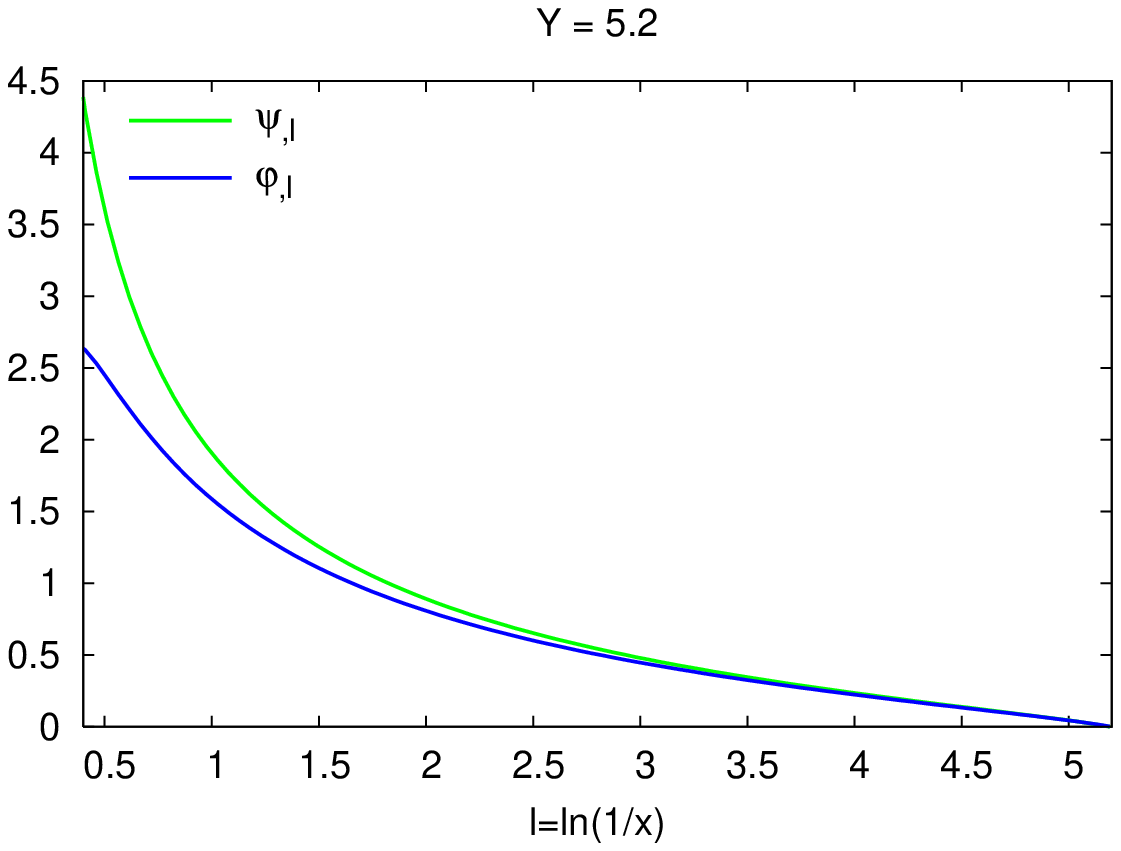, height=6truecm,width=0.45\tw}
\hfill
\epsfig{file=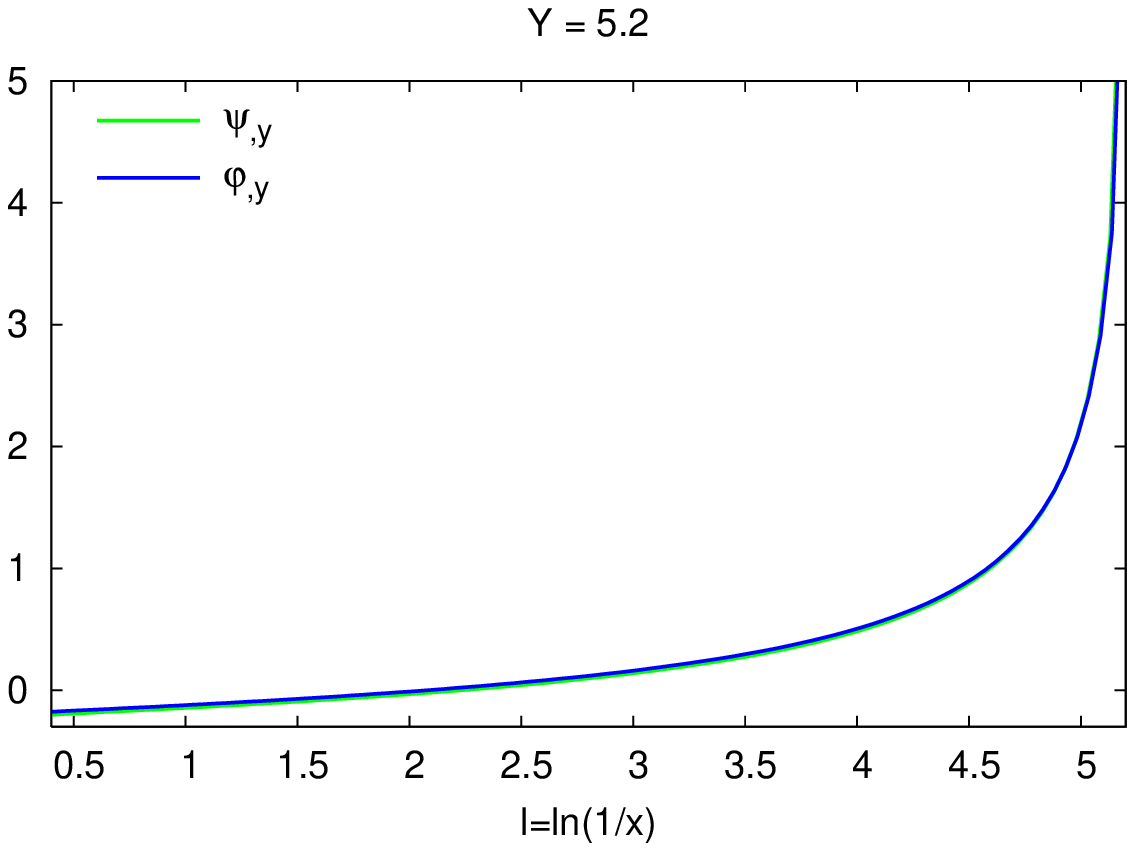, height=6truecm,width=0.45\tw}
\vskip .5cm
\caption{Derivatives $\varphi_{g,\ell}$ and $\varphi_{gy}$ as functions of $\ell$ at fixed 
$Y=5.2$ (left), compared with $\psi_{\ell}$ and $\psi_{y}$}
\label{fig:psiqly}
\end{center}
}
\end{figure}

Fig.~\ref{fig:psiqly} displays the derivatives $\varphi_{\ell}$ and $\varphi_{y}$
together with those $\psi_{\ell}$ and $\psi_{y}$ for the gluon jet, at $Y=5.2$. 
There sizes and shapes are the 
same since the logarithmic derivatives of the single inclusive 
distributions inside a gluon or a quark jet only depend on their shapes (the 
normalizations cancel in the ratio), 
which is the same in both cases. The mismatch at small $\ell$ between
$\varphi_{\ell}$ and $\psi_{\ell}$ stems from the behavior of 
$\psi_{\ell\,\ell}$ $\psi_{\ell\,\ell}\stackrel{\ell\to0}{\longrightarrow}-\infty$. 
Therefore, in the interval of applicability of the soft approximation (\ref{eq:varphil}) and (\ref{eq:varphiy2})
can be approximated by $\psi_{\ell}$ and $\psi_{y}$ respectively.

\subsubsection{$\boldsymbol{\tilde\Delta(\ell_1,\ell_2,Y)}$}
\label{subsub:tildeDelta}

The last statement in \ref{subsub:phinum} numerically supports
the approximation (\ref{eq:Deltatilde}), that is

$$
\tilde\Delta\approx\Delta+{\cal O}(\gamma_0^2).
$$

We get rid of the heavy ${\cal O}(\gamma_0^2)$ factor in (\ref{eq:Deltatilde})
to ease our numerical calculations. Hence, the behavior of $\tilde\Delta$ is
already given in Fig.~\ref{fig:Delta}.

\subsubsection{$\boldsymbol{\Upsilon_q}$ and its derivatives}
\label{subsub:upsqnum}

\begin{figure}
\vbox{
\begin{center}
\epsfig{file=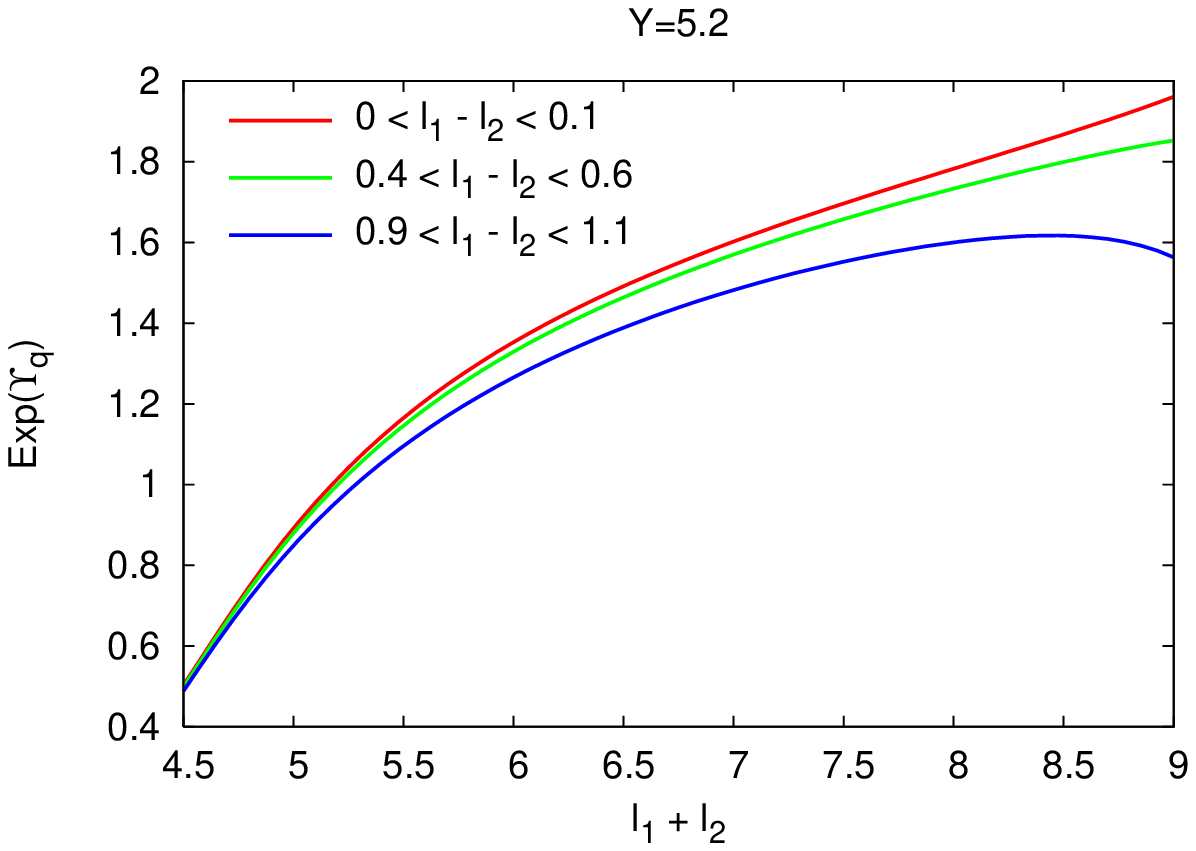, height=6truecm,width=0.47\tw}
\epsfig{file=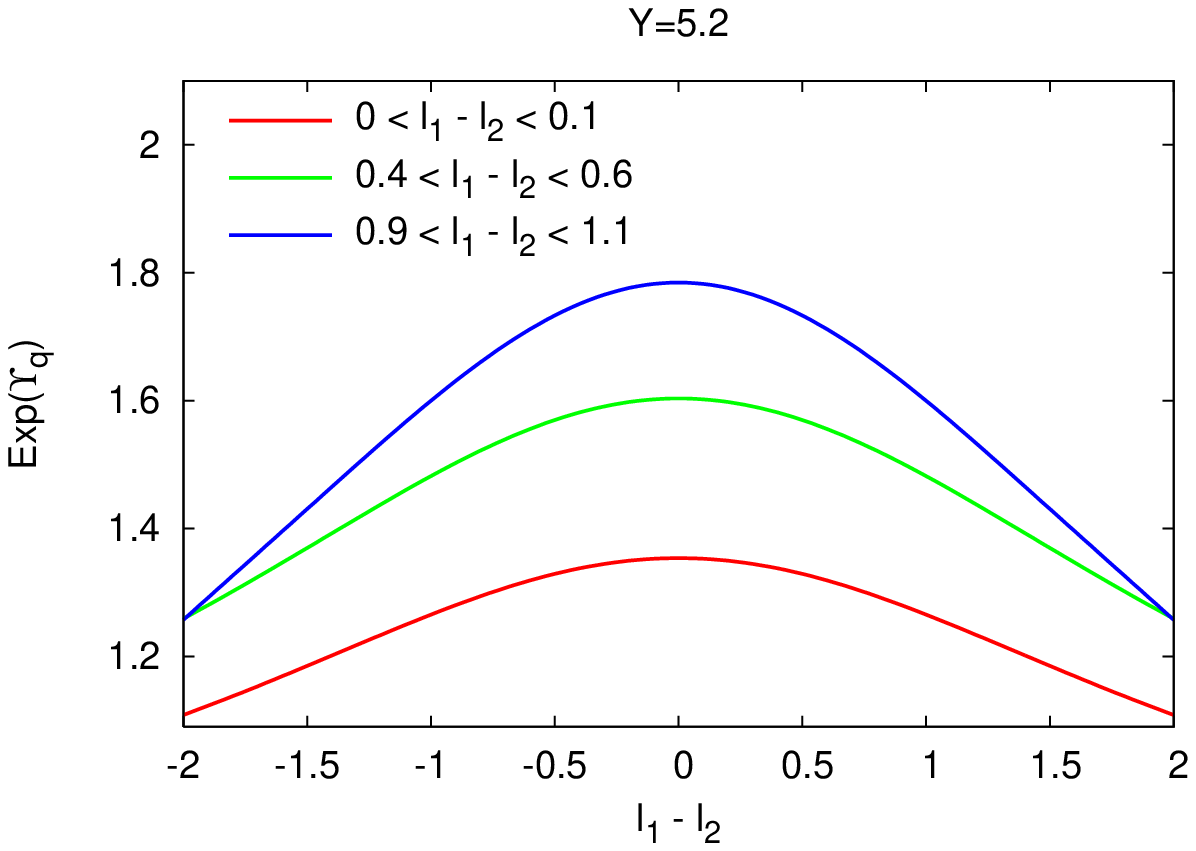, height=6truecm,width=0.47\tw}
\vskip .5cm
\caption{$\exp{(\Upsilon_q)}$ as a function of $\ell_1+\ell_2$ (left) and, 
$\ell_1-\ell_2$ (right) for $Y=5.2$}
\label{fig:Upsilonq}
\end{center}
}
\end{figure}

\begin{figure}
\vbox{
\begin{center}
\epsfig{file=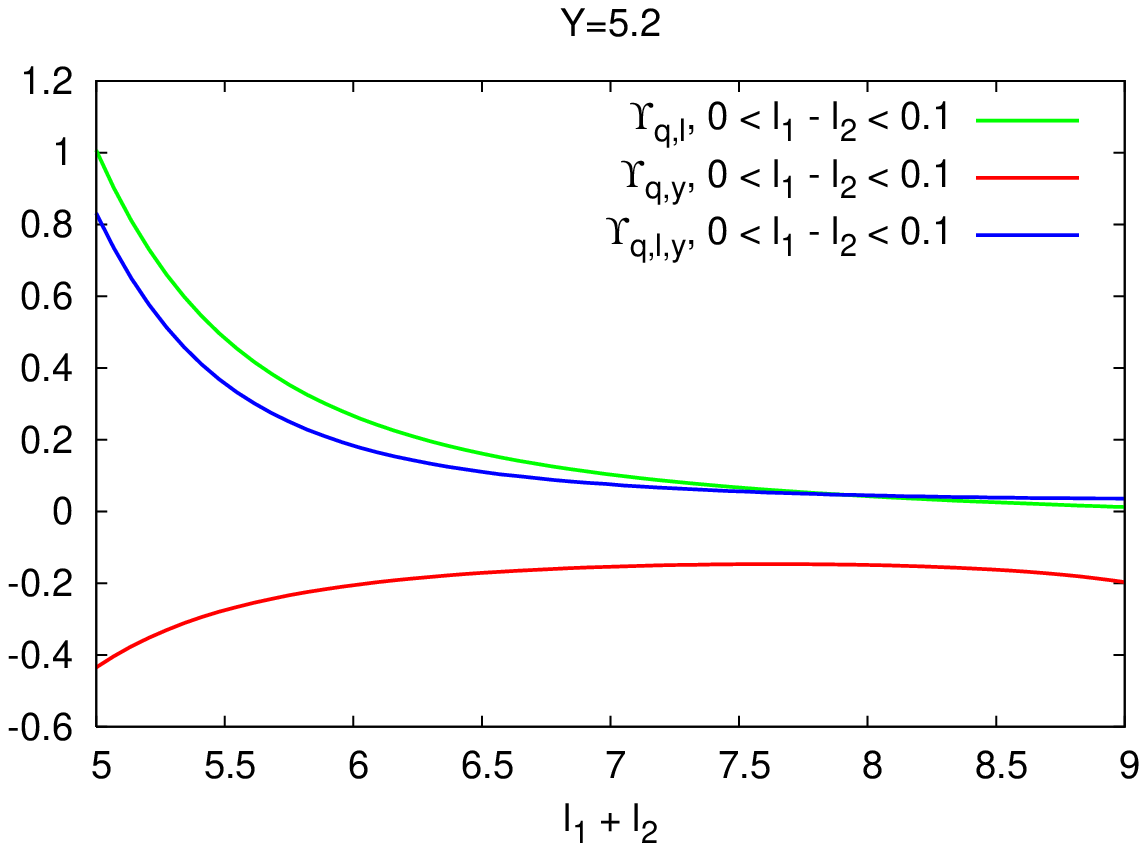, height=6truecm,width=0.48\tw}
\hfill
\epsfig{file=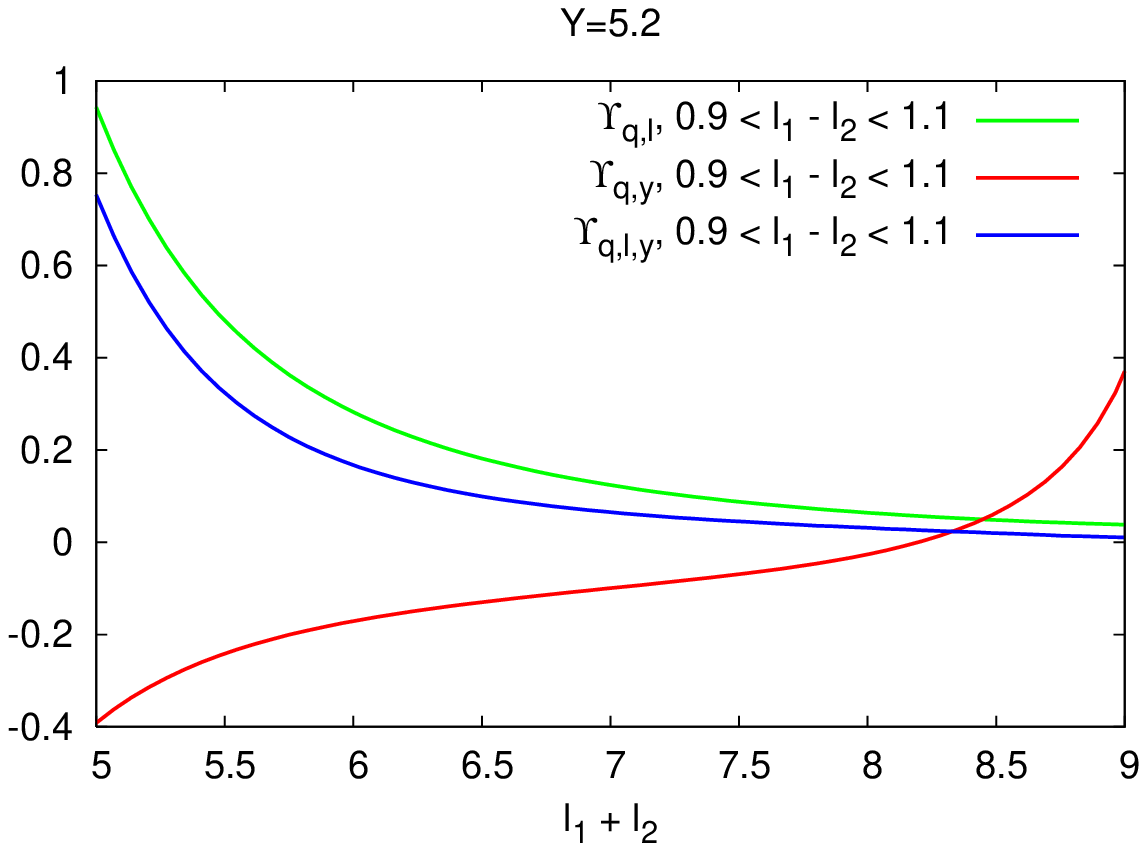, height=6truecm,width=0.48\tw}
\vskip .5cm
\caption{Corrections $\Upsilon_{q,\ell}$, $\Upsilon_{q,y}$ and 
$\Upsilon_{q,\ell,y}$ as functions of $\ell_1+\ell_2$ for $\ell_1-\ell_2=0.1$ (left) 
and $\ell_1-\ell_2=1.0$ (right) at $Y=5.2$} 
\label{fig:derUpsilonq}
\end{center}
}
\end{figure}

The smooth behavior of $\exp\Upsilon_q$ is displayed in Fig.~\ref{fig:Upsilonq}
as a function of the sum $(\ell_1+\ell_2)$ for fixed $(\ell_1-\ell_2)$ and vice versa.
The normalization of $(\exp\Upsilon_q-1)$ is roughly twice larger
($\times\frac{Nc}{CF}\approx2$) than that of $(\exp\Upsilon_g-1)$.
We  then consider derivatives of
this expression to get the corresponding iterative corrections shown in 
Fig.~\ref{fig:derUpsilonq}. The behavior of $\Upsilon_{q,\ell} ({\cal O}(\gamma_0^2))$, $\Upsilon_{q,y} ({\cal O}(\gamma_0^2))$
and $\Upsilon_{q,\ell\,y} ({\cal O}(\gamma_0^4))$ is in good agreement with 
our expectations as far as the order of magnitude and the normalization are concerned
(see also Fig.~\ref{fig:DUpsilon})
\footnote{it is also important to remark that $\Upsilon_{q,\ell},\,
\Upsilon_{q,\ell},\,\Upsilon_{q,\ell\,y}$ are $\times\frac{N_c}{C_F}\Upsilon_{g,\ell},\,
\Upsilon_{g,\ell},\,\Upsilon_{g,\ell\,y}$ .}.

\subsubsection{$\boldsymbol{\tilde{\delta}_1}$,
$\boldsymbol{\tilde{\delta}_2}$ and $\boldsymbol{\tilde{\delta}_c}$}
\label{subsub:tidelq}

\begin{figure}
\vbox{
\begin{center}
\epsfig{file=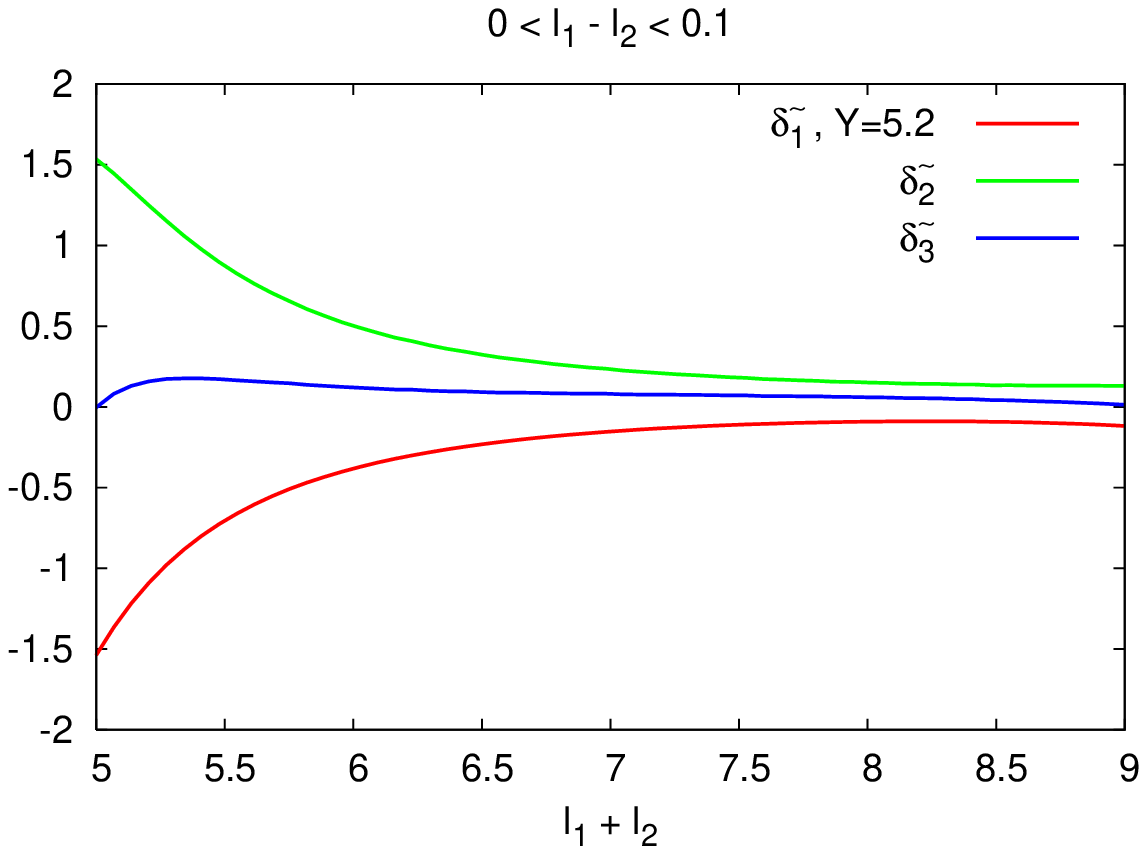, height=6truecm,width=0.48\tw}
\hfill
\epsfig{file=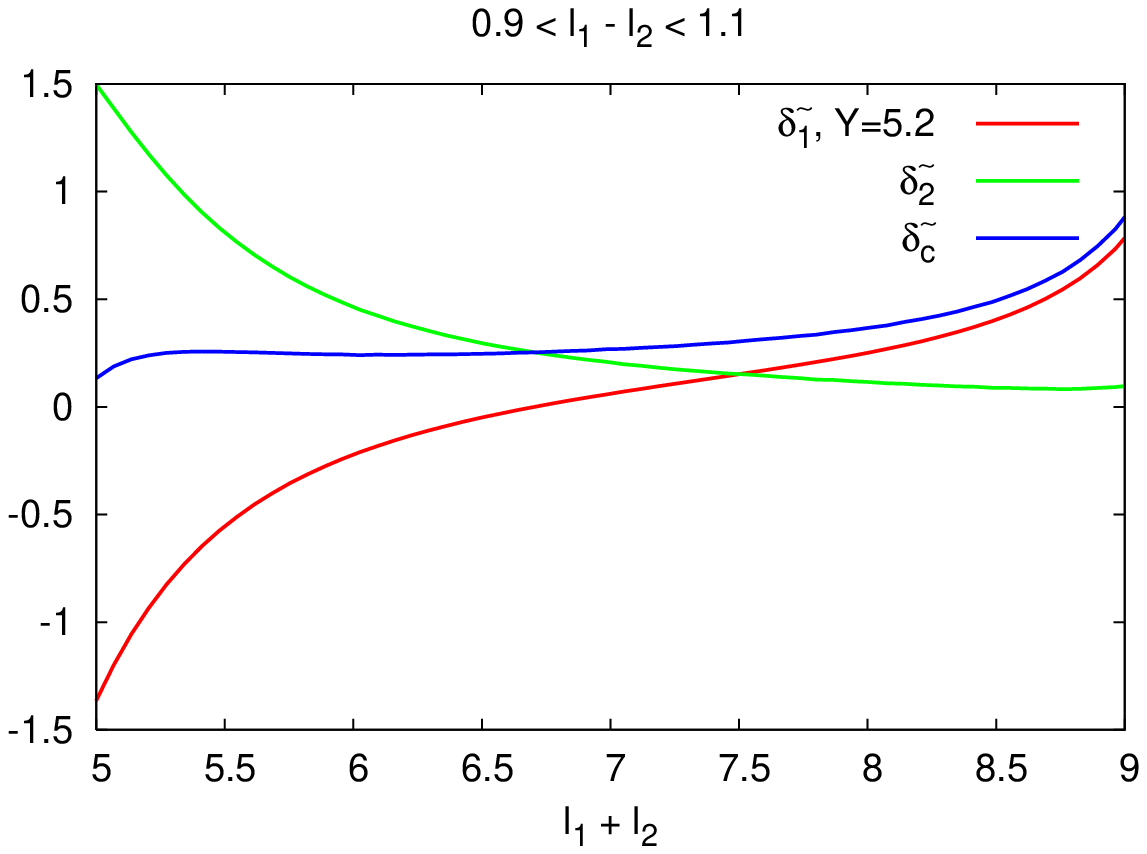, height=6truecm,width=0.48\tw}
\vskip .5cm
\caption{$\tilde\delta_1$, $\tilde\delta_2$ and $\tilde\delta_1+\tilde\delta_2$
as functions of $\ell_1+\ell_2$ for $\ell_1-\ell_2=0.1$ (left) and 
$\ell_1-\ell_2=1.0$ (right) at $Y=5.2$} 
\label{fig:tildedelta}
\end{center}
}
\end{figure}

We define
$$
\tilde{\delta}_{c}=\tilde{\delta}_1 + \tilde{\delta}_2
$$
as it appears in both the numerator and denominator of (\ref{eq:Qcorr}). 
In Fig.~\ref{fig:tildedelta} are displayed $\tilde\delta_1$, $\tilde\delta_2$
and their sum $\tilde\delta_c$ as functions of the sum
$(\ell_1+\ell_2)$ at fixed $(\ell_1-\ell_2)$
($\ell_1-\ell_2=0.1$, left) ($\ell_1-\ell_2=1.0$, right).

At $Y=5.2$, which corresponds to $\gamma_0 \approx 0.5$, the relative
magnitude of $\tilde\delta_1$ and $\tilde\delta_2$ is inverted
\footnote{it has been numerically investigated that the expected relative order of magnitude
of $\tilde\delta_1$ and $\tilde\delta_2$ is recovered for $Y\geq8.0$ (this value can be
eventually reached at LHC).} with respect to what is expected from respectively MLLA and NMLLA corrections
(see subsection \ref{subsection:estimate}). This is the only hint that, at
this energy, the expansion should be pushed to include all NMLLA
corrections to be reliable.

\begin{figure}
\vbox{
\begin{center}
\epsfig{file=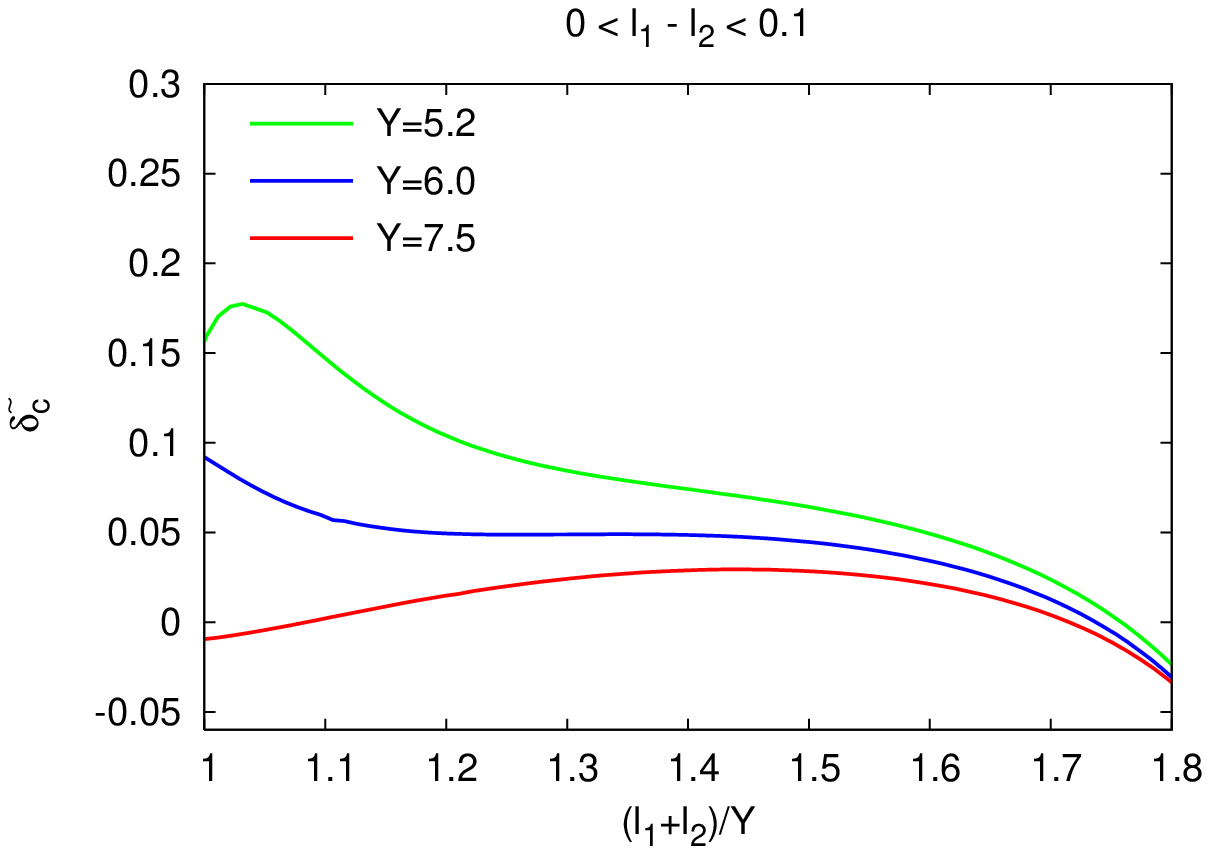, height=6truecm,width=0.48\tw}
\hfill
\epsfig{file=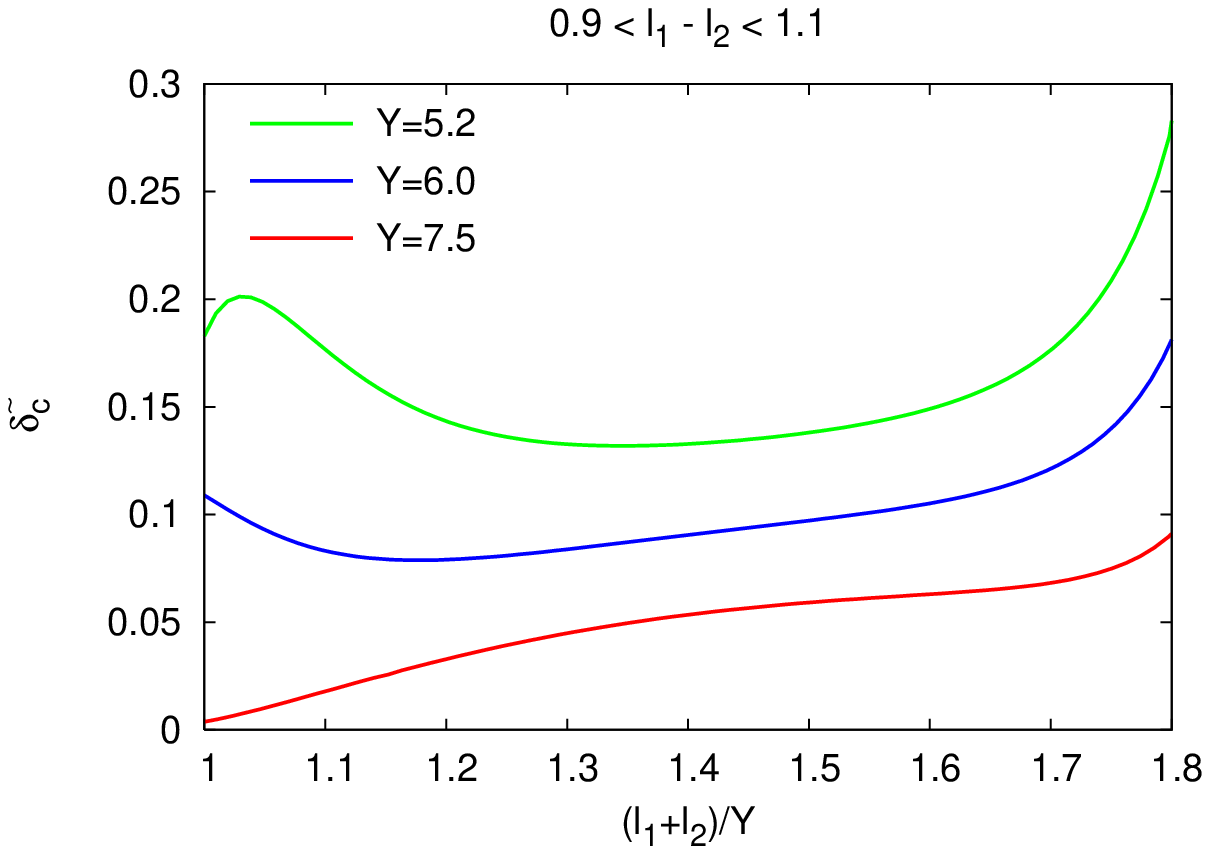, height=6truecm,width=0.48\tw}
\vskip .5cm
\caption{$\tilde\delta_c$ as a function of
$(\ell_1+\ell_2)/Y$ for $\ell_1-\ell_2=0.1$ (left) and $\ell_1-\ell_2=1.0$ (right)
at $Y=5.2,\,6.0,\,7.5$}
\label{fig:delta12chiQ}
\end{center}
}
\end{figure}

Large cancellations are, like for gluons, seen to occur in
$\tilde\delta_c$, making the sum of corrections quite small.
In order to study the behavior of $\tilde\delta_c$ as $Y$
increases, it is enough to look at Fig.~\ref{fig:delta12chiQ} where
we compare $\tilde\delta_c$ at $Y=5.2,\,6.0,\,7.5$.

\subsubsection{Global role of corrections in the iterative procedure}

\begin{figure}
\vbox{
\begin{center}
\epsfig{file=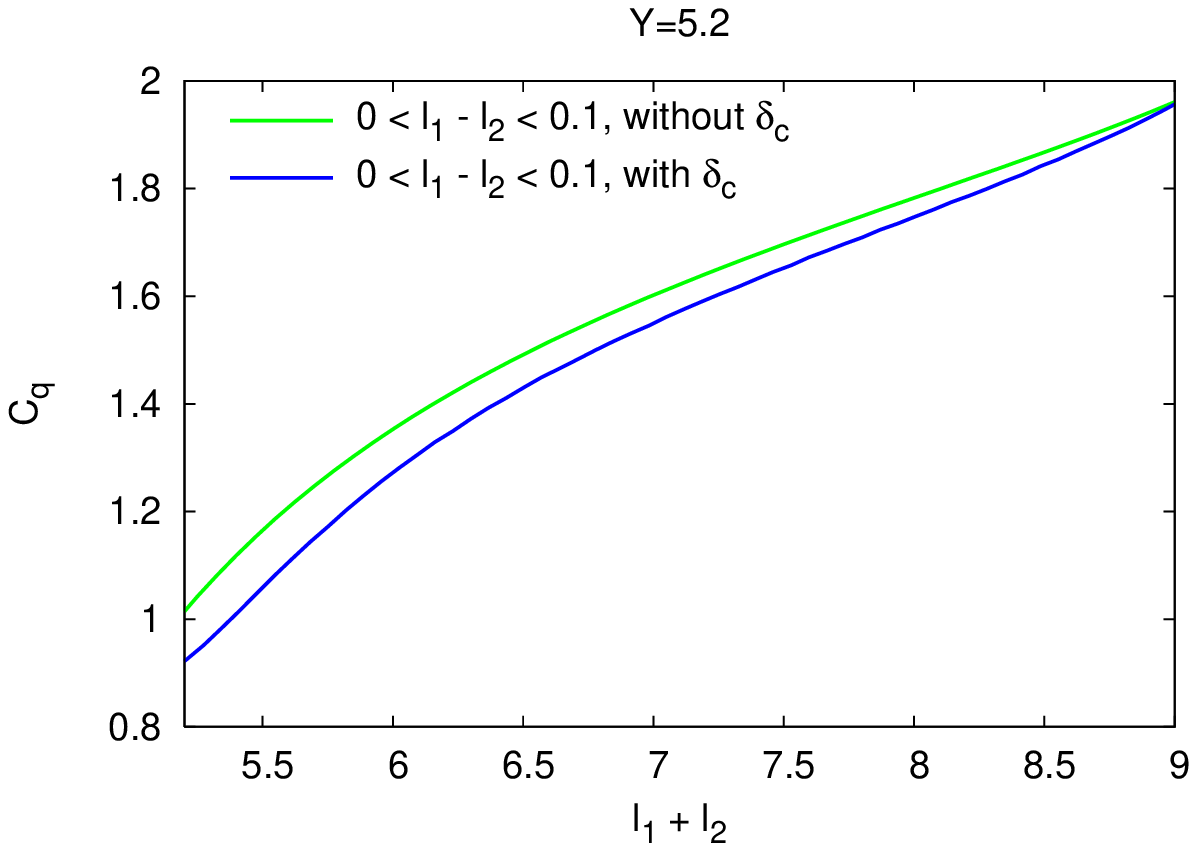, height=6truecm,width=0.48\tw}
\hfill
\epsfig{file=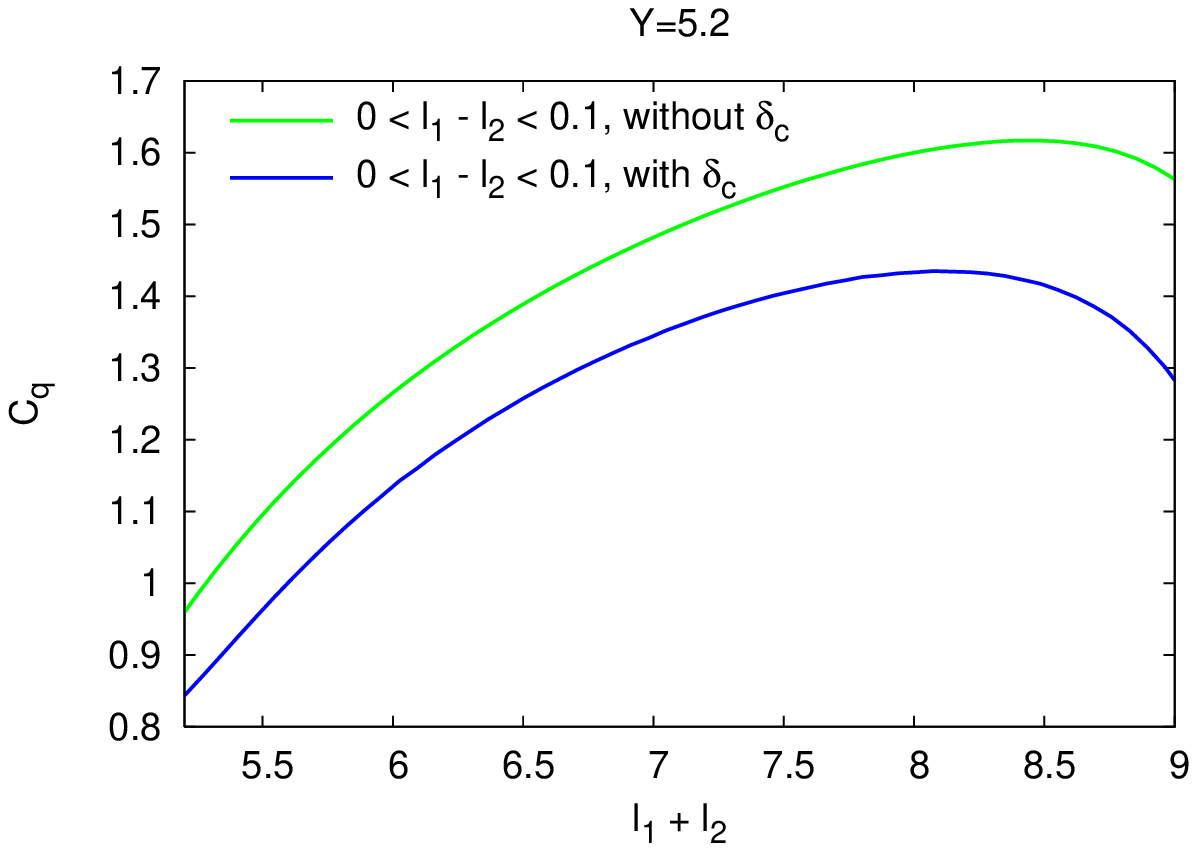, height=6truecm,width=0.48\tw}
\vskip 0.5cm
\epsfig{file=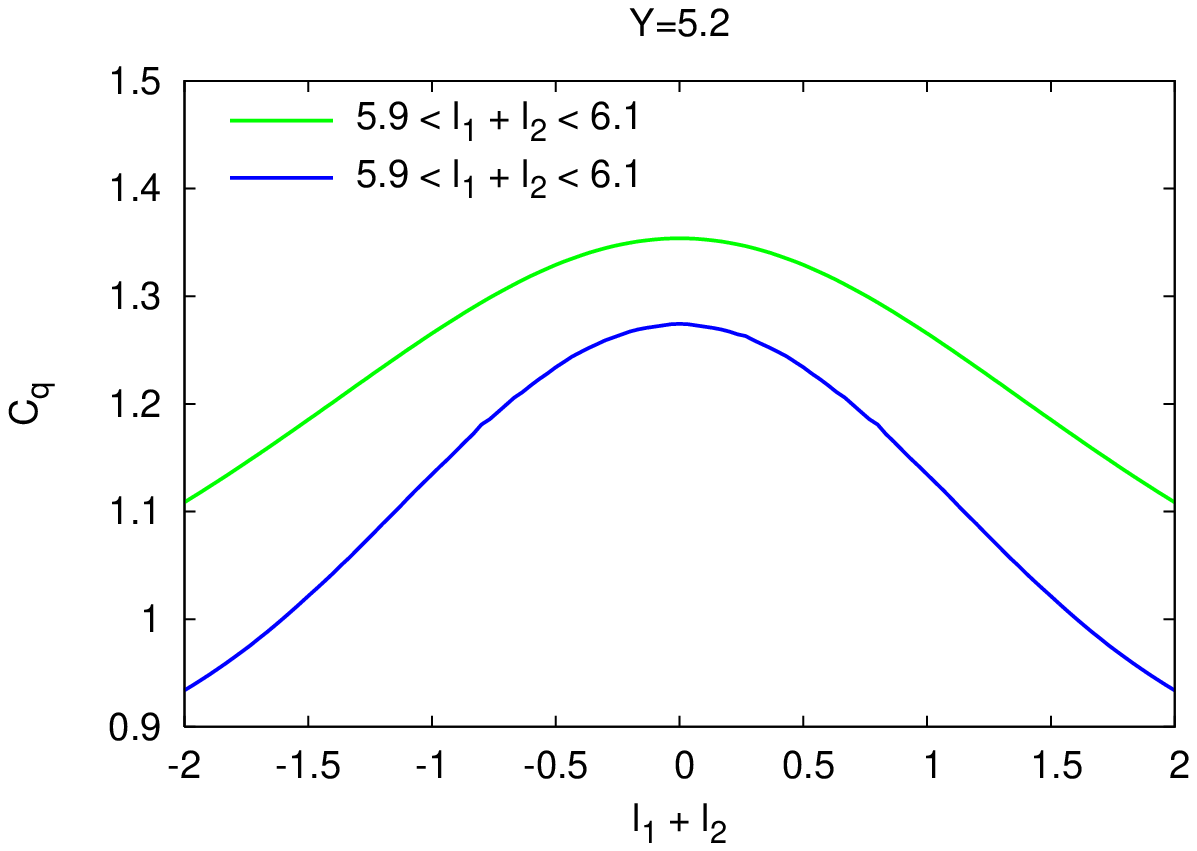, height=6truecm,width=0.48\tw}
\hfill
\epsfig{file=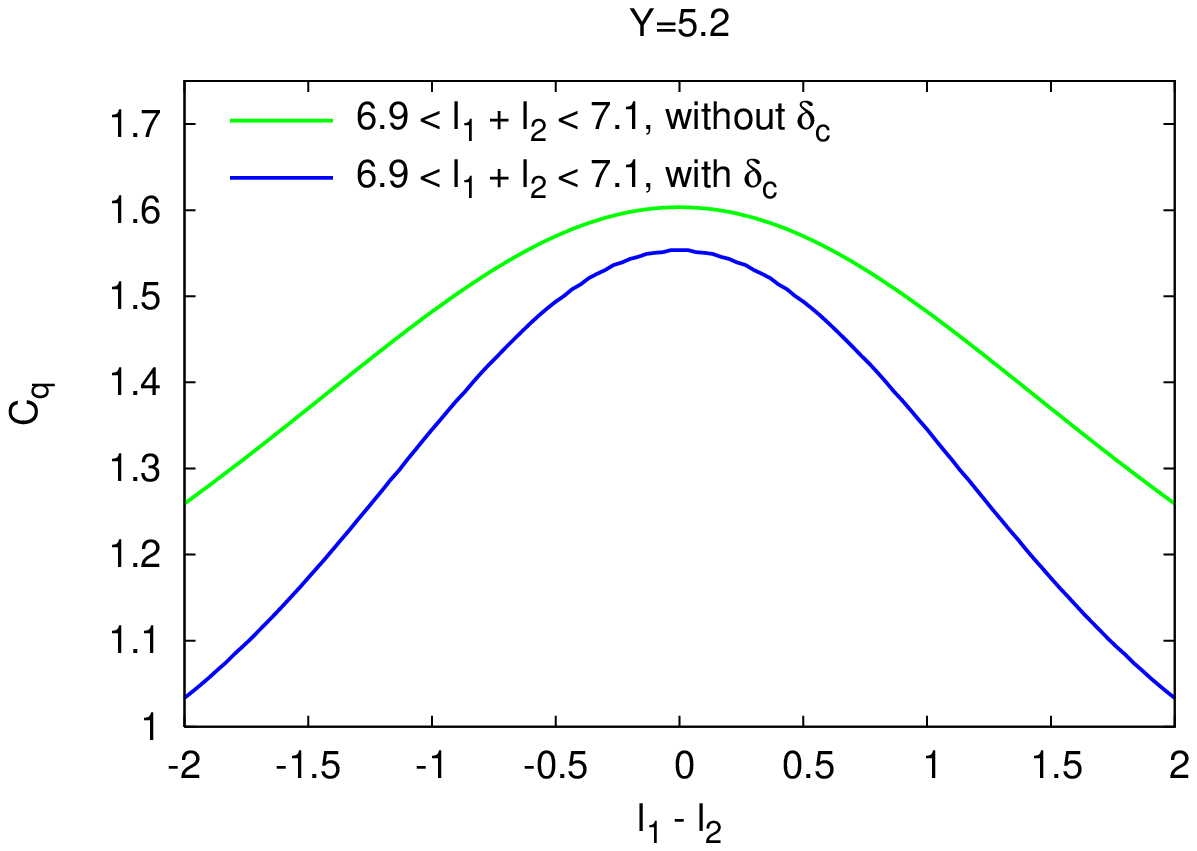, height=6truecm,width=0.48\tw}
\vskip .5cm
\caption{${\cal C}_q$ (blue) compared with $\exp\Upsilon_q$ 
(green)}
\label{fig:corrqUpsilon}
\end{center}
}
\end{figure}

It is displayed in Fig.~\ref{fig:corrqUpsilon}.
${\tilde\delta_c}$ does not affect $\exp\Upsilon_q$ near the main diagonal ($\ell_1=\ell_2$), but it 
does far from it.  We find the same behavior as in the case of a gluon jet.

\section{COMPARING DLA AND MLLA CORRELATIONS}
\label{section:DLAcomp}

\begin{figure}
\vbox{
\begin{center}
\epsfig{file=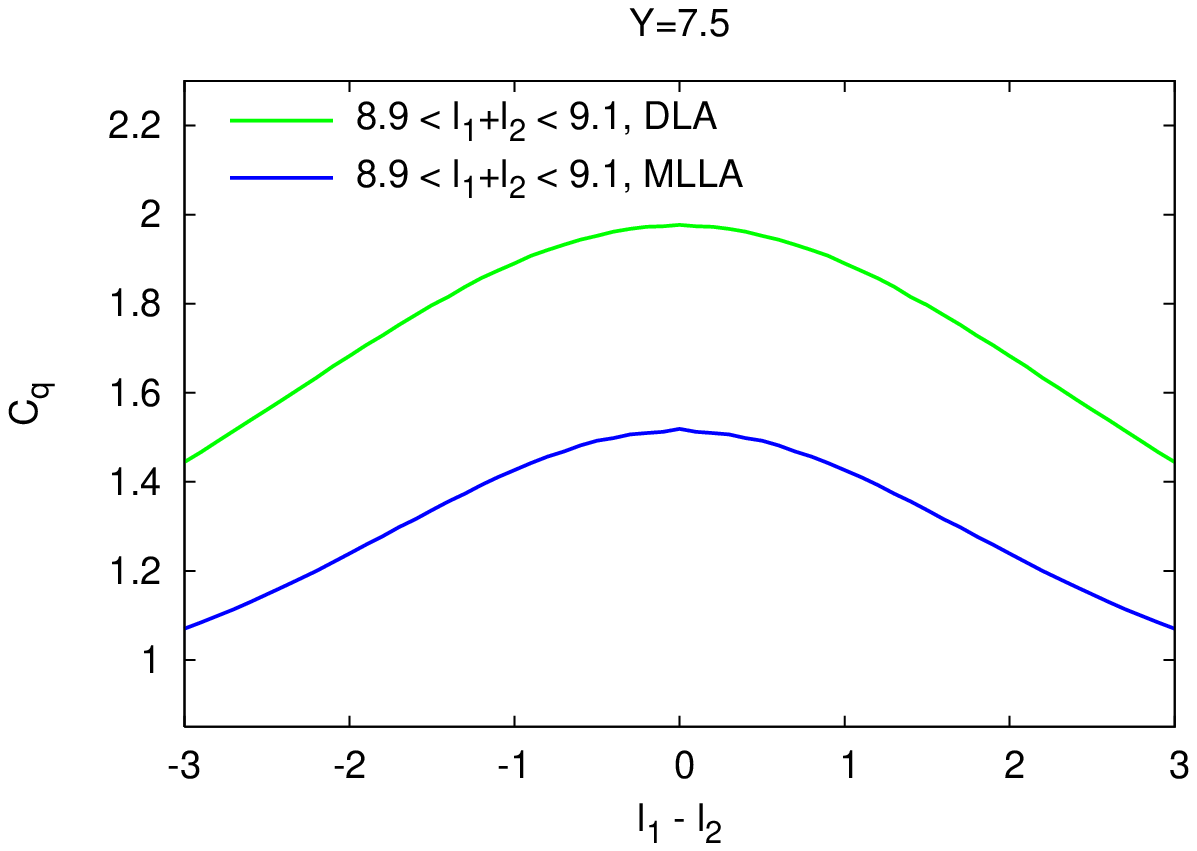, height=6truecm,width=0.48\tw}
\hfill
\epsfig{file=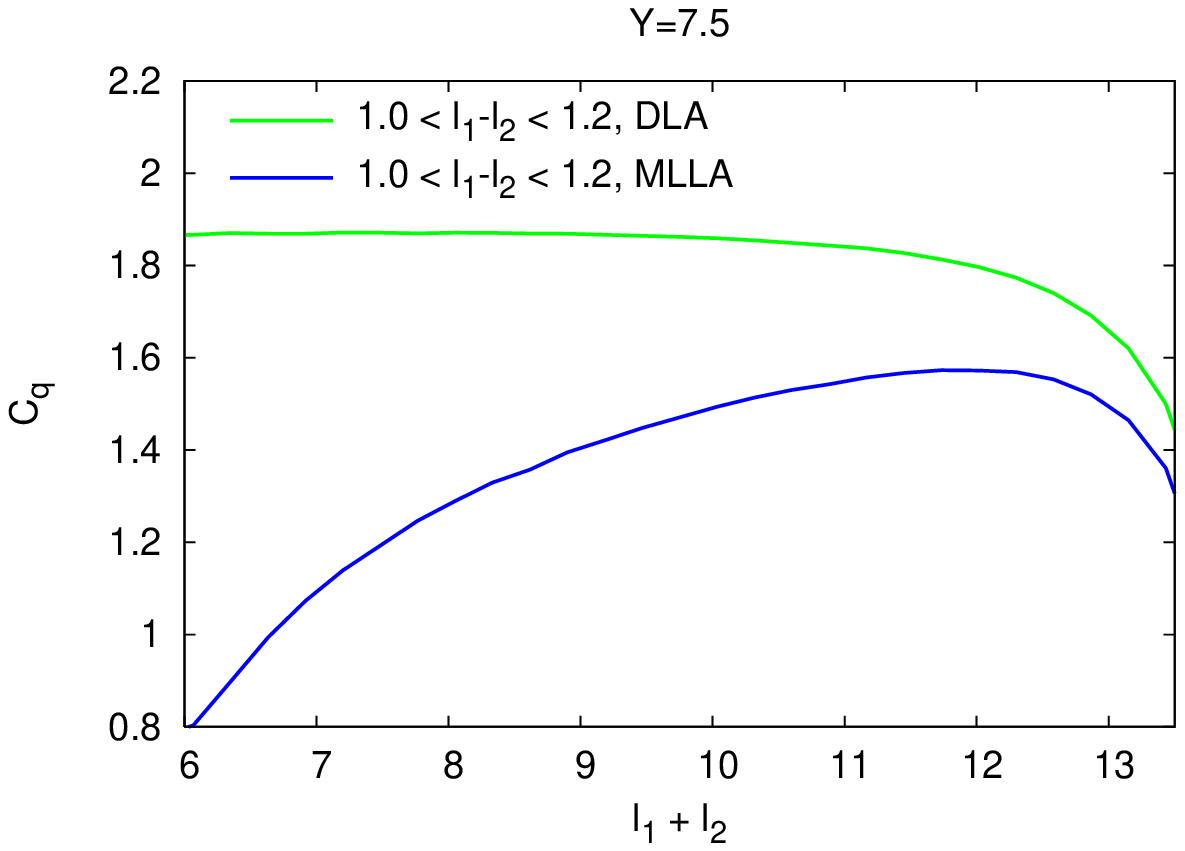, height=6truecm,width=0.48\tw}
\vskip .5cm
\caption{Comparing DLA and MLLA correlations}
\label{fig:DLAMLLA}
\end{center}
}
\end{figure}

In Fig.~\ref{fig:DLAMLLA} we compare the quark correlator at DLA and MLLA.
The large gap between the two curves accounts for the energy balance
that is partially restored in MLLA
by introducing hard corrections in the partonic evolution equations
(terms $\propto$ $a$, $b$ and $\frac34$);
the DLA curve is obtained by setting $a$, $b$ and $\frac34$
to zero in (\ref{eq:CGfull}) and (\ref{eq:Qcorr});
${\cal C}_q$ is a practically constant function of
$\ell_1+\ell_2$ in almost the whole range, and decreases
when $\ell_1+\ell_2\rightarrow 2Y$ by the running of $\alpha_s$.
The MLLA increase of ${\cal C}_q$ with $\ell_1+\ell_2$ follows from energy
conservation.
Similar results are obtained for ${\cal C}_g$.

\null\newpage

\null

\listoffigures

\newpage


\end{document}